\documentclass[structabstract]{aa}
\usepackage{graphicx}
\usepackage[varg]{txfonts}
\usepackage{enumitem}
\usepackage{natbib}   
\bibpunct{(}{)}{;}{a}{}{,}  
\usepackage{amsmath,amssymb}
\usepackage{mathtools}
\usepackage{lscape}
\usepackage{longtable}
\usepackage{enumerate}
\usepackage[usenames,dvipsnames]{color}
\usepackage{hyperref}

\usepackage{float,morefloats} 
\usepackage[above]{placeins}

\begin{document}

\newcommand{\arcm}{$^\prime$}
\newcommand{\arcs}{$^{\prime\prime}$}
\newcommand{\m}{$^{\rm m}\!\!.$}
\newcommand{\D}{$^{\rm d}\!\!.$}
\newcommand{\F}{$^{\rm P}\!\!.$}
\newcommand{\ms}{M$_{\odot}$}
\newcommand{\rs}{R$_{\odot}$}
\newcommand{\ond}{Ond\v{r}ejov}
\newcommand{\kms}{km~s$^{-1}$}
\newcommand{\dvam}{$2 \text{-} \rm{m}$}


\newcommand{\pozn}[1]{\small{\color{red} #1 \color{black}}\normalsize} 
\newcommand{\modre}[1]{{\bf \color{blue} #1 \color{black}}} 
\newcommand{\pozna}[1]{{\bf \color{red} #1 \color{black}}} 
\newcommand{\bude}[1]{\textbf{\textit{\color{blue} #1 \color{black}}}} 
\title{ Time-dependent spectral-feature variations of stars displaying the~B[e] phenomenon 
  \thanks{Based on data from Perek 2~m \, telescope, Ond\v{r}ejov, Czech Republic.}
  \thanks{
   Figs. 
    \ref{mirroring}--\ref{nton}, \ref{SPAHaIIc}, \ref{HaF}, \ref{Cor_FHa}--\ref{Cor_RVEWHa}, \ref{HbF}--\ref{Hb_Ha_corr},
    \ref{RVOI6364}, \ref{O_Ha_corr_fig}, \ref{Fe_col}, \ref{EW_Ha_periods}, \ref{model_disc}
   and Tables
   \ref{observing_log} -- \ref{O_Ha_corr_tab} 
   are available in electronic form at www.aanda.org.
   }
  }

\subtitle{III. HD 50138}
\author{     T. Je\v{r}\'abkov\'a \inst{1,4}
        \and D. Kor\v{c}\'akov\'a \inst{1}
        \and A. Miroshnichenko \inst{2}
	\and S. Danford \inst{2}
	\and S. V. Zharikov \inst {3}
 	\and R. K\v{r}\'{\i}\v{c}ek \inst{1}
        \and P.~Zasche \inst{1}
	\and V.~Votruba \inst{4,5}
	\and M.~\v{S}lechta \inst{4}
        \and P.~\v{S}koda \inst{4}
        \and J.~Jan\'{\i}k\inst{5}
      }
        \institute{
             Astronomical Institute, Charles University in Prague, V 
             Hole\v{s}ovi\v{c}k\'ach 2, CZ-180 00 Praha 8, Czech Republic
        \and 
		Department of Physics and Astronomy, University of North Carolina 
        at Greensboro, Greensboro, NC 27402, USA 
        \and 
			Instituto de Astronom\'ia, Universidad Nacional Aut\'onoma de 
		M\'exico, Apartado Postal 877, 22830, Ensenada, Baja California,
		M\'exico
		\and 
		Astronomical Institute of the~Academy of Science of the~Czech 
        Republic, Fri\v{c}ova 298, CZ-251 65 Ond\v{r}ejov, Czech Republic
	    \and 
		Institute of Theoretical Physics and Astrophysics, Masaryk 
        University, CZ-611 37 Brno, Kotl\'a\v{r}sk\'a 2, Czech Republic
        }
        
        \date{Received 10 May 2015; accepted 6 November 2015} 

\titlerunning{Time Dependent Spectral-Feature Variations of HD~50138}

\abstract
{
B[e] stars are anomalous objects around which extended circumstellar matter is present. The observed 
properties of the central star are significantly affected by the surrounding material. Therefore, 
the use of standard synthetic spectra is disputable in this case and our capability to study 
these objects is limited. One of the~possibilities is to analyse variations of the~spectral features. 
Long-term spectroscopic observations are required for this, but are not found in the literature. \\
For our study we choose the~B[e] star HD~50138 of the FS~CMa type because of the indication
that this star is a~post-main-sequence star, although still not highly evolved. Therefore, it can be a~good object for 
testing evolutionary models. Currently, HD~50138 is the~most extensively observed FS~CMa star 
which makes it an ideal object for modelling. Our observations fill the~gap in the available data. 
}
{
To describe the variability of HD~50138 we have monitored this star spectroscopically over the last twenty years.
To search for the periodicity  on short-term scales, series of night-to-night observations 
were also obtained. We were able to obtain 130 spectra from four different telescopes -- 1.06~m at Ritter Observatory
(\`{e}chelle, $R\sim 26\,000$, 32 spectra, $1994-2003$), the Perek 2~m telescope at Ond\v{r}ejov Observatory 
(slit, $R\sim 12\,500$, 56 spectra, $2004-2013$), the 2.12~m telescope at Observatorio Astronomico Nacional San Pedro Martir
(\`{e}chelle, $R\sim 18\,000$, 16 spectra, $2005-2013$), and the 0.81~m telescope at Three College Observatory
(\`{e}chelle, $R\sim 12\,000$, 26 spectra, $2013-2014$).
} 
{
We describe and analyse variations of the~chosen lines. The~measurements of the~equivalent widths and 
radial velocities of the~H$\alpha$, H$\beta$, and [\ion{O}{i}] $\lambda \lambda$ 6300, 6364~\AA \, lines 
are presented. The~set of obtained spectra allows us to describe the~changes on timescales
from days to years. 
} 
{
The long-term quasi-periodic trend was found in the variations of the H$\alpha$ equivalent width
and confirmed by time dependent studies of the relative flux and equivalent width of the [\ion{O}{i}]~6300~\AA \,
line and radial velocity of the H$\alpha$ violet peak. The two long periods of $3\,000\pm 500$ and $5\,000\pm 1000$~days 
were detected there. We were able to catch moving humps in the H$\alpha$ line, which reveal the rotating 
media around the star. An analysis of the correlation of equivalent widths, radial velocities,
relative fluxes, and $V/R$ ratios for various lines is presented in detail.
} 
{
We describe the spectral variability of HD~50138 over the last twenty years. Based on these data, we determine 
new restrictions for future modelling. We confirm the quasi-periodic behaviour of the object's spectral 
variability, which probably reflects mass transfer in a~binary system. This behaviour also supports 
a~recently introduced explanation of the nature of FS~CMa stars as post-merger systems. 
}  
\keywords
{
 circumstellar matter -- stars: emission line, Be -- stars: mass loss -- binaries: spectroscopic -- stars: individual: MWC~158
}

\maketitle
\section{Introduction}

\label{uvod}

The B[e] phenomenon \cite[][]{zavorky_Conti} is a~designation for hot B-type stars whose spectra show 
forbidden and permitted emission lines of neutral and singly ionised atoms and strong infrared excess  
\citep{Swings71}. These observed properties indicate very extended circumstellar matter and
were discovered in only a~small percentage of known B stars. However, stars of different types
and in different evolutionary stages show the B[e] phenomenon. \cite{Lamers98} were able to 
identify compact planetary nebulae, Herbig stars, supergiants, and symbiotic stars, but they were not 
able to classify half of the stars known at that time. \cite{Miroshnichenko07-FS_CMa} noticed that almost all 
unclassified stars have similar properties and introduced a~new group called FS~CMa objects. 
To date, the nature of FS~CMa stars has not yet been explained. The amount of dust is too large to be 
produced during the evolution of a~single star. One likely explanation is that these are binary stars, 
but a~sufficient number of binaries have not been identified in this group. Detailed discussion of 
the nature of FS~CMa stars can be found in \cite{Miroshnichenko07-FS_CMa,Miroshnichenko13}, 
and \cite{Miroshnichenko_prehled15}.

The study of FS~CMa stars is complicated by the presence of large amounts of circumstellar matter. 
We usually have almost no direct information about the central object. Moreover, due to asymmetry of 
the~envelope and its large extension, commonly used stellar atmosphere models are not appropriate for 
the~analysis. Therefore, in the present study we focus on spectral variability which provides one of 
only a~few limited opportunities to study these objects. Despite the importance of these observations, 
long-term monitoring data are still missing for most group objects. The~first two papers \citep{Polster12,Kucerova13} 
were devoted to systematic spectroscopic observations of the FS CMa objects MWC~623 and MWC~342.
Here we present the~results of a~spectroscopic monitoring of HD~50138, the brightest star of the FS~CMa type.

HD~50138 (MWC~158,  V743~Mon, or IRAS~06491-0654) was discovered to be an emission-line star by \citet{Humanson21} on 
a~photographic plate from December 1920. Soon afterwards, \citet{Merrill25} noticed its spectral variability. 
Following this discovery \cite{Merrill31_ApJ} published the~results from almost ten years of 
spectroscopic observations. He analysed 76 photographic plates obtained from December 1920 
to February 1930 (mostly in the~blue region around the~H$\beta$ line). The radial velocity ($RV$) measurements 
indicated a~period of about 30~days. However, this period did not fit the~measurements well. An~attempt 
to fit with two periods or variable elements of a~binary also failed owing to insufficient data sampling.
Another systematic study was done by \citet{Doazan65}. She analysed 52 spectra ($12.4$~\AA/mm) from 
the Haute Provence observatory obtained between 1960 and 1963. She found that the~envelope was expanding 
continuously. Moreover, the speed of expansion was changing periodically on timescales of 50 days. 
The acceleration of matter was observed during a~half cycle and was followed by its deceleration in the 
following half cycle. A~detailed spectroscopic study was published by \citet{Jaschek98}. Spectra in the~interval 
from $3738-10232$~\AA \, (dispersion 33 \AA/mm) obtained in 1989 and 1996 at the~Haute Provence observatory 
allowed them to determine basic properties of the~envelope. However, the~set of 22 spectra in nine different 
intervals did not permit a~sufficient description of the~variability. Another attempt (and the last so far) 
 to find the~spectral periodicity was made by \citet{Corporon99}. They measured $RV$s on high-dispersion 
spectra ($R=50 000$) obtained during a~three-year campaign started in 1994. They found no trend. 
However, they used an automatic fitting to a~Gaussian profile to measure the~$RV$. Since the~chosen absorption 
lines are asymmetric and variable, this procedure may be questioned. Other studies are based on only a~few 
spectra or on a~short campaign prohibiting detection of variability of this source. 

The~star is also  photometrically variable, but the amplitude of changes is small, no greater than a~tenth of
a~magnitude. HD~50138 as a~variable star was first noted by \citet{Allen73} based on observations in the~IR 
region from February 1971 to January 1972. Ultraviolet observations from 1974 and 1975 were analysed by
\citet{Savage78}. They found that the~changes are wavelength dependent and the~largest change was detected at 
0.25 $\mu$m. Observations in the Johnson photometric system have been done by \citet{Alvarez81}. They found 
no variability, but their analysis was only based on five measurements. More systematic work was done by
\citet{Kilkenny85}. They obtained photoelectric measurements (UBVRIY) from March 1981 to April 1982 and 
detected variations of the order of 0.09 mag. The~magnitude of variations probably differs between 
the~observation epochs (as in another B[e] star -- V1972~Cyg; \cite{Melnikov97}), which is indicated by
the~observations performed by \citet{Winter01}. Unfortunately, the~observations are too sparse to find
periodicity, multiperiodicity, or other regular behaviour of this star.

HD~50138 is variable not only spectroscopically and photometrically, but also polarimetrically. 
Different results have been presented by e.g. \cite{Bjorkman94} and \cite{Yudin98}. Data from \cite{Bjorkman94}
can be explained by a~rotating dusty disc, or binarity of the~object. Further polarimetric 
observations \citep{Bjorkman98} have allowed a picture to be constructed of the circumstellar matter:
a~thin gaseous disc, viewed almost edge-on, together with an optically-thin spherically symmetric dusty 
shell. Detailed polarisation measurements along the~H$\alpha$ line \citep{Harrington07} show 
changes in the~polarisation in the~absorption part of the~line. This leads \citet{Harrington07} to 
suggest that optical pumping (rather than scattering) plays an important role in the~envelope. 

Another technique that gives important information about the~circumstellar material is 
interferometry. The~first interferometric measurements at 10.7 $\mu$m were obtained  in 2005 by
\citet{ Monnier09} using the~Keck telescope. The~fit of the~visibility by a~1D Gaussian function gives 
a~size estimate of $58\pm6$~mas; a~2D Gaussian gives $(66\pm4) \times (46\pm 9)$~mas 
with a~position angle $63\pm 6$~deg. Further observations using the MIDI and AMBER instruments at VLTI 
allowed the study of the dependence of the size of the circumstellar matter on the wavelength \citep{Borges10,Borges11}.
Recently \cite{Ellerbroek15} has published results from the VLTI/AMBER and CHARA/VEGA interferometer
completed by the VLT/CRIRES spectro-astrometry and other spectroscopic and spectropolarimetric data.
They found that the Br$\gamma$ emission originates in a~compact region up to 3~au and the continuum 
emission is produced in a~more extended region. They also proved that Keplerian rotation dominates 
the velocity field. \citet{Marston08} attempted to find signatures of more extended media using CCD images
taken through a~narrow H$\alpha$ filter with a~60-inch telescope. They detected no nebula.

The~circumstellar matter of HD~50138 was modelled first using the~Sobolev approximation by \citet{Doazan65} 
and \citet{Briot81}. \citet{Kuan75} derived the~mass loss rate by this method for the~first time. The~conditions 
in the~envelope suggest the use of a~model involving the~Str\"{o}mgren sphere \citep{Houziaux76}. 
A~more detailed model using the~Monte Carlo method was published by \cite{Bjorkman98}.

In recent years several events have been observed leading to the formation of a~new shell 
\citep{Hutsemekers85,Pogodin97}. The~announcement from \citet{Andrillat91} was followed by 
photometric \citep{Halbedel91} and spectroscopic \citep{Bopp93} observations.

The~detection of such events, or of corotating regions in the~circumstellar disc, as well as the~confirmation 
of the~binary nature of the system, can be done within the~framework of our project. Binarity
is one of the~current key hypotheses regarding stars of the~FS~CMa type. The~amount of circumstellar 
matter in these objects is larger than predicted by the~evolutionary models of single stars, 
but only 30\% of them have been proved or suggested to be binary systems \citep{Miroshnichenko07-FS_CMa}.

In Sect.~\ref{reduction} we describe the~observations and the data reduction process. The~line-profile 
variability, measurements of the equivalent widths ($EW$s), and $RV$s are presented in 
Sects.~\ref{analysis} and \ref{results}. The~observed phenomenon are discussed in detail
in Sect. \ref{diskuze}. Conclusions are presented in Sect.~\ref{conclusion}.


\section{Observations and data reduction} \label{reduction}

Our analysis is based on data from several observatories. The~main properties of 
the equipment used are summarised in Table~\ref{param}. The~main data set presented here was taken 
during years 2004-2013 with the Perek 2~m telescope at Ond\v{r}ejov Observatory (OO), Czech Republic.
For our observations we chose a~spectral interval from 6265~\AA \, to 6770~\AA\,  because of 
a)~the~presence of the~forbidden emission lines [\ion{O}{i}] $\lambda \lambda$ 6300, 6364~\AA, 
which are formed in the~outer parts of the~envelope; b)~the~\ion{Si}{ii} $\lambda \lambda$ 6347, 6371~\AA, 
and \ion{He}{i} 6678~\AA \, lines, which originate in the~inner parts; and c)~the~H$\alpha$ line, which 
forms over a~wide range of distances from the star. Therefore, starlight can be affected by 
a~wide range of phenomena as it passes through the entire envelope.

\begin{table}[!t]
  \caption{Parameters of the spectrographs used.}
  \label{param}
  \centering
  \begin{tabular}{lllll}
  \hline \hline
                           & Ond\v{r}ejov O. & Ritter O.               & SMP O.                               & TCO \\ 
  \hline
   R (H$\alpha$) $\approx$ & 12\,500         & 26\,000                 & 18\,000                              & 12\,000   \\ 
   spectr.                 & slit            & \`{e}chelle             & \`{e}chelle                          & \`{e}chelle\\ 
   PMD $[m]$               & 2.0             & 1.06                    &2.12                                  & 0.81 \\ 
   notation                & OO $\bullet$    & RO {\color{red} $\ast$} & SMP {\color{green}$\blacktriangle$ } & TCO {\color{blue}+} \\ 
  \hline
  \end{tabular}
  \tablefoot{
    The resolution, type of the spectrograph, and diameter of the primary mirror are summarised. The last
    row introduces the notation to be used, which corresponds to the individual instruments.
   }
\end{table}

The~data\footnote{Continuum normalised spectra are available at the CDS database.} are reduced in 
\texttt{IRAF}\footnote{ 
  \texttt{IRAF} is distributed by the~National Optical Astronomy Observatories, 
  operated by the~Association of Universities for Research in Astronomy, 
  Inc., under contract to the~National Science Foundation of the~United States.
}
using standard procedures. To remove cosmic rays, the program \texttt{dcr} \citep{Pych}
is used. This allows us to omit the use of the~optimal extraction during the~spectra subtraction. 
Considering that our spectra are parallel to the~pixel rows, their subtraction is possible by 
individual pixel columns. Therefore, atmospheric night sky lines do not affect 
the~stellar forbidden oxygen lines\footnote{
  The spectrum obtained on 7 February 2005 was reduced by MS using optimal extraction.
}.
Because of the~high intensity of the~H$\alpha$ line, which is almost twenty times higher than the continuum,
normalisation is an important step in the~reduction process. Fortunately, the continuum in the 
interval around the H$\alpha$ line is well defined, which provides a~good fit by Chebyshev polynomials. 
To check the~accuracy of the~normalisation, we changed the~order of the~polynomial ($\pm 1$) and 
the~chosen continuum intervals. The~resulting values ($EW$s, relative fluxes, and $RV$s)
differ within their error estimates.

Another set of spectra were taken in 1993-2005 at the~Ritter observatory (RO), Toledo, Ohio. 
These \`{e}chelle spectra were reduced using standard \texttt{IRAF} and \texttt{IDL} 
routines. The normalisation of individual \`{e}chelle orders was similar to the Ond\v{r}ejov data. 
We decrease the RO spectral resolution to the~Ond\v{r}ejov value using convolution by 
a~Gaussian function.

We also used spectra from Observatorio Astronomico Nacional San Pedro Martir (SPM).
Spectra were reduced using the standard procedures in \texttt{IRAF} without the optimal extraction.
The construction of the spectrograph and reduction process guarantee that the final spectrum is not 
affected by the night-sky lines. MIDAS was used to clean spectra from cosmic rays.

The last set of spectra are from Three College Observatory (TCO, located near Greensboro, 
North Carolina, USA). All the TCO data were reduced using \texttt{IRAF}, specifically its \`{e}chelle 
package to extract the spectral orders, identify lines in the comparison ThAr spectra, and complete 
the wavelength calibration of the objects' spectra.

Time intervals between individual observations were very different across the series. Sequential
spectra were obtained on several nights, as well as the night-to-night series. On the other hand, 
the intervals between some observations exceeded a~month in some seasons. Detailed coverage of 
observations was achieved in the last season when the average temporal step was five days. 
This data set allowed us to find or reject periods in the range from a~few days to years.
The list of observations is printed in Table~\ref{observing_log}$^{e)}$\footnote{ 
Throughout the paper the online material is indicated by the superscript $^{e)}$.
}.

\onllongtab{
\begin{longtable}{lrllrrl}
 \caption{Observing log} \\
    \hline \hline 
 \multicolumn{7}{c}{Ritter Observatory}\\
 date       &  JD-2450000   & filename                    & instrument  & exp. [s] & S/N & observer \\
\hline
1993-Nov-28  &    -680.2251  & 931128Ha([OI]6300).dat	  & \`{e}chelle	& 3600	& 70  &-  \\ 
1994-Feb-06  &    -610.3187  & 940206Ha([OI]6300).dat	  & \`{e}chelle	& 3600	& 50  &-  \\ 
1994-Feb-16  &    -600.3148  & 940216Ha([OI]6300).dat	  & \`{e}chelle	& 3600	& 50  &-  \\ 
1994-Feb-22  &    -594.3742  & 940222Ha([OI]6300).dat	  & \`{e}chelle	& 3600	& 50  &-  \\ 
1994-Mar-02  &    -586.4597  & 940302Ha([OI]6300).dat	  & \`{e}chelle	& 3600	& 50  &-  \\ 
1994-Mar-17  &    -571.4031  & 940317Ha([OI]6300).dat	  & \`{e}chelle	& 3600	& 60  &-  \\ 
1994-Sep-18  &    -386.1090  & 940918Ha([OI]6300).dat	  & \`{e}chelle	& 3600	& 70  &-  \\ 
1994-Sep-21  &    -383.0964  & 940921Ha([OI]6300).dat	  & \`{e}chelle	& 3600	& 60  &-  \\ 
1994-Nov-12  &    -331.0622  & 941112Ha([OI]6300).dat	  & \`{e}chelle	& 3600	& 50  &-  \\ 
1994-Dec-19  &    -294.2986  & 941219Ha([OI]6300).dat	  & \`{e}chelle	& 3600	& 50  &-  \\ 
1994-Dec-22  &    -291.2952  & 941222Ha([OI]6300).dat	  & \`{e}chelle	& 3600	& 40  &-  \\ 
1995-Jan-27  &    -255.3515  & 950127Ha([OI]6300).dat	  & \`{e}chelle	& 3600	& 50  &-  \\ 
1995-Feb-12  &    -239.3537  & 950212Ha([OI]6300).dat	  & \`{e}chelle	& 3600	& 50  &-  \\ 
1995-Feb-19  &    -232.3874  & 950219Ha([OI]6300).dat	  & \`{e}chelle	& 3600	& 70  &-  \\ 
1995-Feb-22  &    -229.4138  & 950222Ha([OI]6300).dat	  & \`{e}chelle	& 3600	& 100  &- \\ 
1995-Feb-24  &    -227.3818  & 950225Ha([OI]6300).dat	  & \`{e}chelle	& 3600	& 100  &- \\ 
1995-Mar-02  &    -221.3767  & 950303Ha([OI]6300).dat	  & \`{e}chelle	& 3600	& 90  &-  \\ 
1995-Mar-14  &    -209.4575  & 950315Ha([OI]6300).dat	  & \`{e}chelle	& 3600	& 80  &-  \\ 
1995-Mar-24  &    -199.4105  & 950325Ha([OI]6300).dat	  & \`{e}chelle	& 3600	& 80  &-  \\ 
1995-Nov-20  &      41.8367  & 951120Ha([OI]6300).dat	  & \`{e}chelle	& 3600	& 50  &-  \\ 
1996-Jan-17  &      99.7329  & 960122Ha([OI]6300).dat	  & \`{e}chelle	& 3600	& 80  &-  \\ 
1997-Nov-27  &     779.8832  & 971127Ha([OI]6300).dat	  & \`{e}chelle	& 3600	& 90  &-  \\ 
1997-Dec-16  &     798.7946  & 971216Ha([OI]6300).dat	  & \`{e}chelle	& 3600	& 40  &-  \\ 
1998-Feb-09  &     853.6453  & 980209Ha([OI]6300).dat	  & \`{e}chelle	& 3600	& 70  &-  \\ 
1998-Feb-15  &     859.6289  & 980215Ha([OI]6300).dat	  & \`{e}chelle	& 3600	& 70  &-  \\ 
1998-Mar-30  &     902.5522  & 980330Ha([OI]6300).dat	  & \`{e}chelle	& 3600	& 70  &-  \\ 
1998-Apr-06  &     909.5389  & 980406Ha([OI]6300).dat	  & \`{e}chelle	& 3600	& 100 &-  \\ 
1998-Apr-13  &     916.5531  & 980413Ha([OI]6300).dat	  & \`{e}chelle	& 3600	& 60  &-  \\ 
1998-Apr-18  &     921.5817  & 980418Ha([OI]6300).dat	  & \`{e}chelle	& 3600	& 50  &-  \\ 
2000-Mar-06  &    1609.5771  & 20000306Ha([OI]6300).dat    & \`{e}chelle	& 3600	& 50  &-  \\ 
2003-Nov-09  &    2952.8989  & 20031109Ha([OI]6300).dat    & \`{e}chelle	& 3600	& 40  &-  \\ 
2003-Nov-21  &    2964.8337  & 20031121Ha([OI]6300).dat    & \`{e}chelle	& 3600	& 40  &-  \\ 
&&&&&& \\  
    \hline \hline
 \multicolumn{7}{c}{Ond\v{r}ejov Observatory}\\
 date      &  JD-2450000     & filename              & instrument    & exp. [s] & S/N & observer \\
\hline
2004-Mar-14          &    3079.3004   &nc140022.dat	& slit;  700~mm focus	&1800	&220	&Budovi\v{c}ov\'{a}\\ 
2004-Mar-22          &    3087.3256   &nc220014.dat	& slit;  700~mm focus	&1800	&140	&DK \\
2005-Jan-07          &    3378.4495   &oa070014.dat	& slit;  700~mm focus	&600	&110	&DK \\  
2005-Jan-16          &    3387.3653   &oa160020.dat	& slit;  700~mm focus	&900	&70	&VV\\ 
2005-Jan-16          &    3387.3784   &oa160021.dat	& slit;  700~mm focus	&900	&70	&VV\\ 
2005-Jan-16          &    3387.3893   &oa160022.dat	& slit;  700~mm focus	&900	&70	&VV\\ 
2005-Jan-16          &    3387.4104   &oa160024.dat	& slit;  700~mm focus	&600	&70	&VV\\ 
2005-Feb-07          &    3409.3459   &ob070006.dat	& slit;  700~mm focus	&900	&120	&MS\\ 
2005-Feb-07          &    3409.3568   &ob070007.dat	& slit;  700~mm focus	&900	&110	&MS\\ 
2005-Feb-07          &    3409.3685   &ob070008.dat	& slit;  700~mm focus	&900	&110	&MS\\ 
2005-Oct-13          &    3656.6557   &oj120037.dat	& slit;  700~mm focus	&300	&60	&PS\\ 
2005-Oct-13          &    3656.6699   &oj120038.dat	& slit;  700~mm focus	&1800	&130	&PS \\ 
2005-Oct-29          &    3672.6472   &oj280051.dat	& slit;  700~mm focus	&1800	&130	&DK\\ 
2006-Jan-08          &    3744.3683   &pa080035.dat	& slit;  700~mm focus	&3600	&140	&DK\\ 
2006-Jan-12          &    3747.5148   &pa110043.dat	& slit;  700~mm focus	&1200	&120	&DK\\ 
2006-Jan-12          &    3747.5343   &pa110045.dat	& slit;  700~mm focus	&1800	&120	&DK\\ 
2006-Feb-06          &    3773.3324   &pb060015.dat	& slit;  700~mm focus	&4500	&110	&DK\\ 
2006-Oct-09          &    4017.6271   &pj080065.dat	& slit;  700~mm focus	&2222	&150	&Netolick\'{y}\\ 
2007-Jan-14          &    4115.4835   &qa140071.dat	& slit;  700~mm focus	&2345	&150	&Netolick\'{y}\\ 
2007-Mar-11          &    4171.3618   &qc110030.dat	& slit;  700~mm focus	&1200	&150	&DK\\ 
2007-Mar-11          &    4171.3786   &qc110034.dat	& slit;  700~mm focus	&300	&60	&DK\\ 
2007-Mar-13          &    4173.3806   &qc130037.dat	& slit;  700~mm focus	&500	&100	&DK\\ 
2007-Mar-14          &    4174.3520   &qc140004.dat	& slit;  700~mm focus	&600	&60	&VV\\ 
2007-Mar-25          &    4185.3377   &qc250026.dat	& slit;  700~mm focus	&2100	&140	&DK\\ 
2007-Mar-30          &    4190.3325   &qc300022.dat	& slit;  700~mm focus	&1800	&140	&Polster\\ %
2007-Apr-15          &    4206.2997   &qd150013.dat	& slit;  700~mm focus	&1200	&140	&DK\\ 
2008-Mar-09          &    4535.3358   &rc090021.dat	& slit;  700~mm focus	&1800	&160	&Ceniga\\ 
2008-Mar-28          &    4554.3230   &rc280013.dat	& slit;  700~mm focus	&2600	&140	&Polster\\ %
2008-Mar-29          &    4555.3173   &rc290019.dat	& slit;  700~mm focus	&1400	&160	&Polster\\ %
2008-Apr-05          &    4562.2935   &rd050012.dat	& slit;  700~mm focus	&3000	&90	&Polster\\ 
2009-Jan-09          &    4841.4609   &sa090017.dat	& slit;  700~mm focus	&2000	&50	&Polster\\ 
2009-Jan-09          &    4841.4869   &sa090018.dat	& slit;  700~mm focus	&2250	&50	&Polster\\ 
2009-Jan-11          &    4843.4355   &sa110048.dat	& slit;  700~mm focus	&2400	&140	&Polster\\ %
2010-Mar-19          &    5275.3812   &tc190022.dat	& slit;  700~mm focus	&900	&200	&PS\\ 
2010-Mar-31          &    5287.3706   &tc310009.dat	& slit;  700~mm focus	&900	&100	&PS\\ 
2010-Mar-31          &    5287.3868   &tc310010.dat	& slit;  700~mm focus	&1635	&80	&PS\\ 
2010-Apr-07          &    5294.2868   &td070009.dat	& slit;  700~mm focus	&1200	&270	&PS\\  
2010-Sep-20          &    5459.6449   &ti190046.dat	& slit;  700~mm focus	&1500	&150	&DK\\ 
2011-Feb-08          &    5601.4416   &ub080062.dat	& slit;  700~mm focus	&1726	&240	&MS\\ 
2011-Feb-09          &    5602.3830   &ub090044.dat	& slit;  700~mm focus	&1200	&190	&DK\\ 
2011-Feb-23          &    5616.3552   &ub230037.dat	& slit;  700~mm focus	&1800	&150	&DK\\ 
2011-Mar-21          &    5642.3310   &uc210020.dat	& slit;  700~mm focus	&900	&160	&DK\\ 
2011-Mar-21          &    5642.3472   &uc210022.dat	& slit;  700~mm focus	&1200	&160	&DK\\ 
2011-Nov-12          &    5877.5425   &uk110042.dat	& slit;  700~mm focus	&2700	&220	&DK\\ 
2011-Nov-15          &    5880.5281   &uk140029.dat	& slit;  700~mm focus	&1800	&160	&PS\\ 
2012-Feb-10          &    5968.4133   &vb100041.dat	& slit;  700~mm focus	&3600	&110	&DK\\ 
2012-Feb-11          &    5969.3987   &vb110037.dat	& slit;  700~mm focus	&1835	&130	&DK\\ 
2012-Feb-12          &    5970.4331   &vb120022.dat	& slit;  700~mm focus	&3600	&90	&DK\\ 
2012-Mar-24          &    6011.3095   &vc240017.dat	& slit;  700~mm focus	&1500	&160	&DK\\ 
2012-Mar-25          &    6012.3051   &vc250026.dat	& slit;  700~mm focus	&1200	&130	&DK\\ 
2012-Sep-17          &    6187.6396   &vi160055.dat	& slit;  700~mm focus	&1200	&160	&DK\\ 
2013-Feb-06          &    6330.4010   &wb060036.dat	& slit;  700~mm focus	&2261	&190	&PM\\ 
2013-Feb-08          &    6332.4040   &wb080039.dat	& slit;  700~mm focus	&1538	&120	&DK\\ 
2013-Mar-05          &    6357.3923   &wc050028.dat	& slit;  700~mm focus	&800	&130	&Zasche\\ 
2013-Apr-12          &    6395.2883   &wd120013.dat	& slit;  700~mm focus	&900	&200	&RK \& TJ\\ %
2013-Apr-12          &    6395.2883   &wd120013.dat	& slit;  700~mm focus	&900	&200	&RK \& TJ\\ %
&&&&&& \\  
    \hline \hline
 \multicolumn{7}{c}{Observatorio Astronomico Nacional San Pedro Martir}\\
 date       &  JD-2450000       & filename              & instrument    & exp. [s] & S/N & observer \\
\hline
2005-Oct-10  &   3653.9990 	&20051010Ha(Hb).dat	& \`{e}chelle	&300	&70	&AM \& SZ\\ 
2006-Dec-12  &   4082.0270 	&20061212Ha(Hb).dat	& \`{e}chelle	&1980	&200	&AM \& SZ\\ 
2006-Dec-13  &   4082.9700 	&20061213Ha(Hb).dat	& \`{e}chelle	&900	&160	&AM \& SZ\\ 
2006-Dec-14  &   4084.0040 	&20061214Ha(Hb).dat	& \`{e}chelle	&900	&180	&AM \& SZ\\ 
2007-Nov-10  &   4414.8920 	&20071110Ha(Hb).dat	& \`{e}chelle	&2220	&130	&SZ   \\ 
2007-Nov-19  &   4423.9820 	&20071119Ha(Hb).dat	& \`{e}chelle	&1900	&170	&SZ   \\ 
2008-Oct-05  &   4745.0190 	&20081004Ha(Hb).dat	& \`{e}chelle	&600	&120	&AM \& SZ\\ 
2008-Oct-11  &   4750.9670 	&20081010Ha(Hb).dat	& \`{e}chelle	&840	&180	&AM   \\ 
2009-Nov-05  &   5140.9370 	&20091104Ha(Hb).dat	& Otro    	&600	&110	&AM \& SZ\\ 
2009-Nov-11  &   5146.9590 	&20091110Ha(Hb).dat	& Otro	        &900	&130	&AM \& SZ\\ 
2010-Oct-14  &   5484.0000 	&20101014Ha(Hb).dat	& \`{e}chelle	&1200	&100	&SZ   \\ 
2011-Nov-01  &   5867.0130 	&20111031Ha(Hb).dat	& \`{e}chelle	&1800	&180	&JLOA \& ING\\ 
2011-Nov-04  &   5870.0030 	&20111103Ha(Hb).dat	& \`{e}chelle	&2700	&180	&SZ    \\ 
2012-Nov-13  &   6244.9930 	&20121112Ha(Hb).dat	& \`{e}chelle	&5400	&210	&AM \& SZ \\ 
2012-Nov-14  &   6245.9530 	&20121113Ha(Hb).dat	& \`{e}chelle	&2400	&200	&AM \& SZ \\ 
2013-Oct-17  &   6583.0210 	&20131016Ha(Hb).dat	& \`{e}chelle	&600	&150	&SZ    \\ 
&&&&&& \\  
    \hline \hline
 \multicolumn{7}{c}{Three College Observatory}\\
 date       &   JD-2450000      & filename              & instrument    & exp. [s] & S/N & observer \\
\hline
2013-Nov-29  &   6625.8120 	&20131128Ha(Hb.[OI]6300.6364).dat	& \`{e}chelle	&1200	&80	&AM\\
2013-Nov-30  &   6626.7420 	&20131129Ha(Hb.[OI]6300.6364).dat	& \`{e}chelle	&1440	&100	&AM\\
2013-Dec-01  &   6627.7780 	&20131130Ha(Hb.[OI]6300.6364).dat	& \`{e}chelle	&900	&80	&AM\\
2013-Dec-02  &   6628.7800 	&20131201Ha(Hb.[OI]6300.6364).dat	& \`{e}chelle	&720	&100	&AM\\
2013-Dec-11  &   6637.7500 	&20131210Ha(Hb.[OI]6300.6364).dat	& \`{e}chelle	&780	&80	&AM\\
2013-Dec-12  &   6638.7680 	&20131211Ha(Hb.[OI]6300.6364).dat	& \`{e}chelle	&450	&80	&SD\\
2013-Dec-13  &   6639.7920 	&20131212Ha(Hb.[OI]6300.6364).dat	& \`{e}chelle	&780	&90	&AM\\
2013-Dec-16  &   6642.7780 	&20131215Ha(Hb.[OI]6300.6364).dat	& \`{e}chelle	&1200	&80	&AM \& SD\\
2013-Dec-17  &   6643.7320 	&20131216Ha(Hb.[OI]6300.6364).dat	& \`{e}chelle	&720	&90	&SD\\ 
2013-Dec-19  &   6645.7220 	&20131218Ha(Hb.[OI]6300.6364).dat	& \`{e}chelle	&660	&100	&SD\\ 
2013-Dec-27  &   6653.7270 	&20131226Ha(Hb.[OI]6300.6364).dat	& \`{e}chelle	&1050	&90	&AM\\ 
2013-Dec-28  &   6654.7280 	&20131227Ha(Hb.[OI]6300.6364).dat	& \`{e}chelle	&1200	&120	&AM\\ 
2014-Jan-04  &   6661.7110 	&20140103Ha(Hb.[OI]6300.6364).dat	& \`{e}chelle	&1260	&100	&AM\\ 
2014-Jan-13  &   6670.7080 	&20140112Ha(Hb.[OI]6300.6364).dat	& \`{e}chelle	&1560	&100	&AM\\ 
2014-Jan-17  &   6674.6870 	&20140116Ha(Hb.[OI]6300.6364).dat	& \`{e}chelle	&1440	&90	&AM \& SD\\
2014-Jan-20  &   6677.7040 	&20140119Ha(Hb.[OI]6300.6364).dat	& \`{e}chelle	&1260	&100	&SD\\ 
2014-Jan-21  &   6678.6400 	&20140120Ha(Hb.[OI]6300.6364).dat	& \`{e}chelle	&1260	&110	&AM\\ 
2014-Jan-25  &   6682.6460 	&20140124Ha(Hb.[OI]6300.6364).dat	& \`{e}chelle	&1260	&80	&AM\\ 
2014-Feb-01  &   6689.6530 	&20140131Ha(Hb.[OI]6300.6364).dat	& \`{e}chelle	&1500	&80	&AM\\ 
2014-Feb-22  &   6710.5710 	&20140221Ha(Hb.[OI]6300.6364).dat	& \`{e}chelle	&1680	&90	&AM\\ 
2014-Feb-23  &   6711.5730 	&20140222Ha(Hb.[OI]6300.6364).dat	& \`{e}chelle	&1500	&70	&AM\\ 
2014-Feb-24  &   6712.6980 	&20140223Ha(Hb.[OI]6300.6364).dat	& \`{e}chelle	&1500	&60	&SD\\ 
2014-Feb-25  &   6713.5650 	&20140224Ha(Hb.[OI]6300.6364).dat	& \`{e}chelle	&1500	&90	&AM\\ 
2014-Feb-28  &   6716.5500 	&20140227Ha(Hb.[OI]6300.6364).dat	& \`{e}chelle	&1800	&100	&AM\\ 
2014-Mar-14  &   6730.5280 	&20140313Ha(Hb.[OI]6300.6364).dat	& \`{e}chelle	&1800	&120	&AM\\ 
2014-Mar-23  &   6739.5310 	&20140322Ha(Hb.[OI]6300.6364).dat	& \`{e}chelle	&1500	&80	&AM\\ 
%
 \label{observing_log}
\end{longtable}
 \tablefoot{
``AM'' Anatoly Miroshnichenko, ``SZ'' Sergey Zharikov, ``SD'' Steve Danford, ``DK'' Daniela Kor\v{c}\'{a}kov\'{a},
   ``MS'' Miroslav \v{S}lechta, ``PS'' Petr \v{S}koda, ``VV'' Viktor Votruba, ``RK'' Radek K\v{r}\'{\i}\v{c}ek, ``TJ''
   Tereza Je\v{r}\'{a}bkov\'{a}
} 
}



\section{Analysis}
\label{analysis}

To describe the observed changes of the spectral lines, we measure their $EW$s,
$RV$s, and relative fluxes. We summarise the procedures used in this section because of the 
specification of the error estimate, which is as important as the value itself for the rigorous 
analysis of temporal variability.  

\renewcommand{\labelenumi}{\roman{enumi})}
\begin{enumerate}

  \item $EW$ measurements using line-profile integration: \\
    Numerical integration by the trapezium method is used on the interpolated data
    \citep{Steffen}. Errors are estimated according to \cite{Vollmann}
    where the signal-to-noise ratio ($S/N$) is taken into account. The value of $S/N$
    is estimated by a~linear fit of the appropriate part of the continuum. 
    This straightforward integration is used for the H$\alpha$ line.
 
  \item 	$EW$ determined by fitting of a~Gaussian function: \\
    Narrow symmetric lines ([\ion{O}{i}] $\lambda \lambda$ 6300~\AA, 6364 \AA\,) are fitted by 
    a~Gaussian function. The least squares method is used for the procedure and the error 
    estimate is calculated following \cite{Vollmann}. 
 
  \item 	$RV$ and line intensities obtained by Gaussian fitting: \\
   The procedure is identical to the one described in item \textit{ii$\left.\right)$}, except for 
   the error estimate, which is determined by both the formal error of the least 
   squares method and Monte Carlo simulation based on the $S/N$.  
 
  \item 	polynomial fitting: \\
    $RV$, line intensities, or extrema of several functions are fitted by
    a~polynomial ($3^{rd}$- $6^{th}$ order) using the least squares
    method. The error is a~formal error of this method combined with the error estimated 
    from the $S/N$ using the Monte Carlo method. 
 
  \item $RV$ determined by the line-profile mirroring: \\
    The flipped line is shifted automatically using the least squares
    method to fit the chosen part of the line (Fig.~\ref{mirroring}$^{e)}$). The uncertainty 
    is computed as a~combination of the error of the fit and the $S/N$ ratio.
    \onlfig{
      \begin{figure}[!h]
        \includegraphics[width=0.5\textwidth]{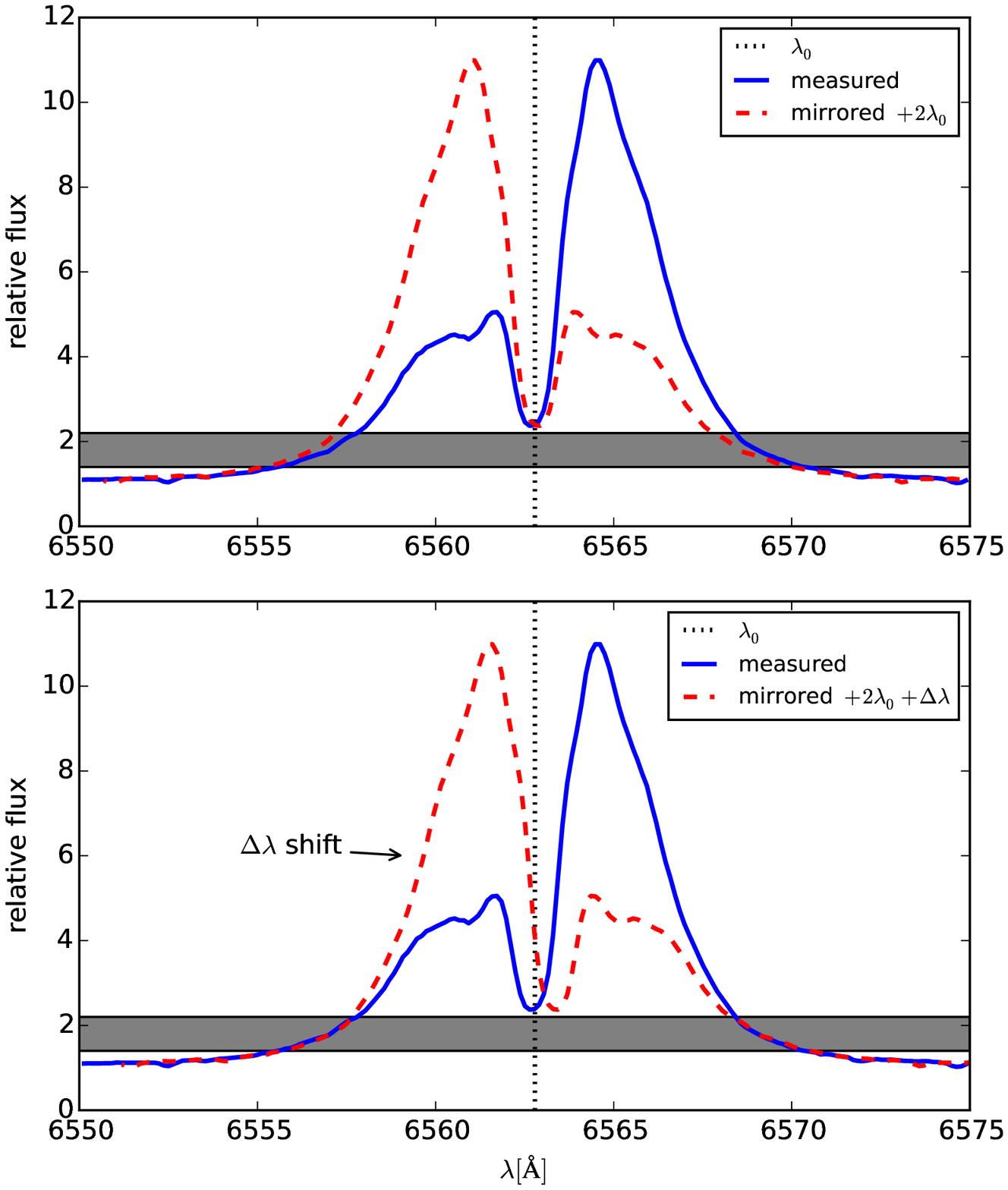}
        \caption{
           Demonstration of the $RV$ measurements using the line-profile mirroring. 
           The best agreement of the original (blue solid line) and flipped spectra (red dashed line)
           is obtained by the least squares method. The grey belt defines the interval of the relative 
           flux where the $RV$ are measured. The plotted spectrum is from 14 October 2010 (SPM).
                 }
        \label{mirroring}
      \end{figure}
    \FloatBarrier
    }
  \item period analysis: \\
   We use a~Scargle periodogram analysis code written by K.~Bjorkman, which is based on the Fourier 
   transformation for data that are not equally spaced using the Lomb-Scargle method \citep{Press}. 
   All the significant peaks are then independently checked by the code 
   \texttt{HEC27}\footnote{http://astro.troja.mff.cuni.cz/ftp/hec/HEC27/},
   written by Petr Harmanec, based on Stellingwerf's method \citep{Stellingwerf78}.

\end{enumerate}


\section{Results}
\label{results}

We discuss the~observed line-profile variations and present $RV$ and $EW$ measurements for the~H$\alpha$, 
H$\beta$, and  [\ion{O}{i}] lines. We did not measure $RV$s and $EW$s of \ion{Fe}{ii}, \ion{Si}{ii}, 
and \ion{He}{i}. The~\ion{Fe}{ii} lines are too faint to be measured. The \ion{Si}{ii} and \ion{He}{i} 
lines have a~complicated structure; therefore, neither $EW$ nor $RV$ have any clear physical meaning. 

To show temporal variability in the~chosen spectral lines we use the grey-scale representation. 
Spectra are rearranged in wavelengths by interpolation  \citep{Steffen} and plotted chronologically in 
a~row. The~value of relative flux is shown by the shades. We plot relative fluxes $F$ (dynamical spectra), 
variance $\left( \frac{F-\bar{F}}{\bar{F}}\right)$, and absolute variance 
$\left(\lvert\frac{F-\bar{F}}{\bar{F}}\rvert\right)$ from the~mean relative flux 
$\bar{F} = \frac{\sum_{1,n} F_{i}}{\sum_{1,n} i}$, where $n$ is the number of observations. We show the~dynamical spectra 
in the printer version of this paper, while the~other figures are only shown in the electronic version.

\subsection{Line identification}

To identify spectral lines, we use the previous identification done by \cite{Doazan65} and 
the~\texttt{NIST}\footnote{P.J. Linstrom and W.G. Mallard, Eds., NIST Chemistry WebBook, 
  NIST Standard Reference Database Number 69, National Institute of Standards and Technology, 
  Gaithersburg MD, 20899, http://webbook.nist.gov, (retrieved 12 June 2014)
} 
database. The~spectrum of HD~50138 in the~chosen interval contains both emission and absorption lines, 
forbidden emission lines of neutral oxygen [\ion{O}{i}] $\lambda \lambda$ 6300, 6364 \AA, and weak emission lines 
of permitted singly ionised iron lines \ion{Fe}{II} $\lambda \lambda$ 6318, 6384 \AA. The~chosen spectral 
region together with the~identification is shown in Fig.~\ref{ident}$^{e)}$. 
\onlfig{
  \begin{figure*}[!h]
    \includegraphics[width=17cm]{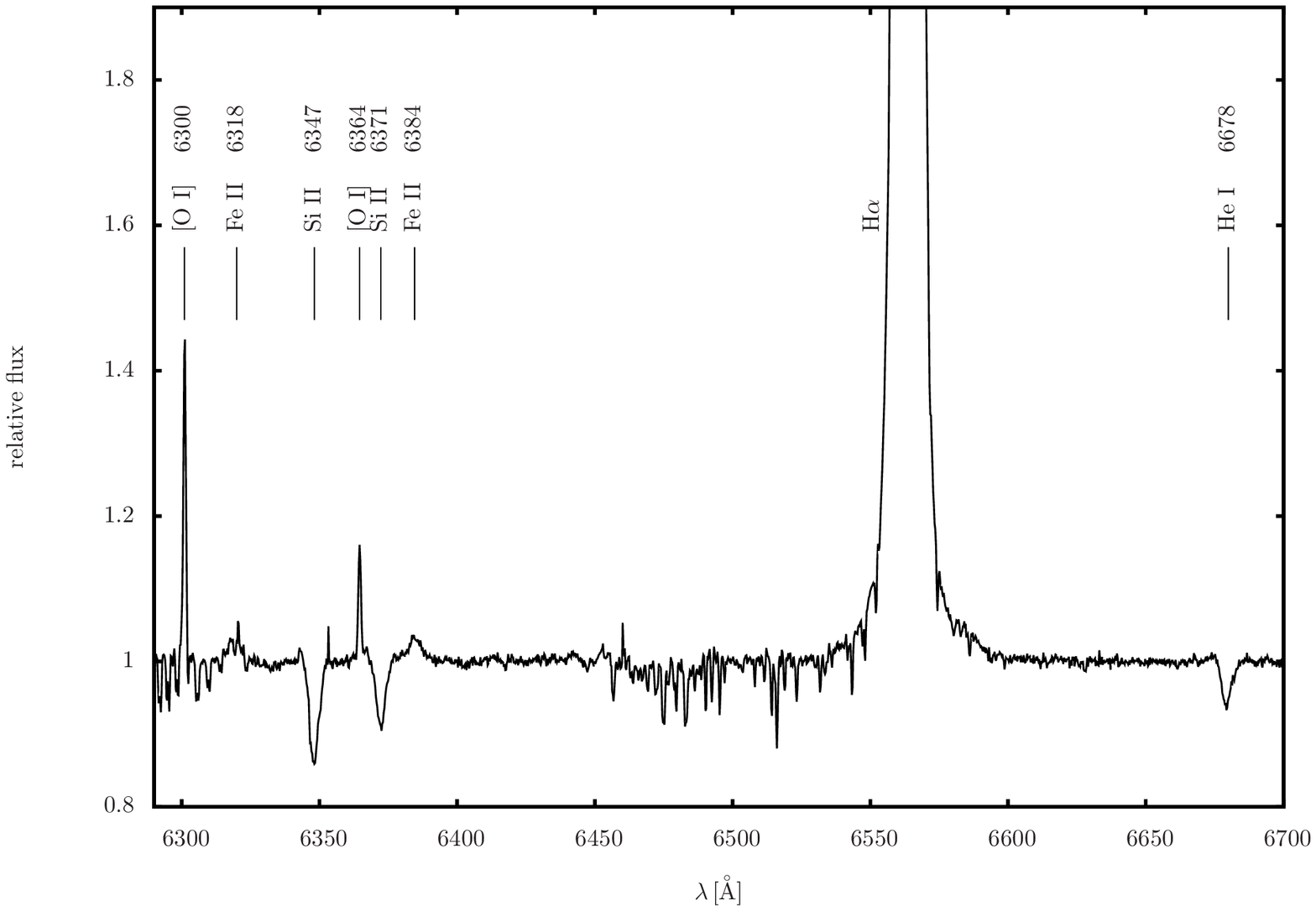}
    \caption{
          Identification of the spectral lines based on the work of \cite{Doazan65} and the~\texttt{NIST} database.
          The spectral interval shows the important lines chosen for the long-term study at the
          Ond\v{r}ejov slit spectra ($R\sim 12\,500$\,\AA).
        }
   \label{ident}
  \end{figure*}
  \FloatBarrier
}

\subsection{H$\alpha$ line}

The~H$\alpha$ line is the~most intense line in the~spectrum of HD~50138. Though one might expect this 
to be an advantage, it was not in this case. It was very difficult to obtain good-quality spectra with 
photographic plates, and the~high-resolution \`{e}chelle spectra usually do not cover enough  
of the~surrounding continuum to normalise it. Moreover, the~H$\alpha$ line forms in an extensive part of 
the~circumstellar medium and hence reflects a~broad range of physical conditions. Analysis of this line 
is very complicated, especially in the extended inhomogeneous circumstellar environments that 
exist around B[e] stars. Recently, numerical models became available for the ana\-lysis. 
It is possible to use a~combination of multidimensional radiative transfer codes 
\citep{Zsargo08, Carciofi06, Carciofi08, Korcakova05} with multidimensional \citep{Cure04} 
and time-dependent \citep{Votruba07} hydrodynamics.
The situation is particularly complicated in the case of HD~50138. The~polarimetric measurements of 
\cite{Oudmaijer99} show that there must be two distinct line-forming regions for the H$\alpha$ line. 
They found that the~red peak shows strong depolarisation, the~violet peak only slight depolarisation, 
but the~polarised 
fluxes are almost equal. They explain this behaviour as a~rotating disc located close to the~star and 
a~single-peaked emission originating in an~extended region.

The~H$\alpha$ line variations can give important supplementary information for the~construction of 
a~realistic model in the~future. Therefore, we present here our observations, and 
also summarise the~measurements from previous studies (Tables~\ref{VR_Halpha_tab_st}$^{e)}$, 
\ref{EW_Ha_tab_st}$^{e)}$, \ref{RV_Halpha_tab_st}$^{e)}$, and \ref{peak_sep_Halfa_tab_st}$^{e)}$).

\subsubsection{H$\alpha$ line-profile variations}

Since the discovery of HD~50138 as an emission-line star in 1920, the H$\alpha$ line has always been observed 
to be very intense and double peaked (see Fig.~\ref{nton}$^{e)}$). The violet peak was smaller than 
the red peak in all previous studies (Table~\ref{VR_Halpha_tab_st}$^{e)}$). However, we measured
the flux ratio of the violet and red peaks ($V/R$, Fig.~\ref{EWHa}, bottom panel) to exceed 1.0
in two spectra (4 and 10 October 2008 at SPM).

The profile of the violet peak is more complicated than the red one. Moving humps were 
often detected in high-resolution data, sometimes so strong that it can be called
peak splitting (\citealp{Dachs92}, Fig.~6; our observations on 6 and 16 February 1994, RO,
and 10 November 2007, SPM). Our data sample is sufficiently rich to be able to describe the 
behaviour of the moving humps (Figs.~\ref{SHaIIc} and \ref{SPAHaIIc}$^{e)}$) and measure their 
$RV$s (Fig.~\ref{HaRVcomplet}, lowermost panel). This could be a~good tracer of the circumstellar 
environment, e.g. the rotation or expansion of the matter. The humps were also detected in the red peak; 
however, their appearance is very rare. We were able to capture it at the high-resolution data 
from the RO on 11 and 21 September 1994, and several times it appeared only as a~small deformation 
of the red peak in data sets from every observatory.

Very important is the detection of a~small emission peak in the central absorption (Fig.~\ref{cqepHa}$^{e)}$), 
which could be a~signature of the binarity of the object. This emission can originate 
near the Lagrangian point L1 and hence appears at the 
zero velocity of the system. It has only been observed once before, on 22
March 1991 \citep{Bopp93}. We were able to catch it at 
seven spectra from RO (22/12/1994, 9/2/1998, 15/2/1998, 30/3/1998, 6/4/1998, 13/4/1998, 6/3/2000), 
and three spectra from TCO (10/12/2013, 15/12/2013, and 26/12/2013). 
The spectra taken on 15 February 1998, 30 March 1998 exclude possible contamination by the water absorption.

\onlfig{
 \begin{figure}[!h]
   \resizebox{0.5\textwidth}{!}{\includegraphics{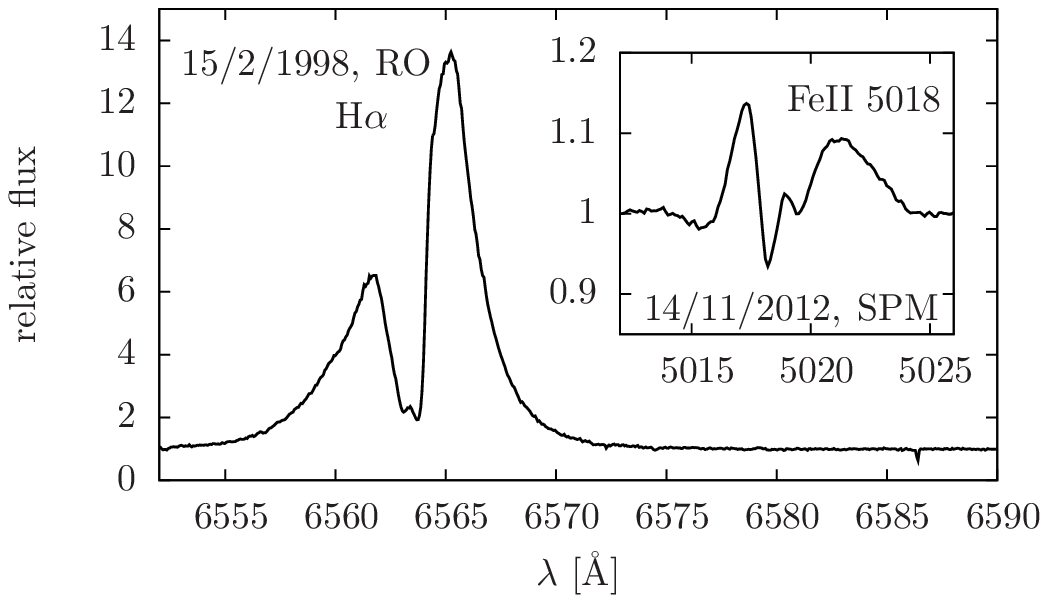}}
   \caption{Central quasiemission peak (CQEP) of the H$\alpha$ (15 February 1998, RO), and
     \ion{Fe}{ii} 5018~\AA\, line (14 November 2012, SPM). The CQEP were not detected
     simultaneously in the \ion{Fe}{ii} lines and H$\alpha$.
     }
  \label{cqepHa}
 \end{figure}
\FloatBarrier
}
The H$\alpha$ line is strongly variable. Both night-to-night changes as well as long-term changes have 
been mentioned in the literature many times. The complex study of night-to-night variations is presented in
\cite{Pogodin97}. He analysed 71 spectra obtained from 15 to 18 March 1994. The amplitude of $EW$ 
changes is up to 15\%. The residuals of the line profiles, tracing important features of the circumstellar 
matter, are based on a~few average spectra from each night. This observation strategy is able to reveal 
the variations on the timescale of hours; however, no note is mentioned there. In order to 
describe the variability on timescales from hours up to years, we acquired sequential spectra during 
one night, and several following nights on a~number of different occasions. The H$\alpha$ line profiles 
from one such run (March 2007, OO) is plotted in Fig.~\ref{nton}$^{e)}$. We found that the changes 
in the H$\alpha$ line during one night are negligible. This allows us to study the long-term variability, 
which to date has not been described.

\onlfig{
  \begin{figure*}[!h]
    \includegraphics[width=18cm]{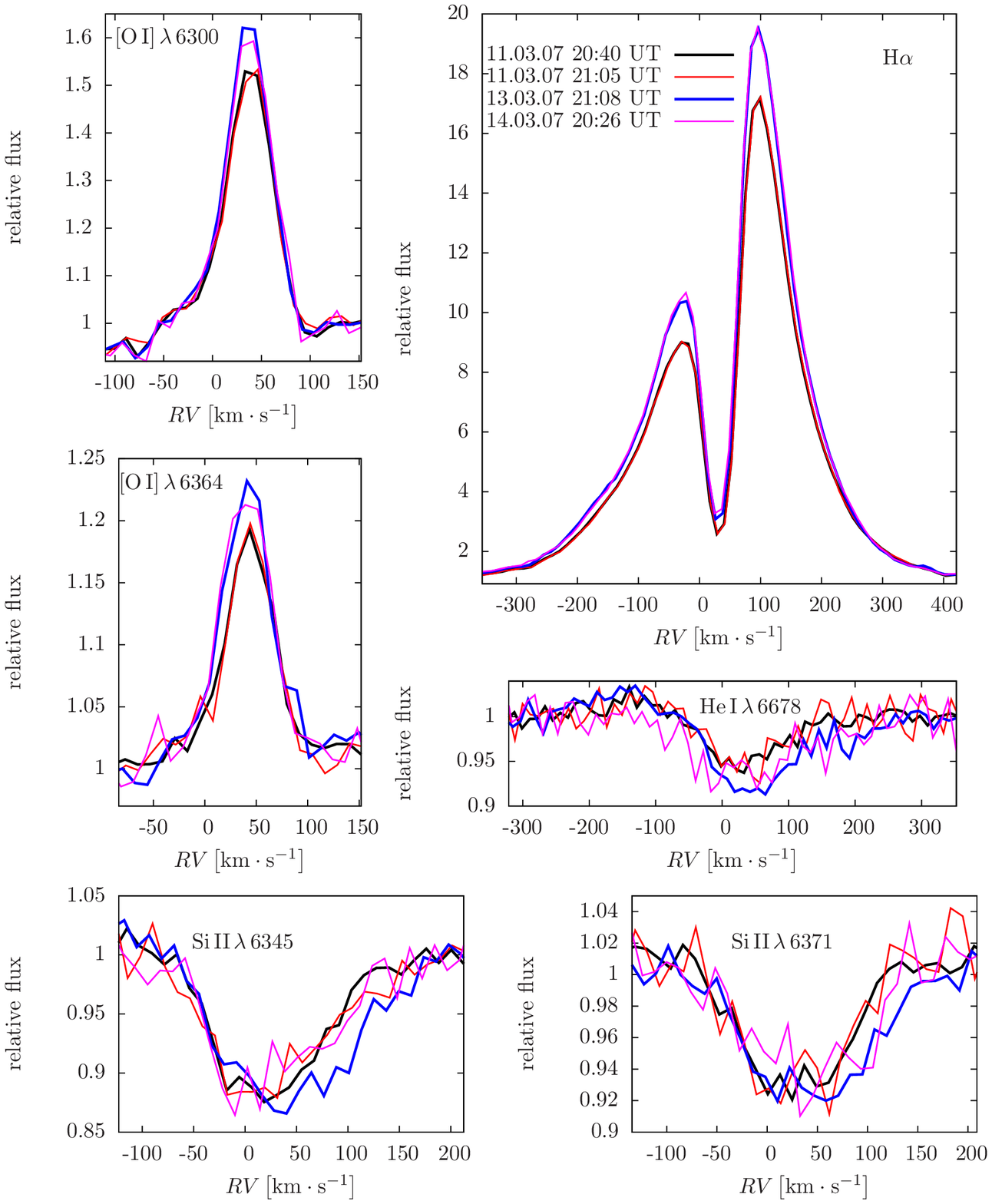}
    \caption{Night-to-night spectral line changes.}
    \label{nton}
  \end{figure*}
  \FloatBarrier
   }

\begin{figure}[!Htp]
    \resizebox{\hsize}{!}{\includegraphics{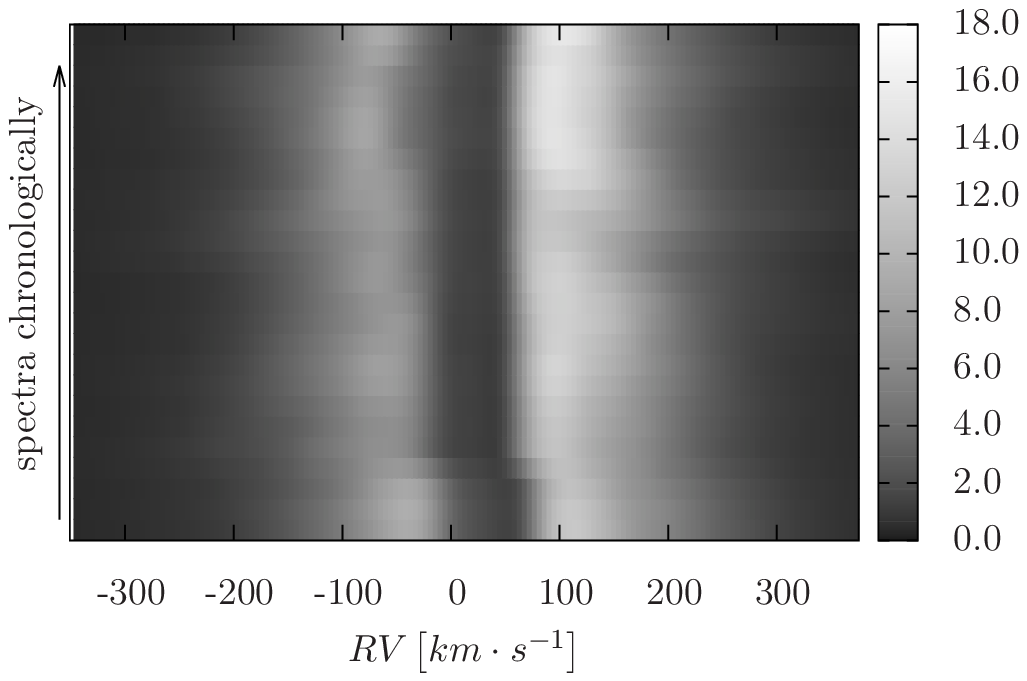}}
    \resizebox{\hsize}{!}{\includegraphics{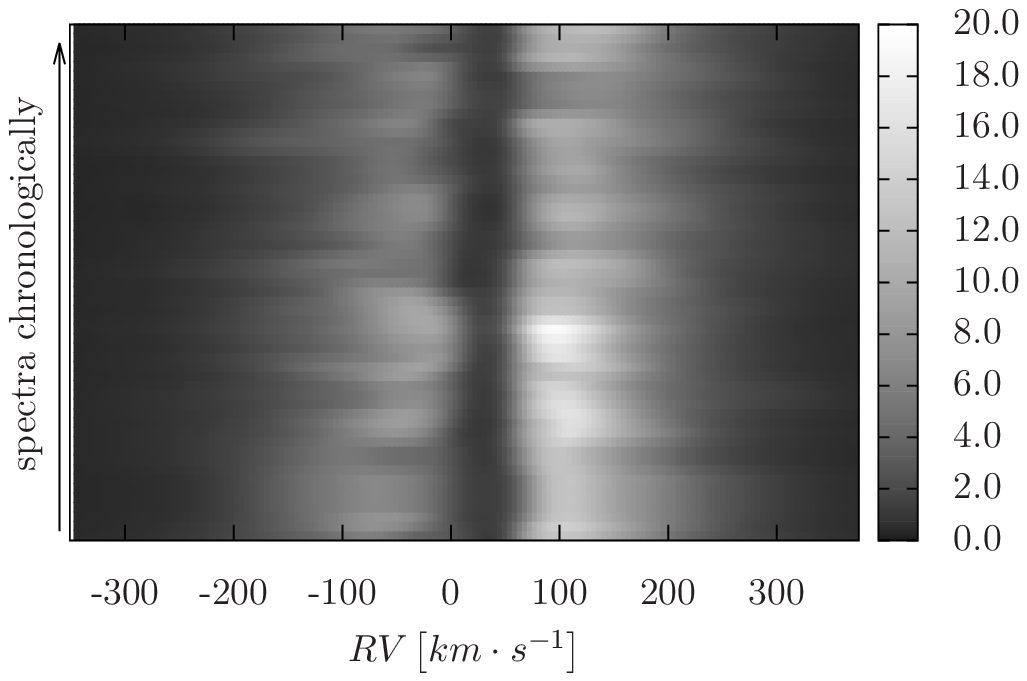}}
    \resizebox{\hsize}{!}{\includegraphics{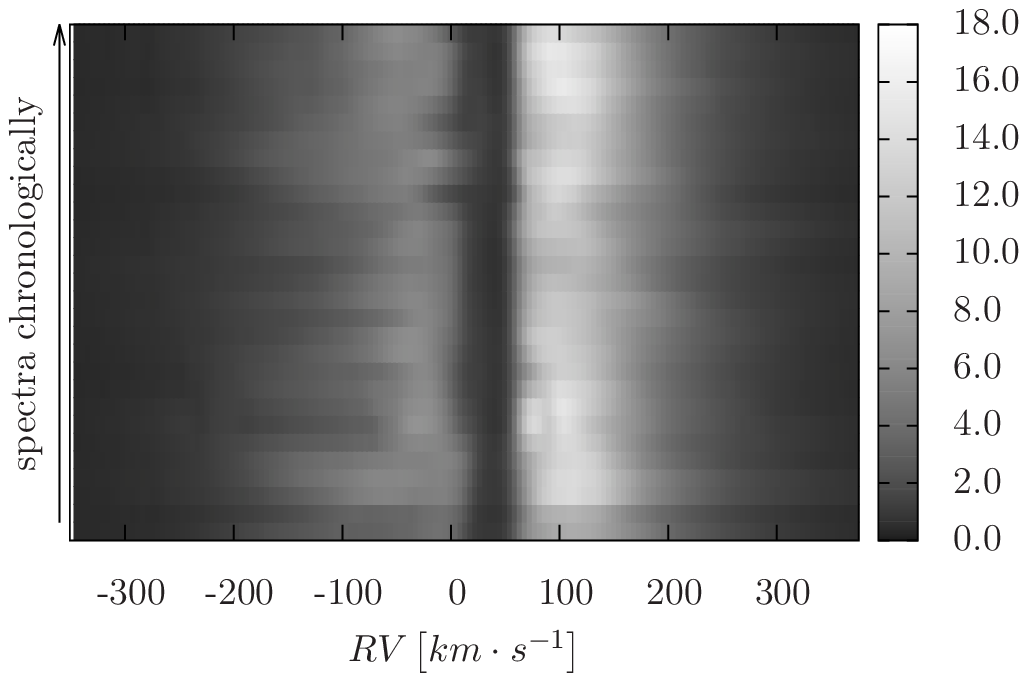}}
    \caption{Grey-scale representation of the~H$\alpha$ line. 
	Top: spectra from TCO, Middle: spectra 
	from OO, Bottom: spectra from RO.}
    \label{SHaIIc}
\end{figure}

\onlfig{
  \begin{figure}[!h]
     \resizebox{\hsize}{!}{\includegraphics{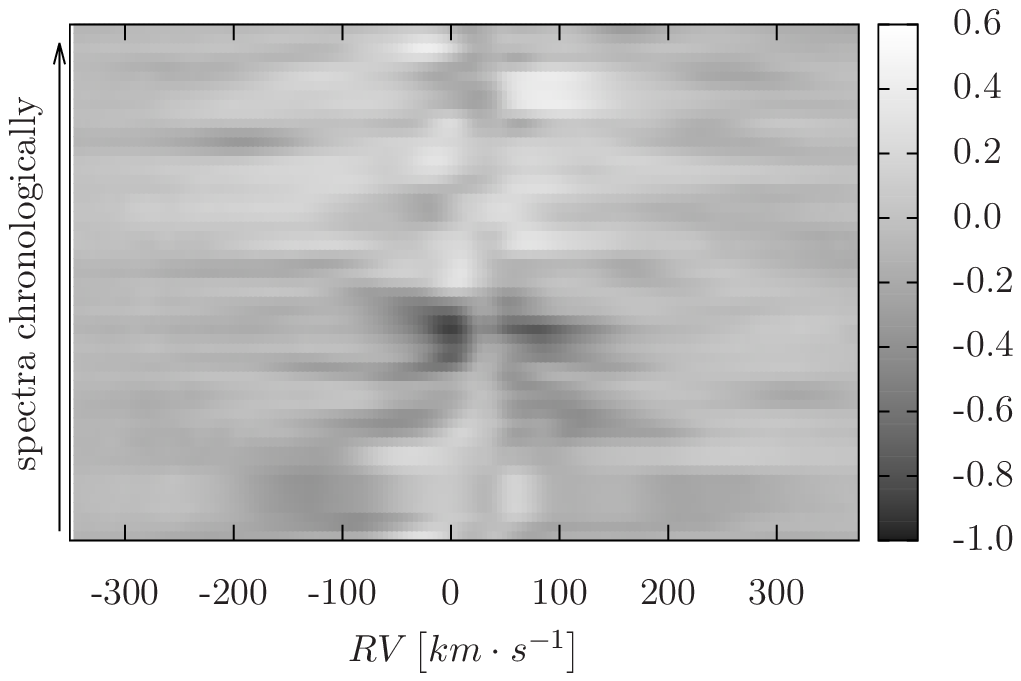}}
     \resizebox{\hsize}{!}{\includegraphics{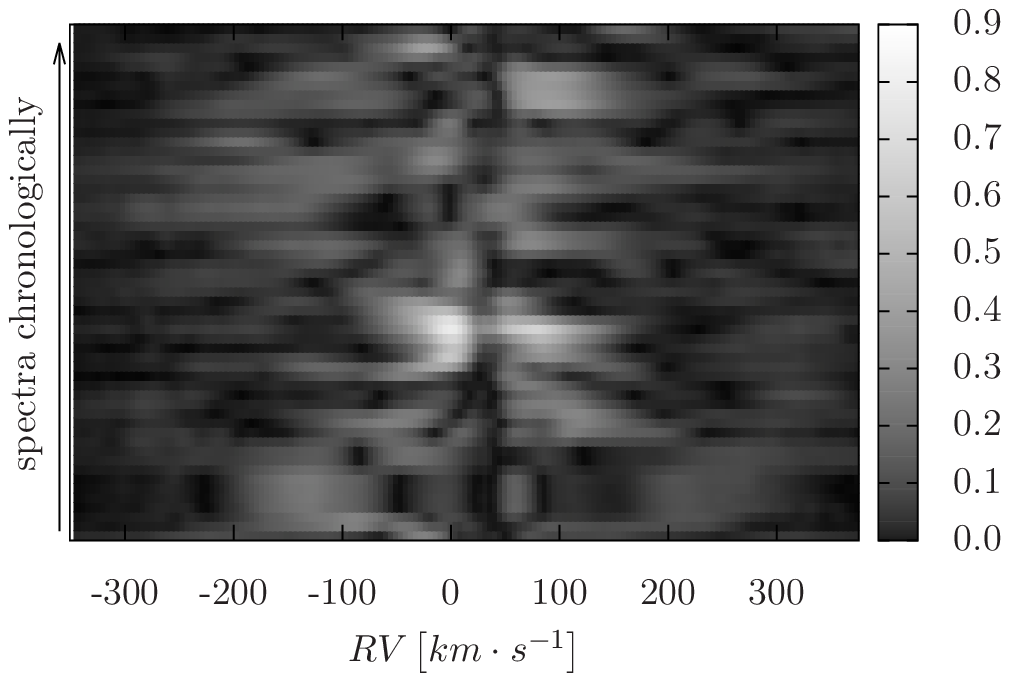}}
     \caption{Grey-scale representation of the~H$\alpha$ line on the spectra from OO.  
     The residuals between the given line profile and averaged line profile 
     (upper panel). To emphasise the structures, the absolute value 
     of the differences is plotted on the lower panel.
     } 
     \label{SPAHaIIc}
  \end{figure}
  \FloatBarrier
        }

\FloatBarrier

\begin{figure*}
\centerline{
  \resizebox{0.5\textwidth}{!}{\includegraphics{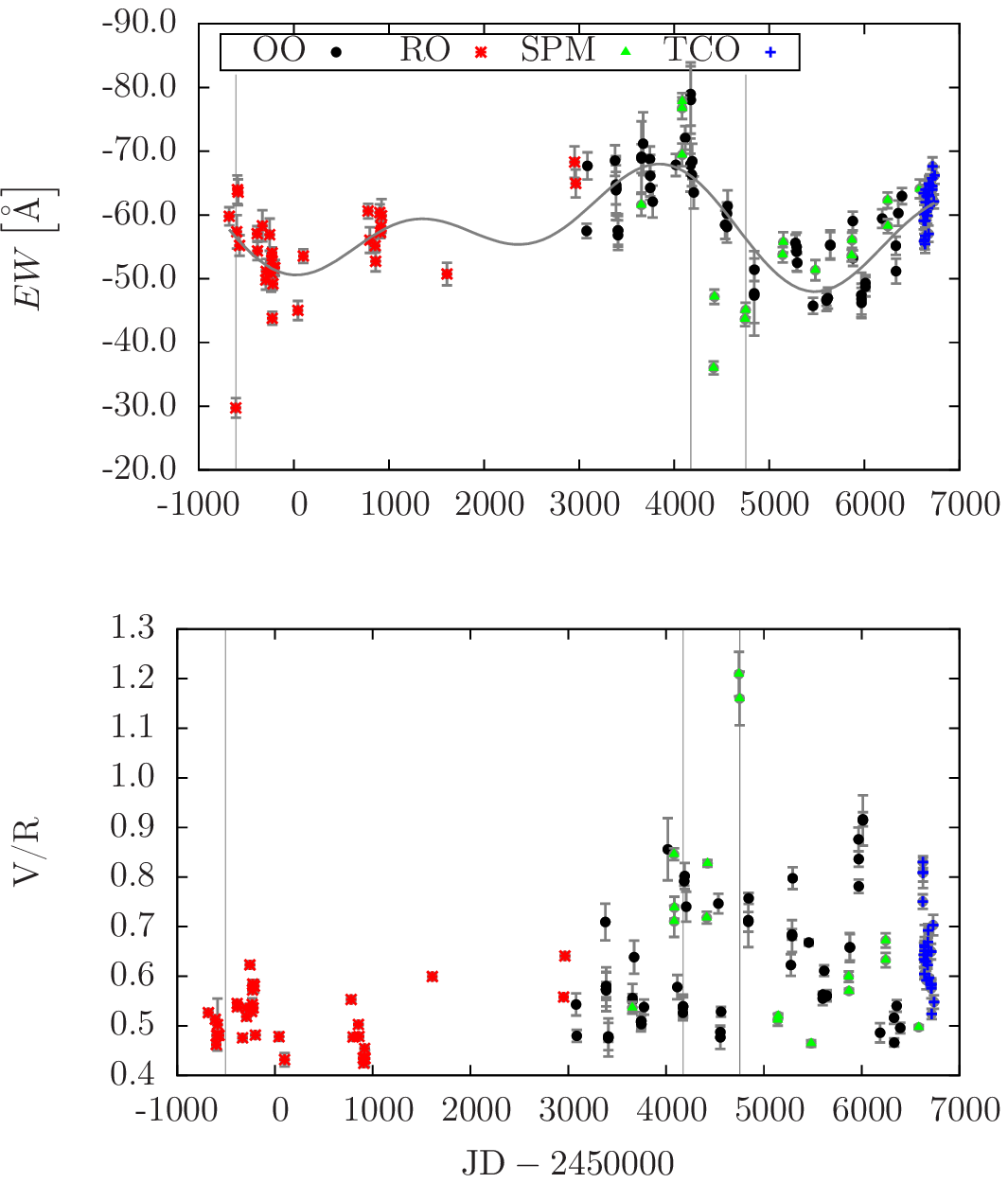}}
  \resizebox{0.5\textwidth}{!}{\includegraphics{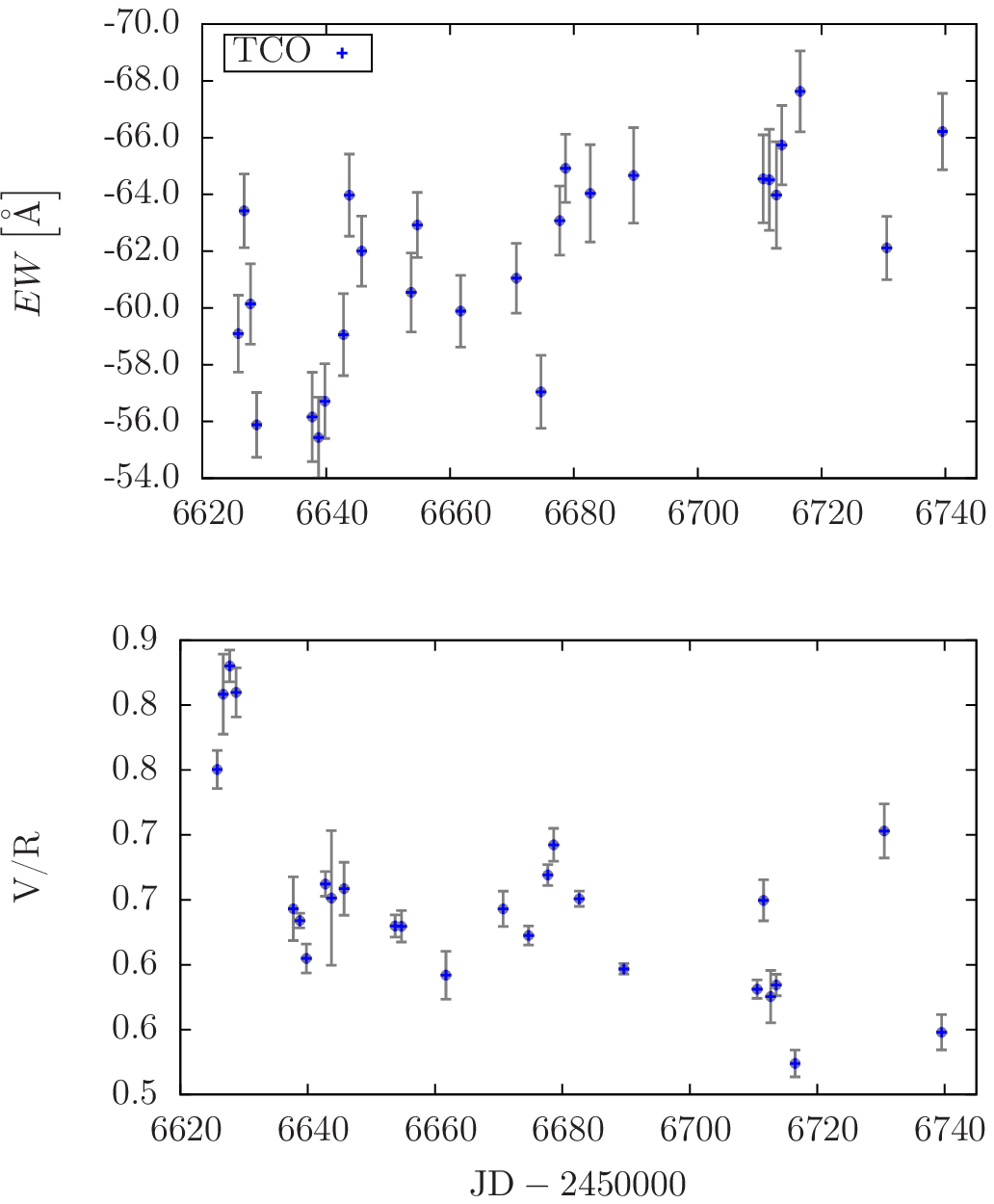}}
  }
  \caption{$EW$ and $V/R$ ratio of relative fluxes  of the~ H$\alpha$ line.
     Vertical lines indicate the minimum and maximum values 
     of $|EW|$ ($JD = 2\,449\,389.68$, and $2\,454\,174.28$), and
     time when $V/R > 1$ ($JD=2\,454\,752.92$), respectively. 
     }
  \label{EWHa}
\end{figure*}

\subsubsection{Relative fluxes, $V/R$ ratio}

To describe the variations of the H$\alpha$ line, we measure the intensity of both 
of the peaks and of the central depression. 
The values obtained by the polynomial fitting (Sect.~\ref{analysis},\,\textit{iv})
are presented in Fig.~\ref{HaF}$^{e)}$. The ratio of the violet and red peak is plotted 
in the bottom panel of Fig.~\ref{EWHa}. To see the variability on longer timescales 
than our observations, we summarise all the measurements presented in the previous works 
in Table~\ref{VR_Halpha_tab_st}$^{e)}$. Even if the $V/R$ ratio does not conserve with 
respect to different resolutions for asymmetric lines, it can be used for the rough 
guess of the variability. The values of $V/R$ were observed in the interval from 0.36 \citep{Borges09}
up to 1.21 (our data, Fig.~\ref{EWHa}, 4 October 2008). The violet peak was always smaller
the red one, with the exception of a~short period in October 2008 caught in our spectra.

\onlfig{
 \begin{figure*}[!h]
 \centerline{  
   \resizebox{0.5\textwidth}{!}{\includegraphics{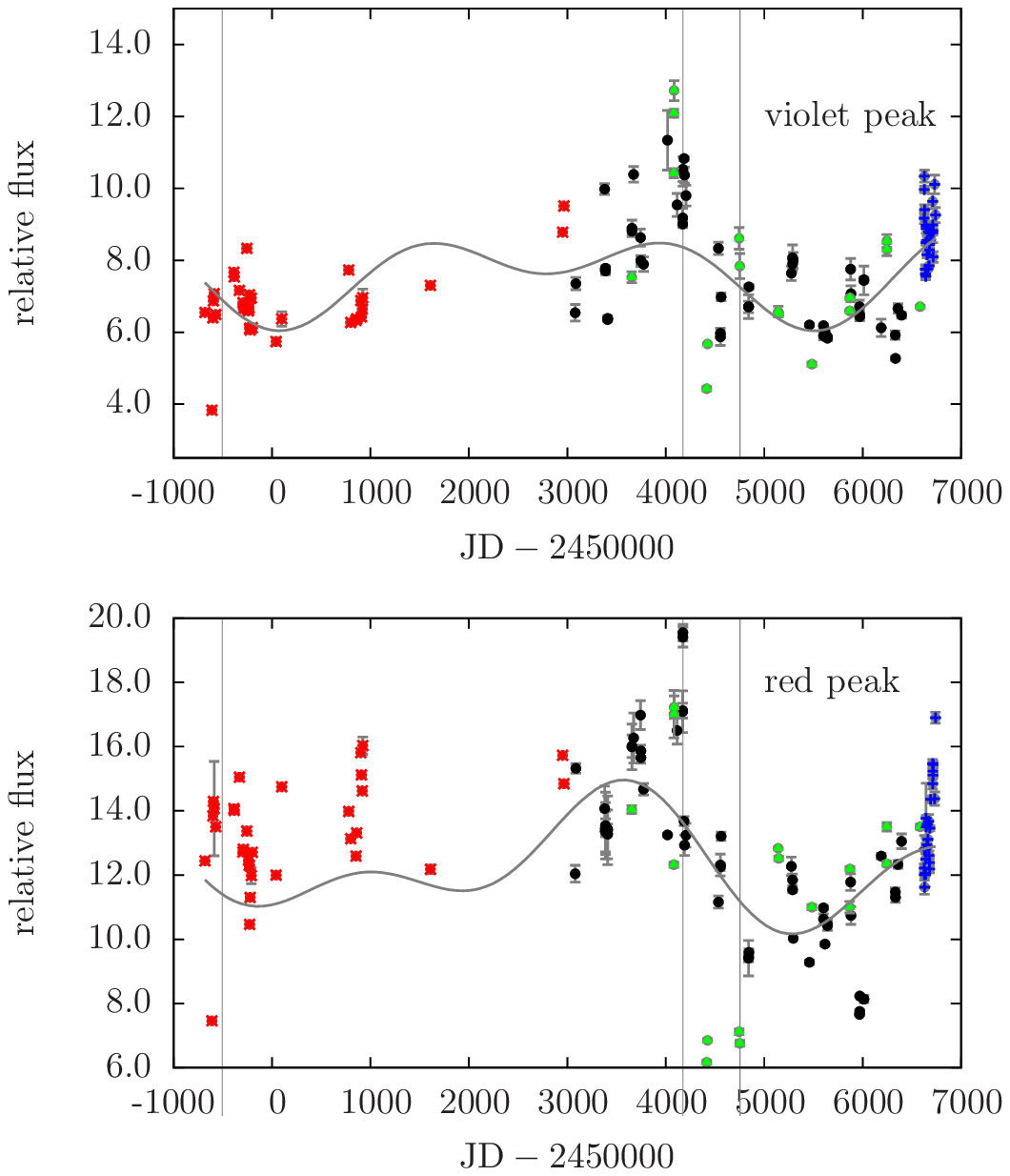}}
   \resizebox{0.5\textwidth}{!}{\includegraphics{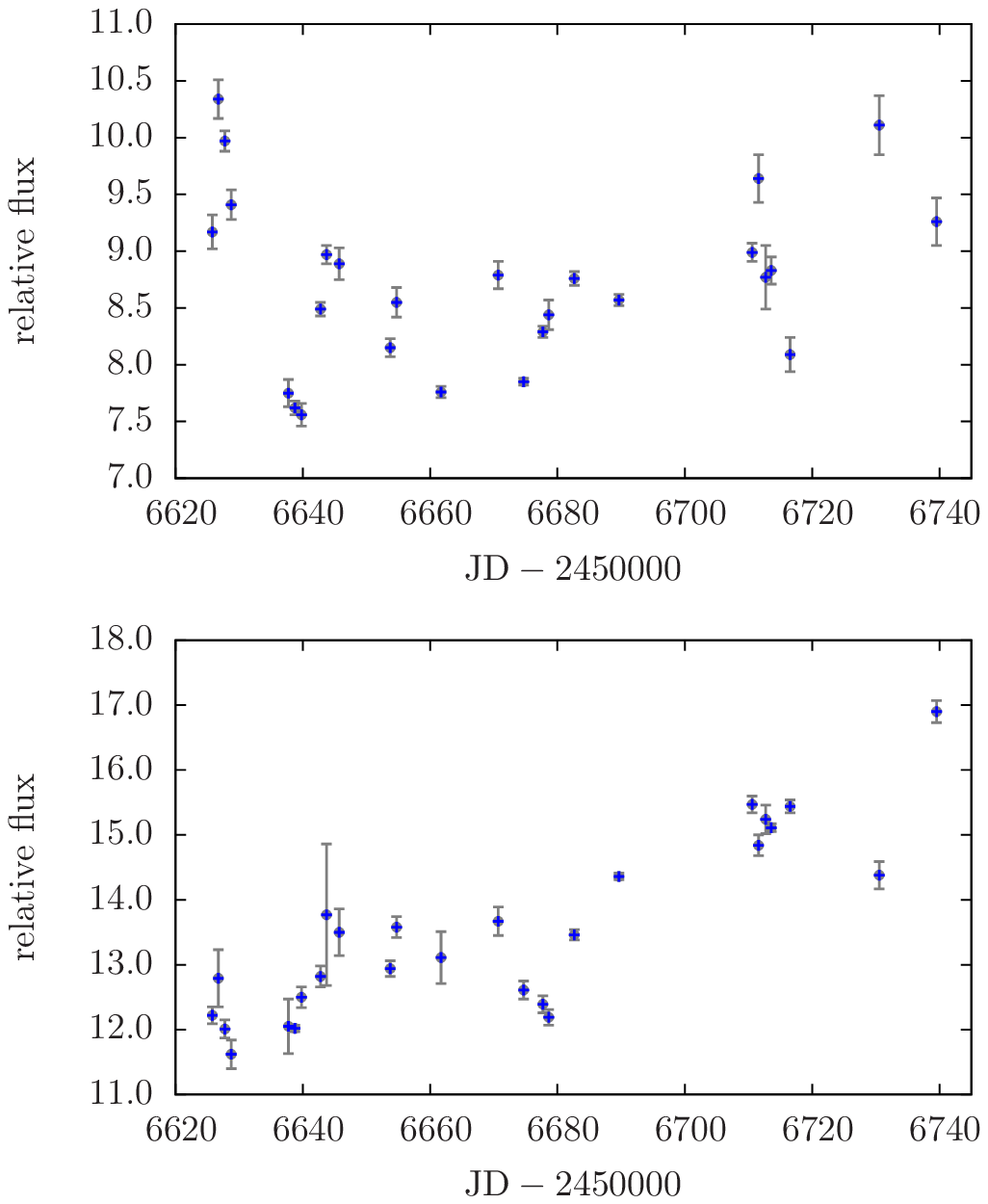}}
 }
 \centerline{
   \resizebox{0.5\textwidth}{!}{\includegraphics{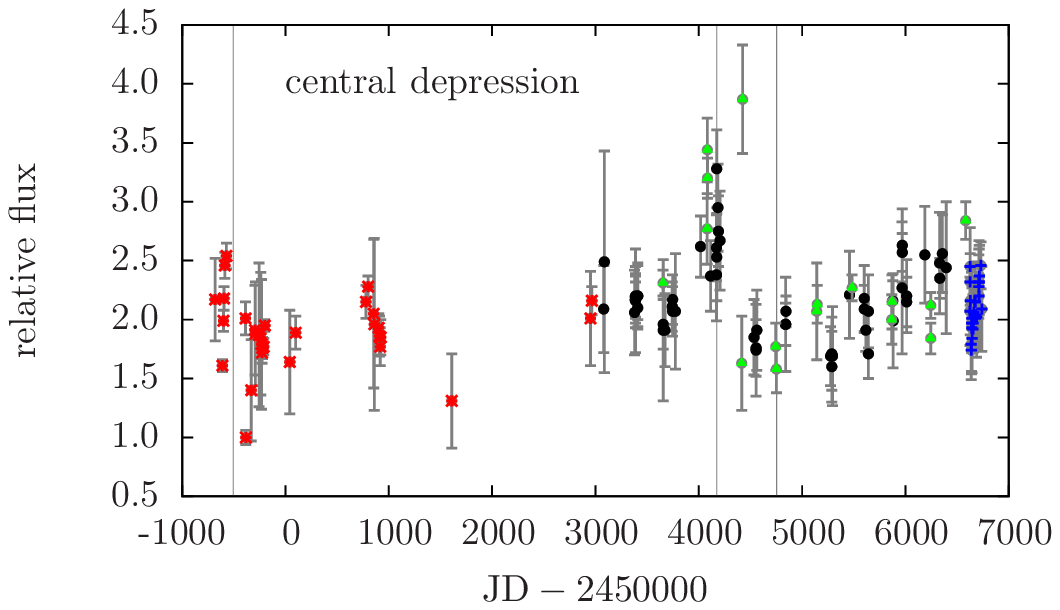}}
   \resizebox{0.5\textwidth}{!}{\includegraphics{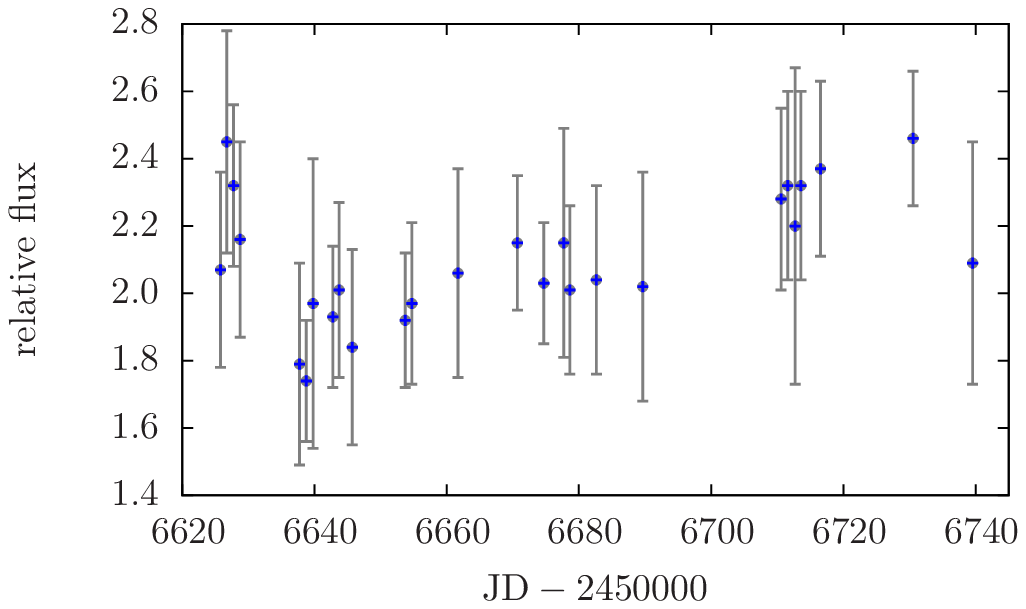}}
 } 
 \caption{Relative fluxes of violet and red peak (upper) and of the central depression 
   of the H$\alpha$ line. The minimum and maximum value of $|EW|$ of the H$\alpha$ line 
   and its $V/R>1$ are shown by the vertical lines. The curve at two upper graphs is
   a~fit of linear combinations of two sine functions with the fixed periods obtained from 
   the period analysis of the H$\alpha$ $EW$s.
 }
 \label{HaF}
 \end{figure*}
 \FloatBarrier
}

\onltab{
  \begin{table}[!h]
    \caption{  $V/R$ changes of the~H$\alpha$ line}
    \label{VR_Halpha_tab_st}
    \begin{center}
      \begin{tabular}{llll}
        \hline \hline
               date         &  V/R                & Dispersion/   & Ref.\\
                    &                     & Resolution   & \\
    \hline
       1981-Feb-01   &  0.65                & 5.5 \AA/mm         & 1 \\
       1987-Feb-09   &  0.388$^{a}$         & $R=50\,000$        & 2 \\
                    &  0.384               &                    &\\
       1987-Mar-22   &  0.6                 & $R=7\,300$    & 3 \\
       1999-Oct-17   & $0.36$               & $R = 55\,000$     & 4\\
       2007-Oct-04   & $0.88$               & $R = 55\,000$     & 4\\
   \hline

      \end{tabular}
    \end{center}
    \tablebib{(1)~\citet{Andrillat82}; (2) \citet{Dachs92}; (3) \citet{Halbedel91}; (4)~\citet{Borges09}.}      
    \tablefoot{\tablefoottext{a}{double peak violet part with almost the same intensity}}
  \end{table}
}

\subsubsection{Equivalent width} 

We measure the $EW$ of the~H$\alpha$ line by numerical integration (Sect.~\ref{analysis},\,\textit{i}).
Results of our measurements are plotted in Fig.~\ref{EWHa}. The  maximum value (minimum line strength) 
$-29.74 \pm 1.54$~\AA\, is obtained on 6 February 1994. The minimum value (maximum line strength) in our data 
$-78.96 \pm 1.65$~\AA\, corresponds to 13 March 2007. The previous observations, summarised in  
Table~\ref{EW_Ha_tab_st}$^{e)}$, are in the interval defined by these limits with only one exception; 
the first measurement on 20 February 1960 \citep{Doazan65}. The $EW$ at the beginning of 
the 1960s was $-116$~\AA. Unfortunately, we can find no published data for more than twenty years 
following the 1960s. We can only guess that the variability of the object is greater than is shown in our spectra.

\onltab{
  \begin{table}[!h]
    \caption{$EW$ of the~H$\alpha$ line}
    \label{EW_Ha_tab_st}
    \begin{center}
    \begin{tabular}{llll}
        \hline \hline
               date         &  EW             & Dispersion/   & Ref.\\
                    &  (\AA)         & Resolution &  \\
    \hline
       1960-Feb-20   & $-116$         & 12.4 \AA/mm       & 1 \\
       1981-Feb-01   & $-47.2$        & 5.5 \AA/mm        & 2 \\
       1987-Feb-09   & $-50.1 \pm 3$  & $R=50\,000$       & 3 \\
       1987-Mar-22   & $-58.4$        & $R= 7\,300$       & 4 \\  
       1994-Mar-15   & $-49.0$        & $R =40\,000$      & 5 \\
       1994-Mar-16   & $-56.2$        & $R =40\,000$      & 5 \\
       1995-Jan-11   & $-67 \pm 4$    & $R = 5\,000$      & 6 \\ 
       1997-Jan-01   & $-58 \pm 3$    & $R = 5\,000$      & 6 \\ 
       1999-Dec-18   & $-54 \pm 3$    & $R = 8\,500$      & 7\\     
       2002-Jan-28   & $-71 \pm 7$    & $R=  7\,500$      &  8 \\
       2002-Sep-20   & $-77 \pm 7$    & $R= 15\,000$      & 8 \\ 
       1999-Oct-17   & $-61.35$       & $R= 55\,000$      & 9\\
       2007-Oct-04   & $-57.56$       & $R= 55\,000$      & 9\\
    \hline

    \end{tabular}
    \end{center}
    \tablebib{(1)~\citet{Doazan65}; (2)~\citet{Andrillat82}; 
    (3)~\citet{Dachs92}; (4)~\citet{Halbedel91}; (5)~\citet{Pogodin97}; 
    (6)~\citet{Oudmaijer99}; (7)~\citet{Vink02}; (8)~\citet{Baines06}; 
    (9)~\citet{Borges09}.}
   \end{table}
}

\subsubsection{Radial velocities}
\label{rv_Halpha}

\begin{figure*}
  \centerline{
    \resizebox{0.46\textwidth}{!}{\includegraphics{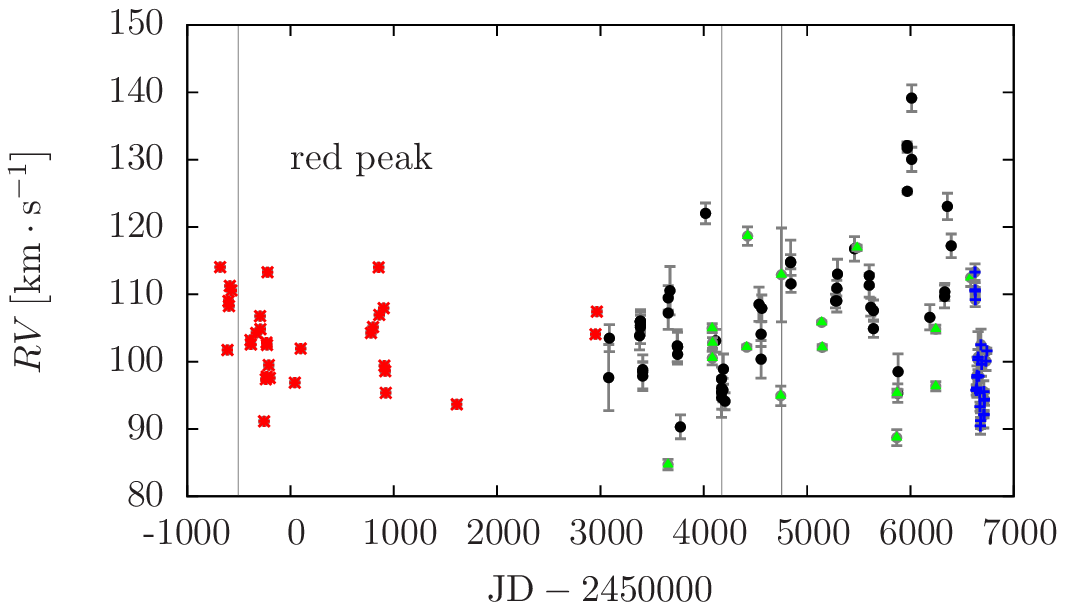}}
     \resizebox{0.46\textwidth}{!}{\includegraphics{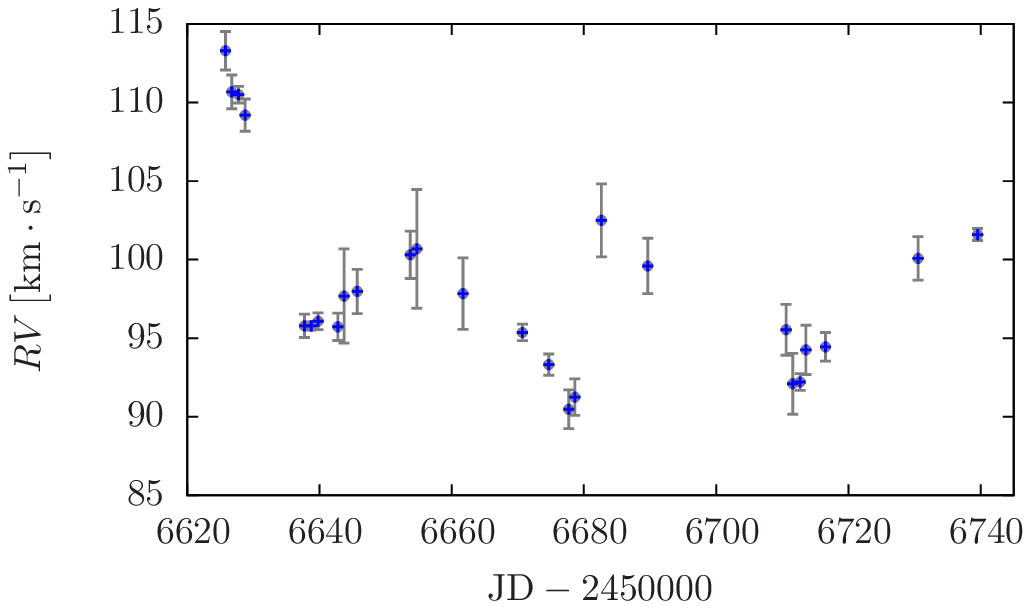}}
  }
  \centerline{
    \resizebox{0.46\textwidth}{!}{\includegraphics{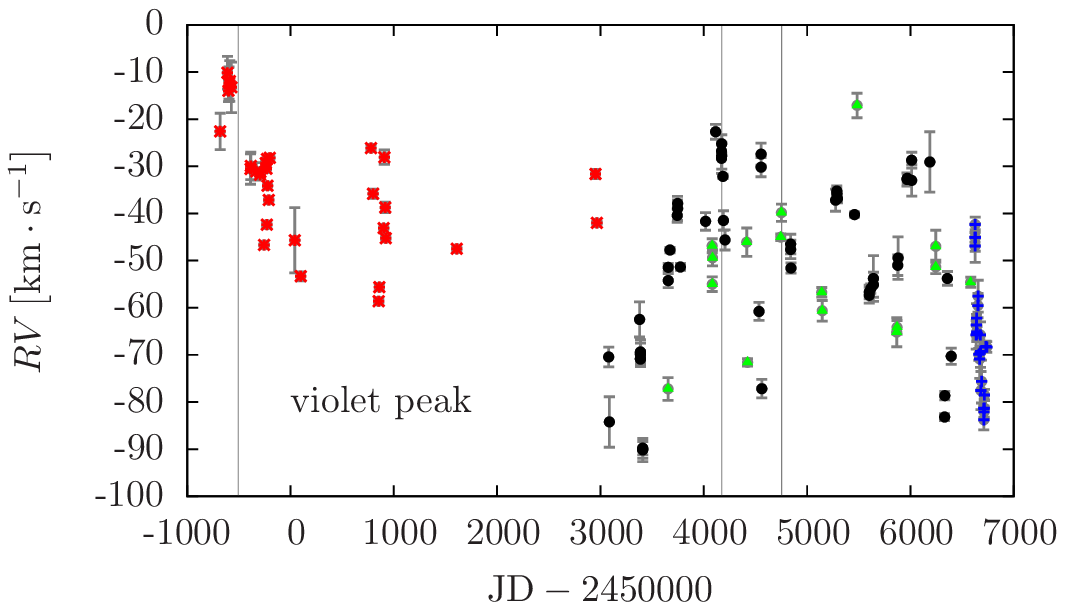}}
    \resizebox{0.46\textwidth}{!}{\includegraphics{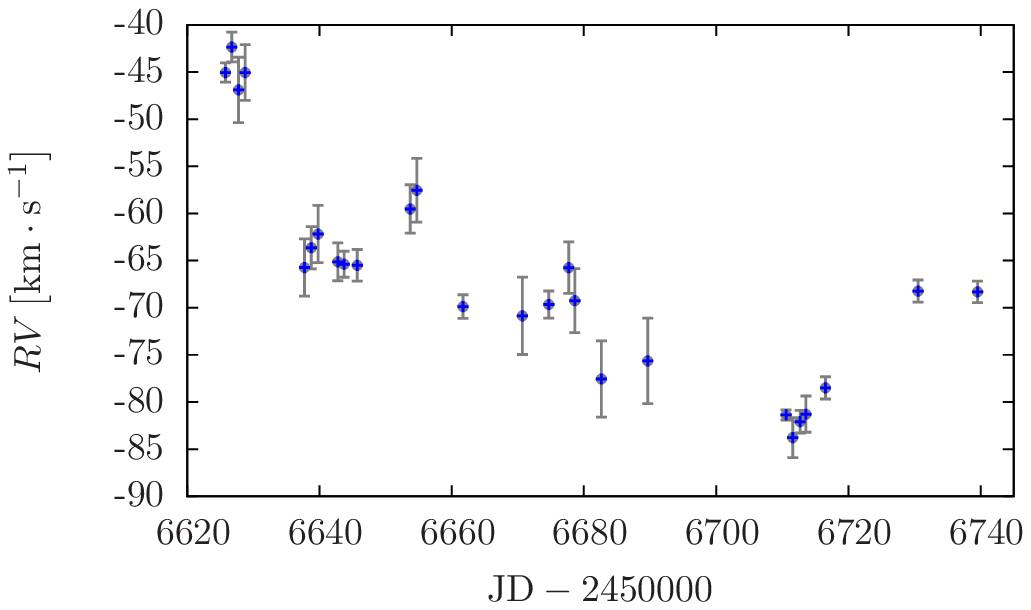}}
  }
  \centerline{
    \resizebox{0.46\textwidth}{!}{\includegraphics{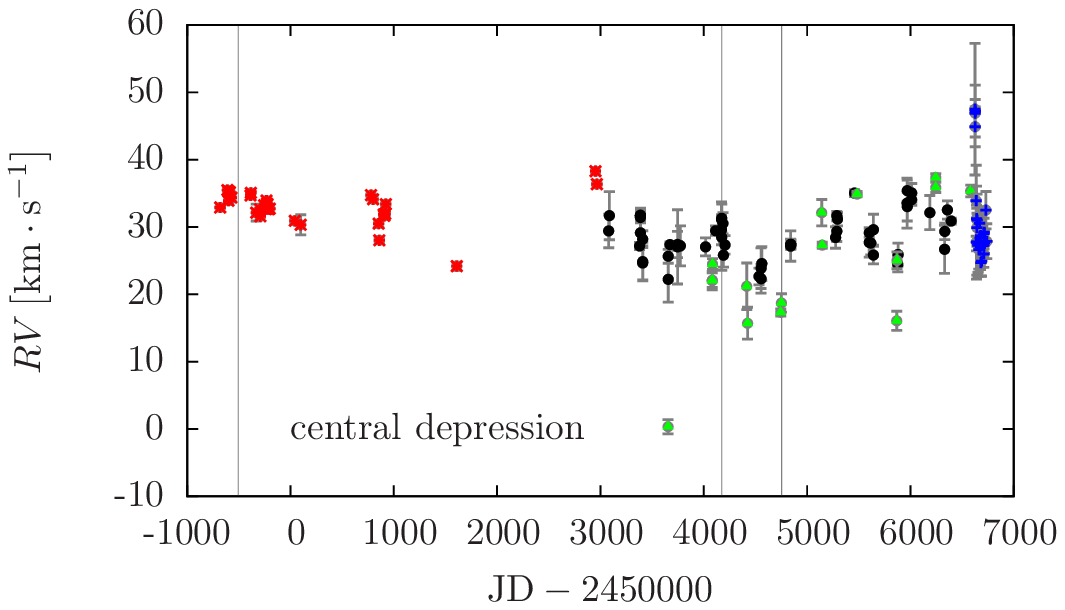}}
    \resizebox{0.46\textwidth}{!}{\includegraphics{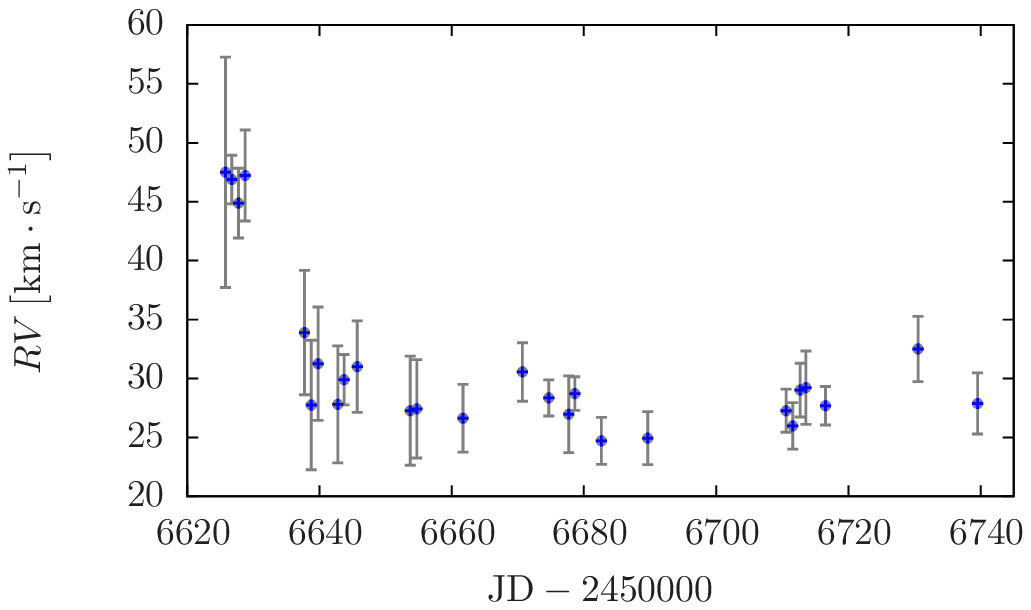}}
  }
  \centerline{
    \resizebox{0.46\textwidth}{!}{\includegraphics{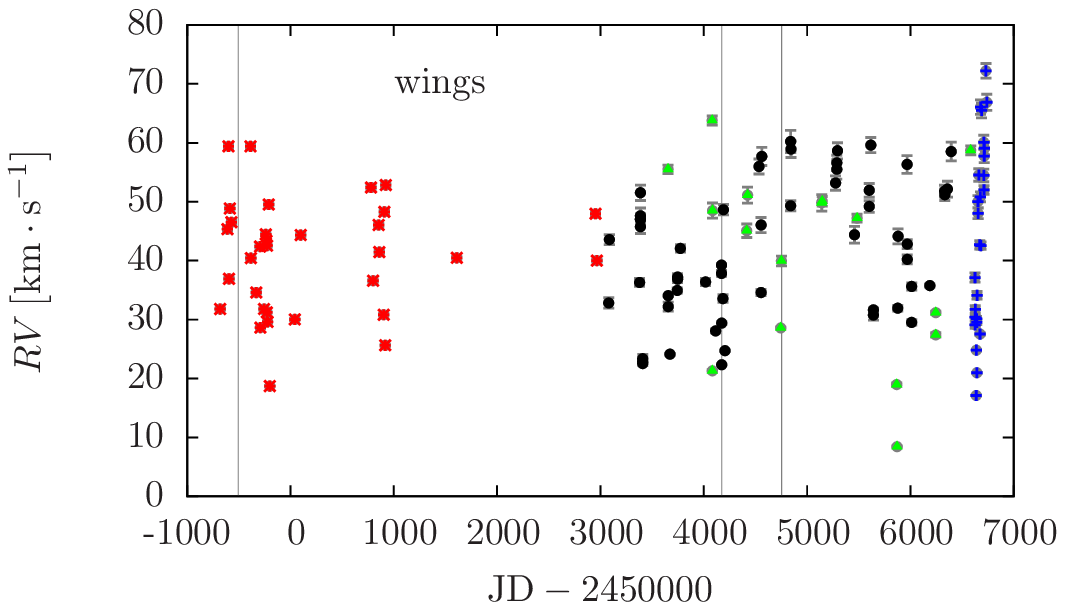}}
    \resizebox{0.46\textwidth}{!}{\includegraphics{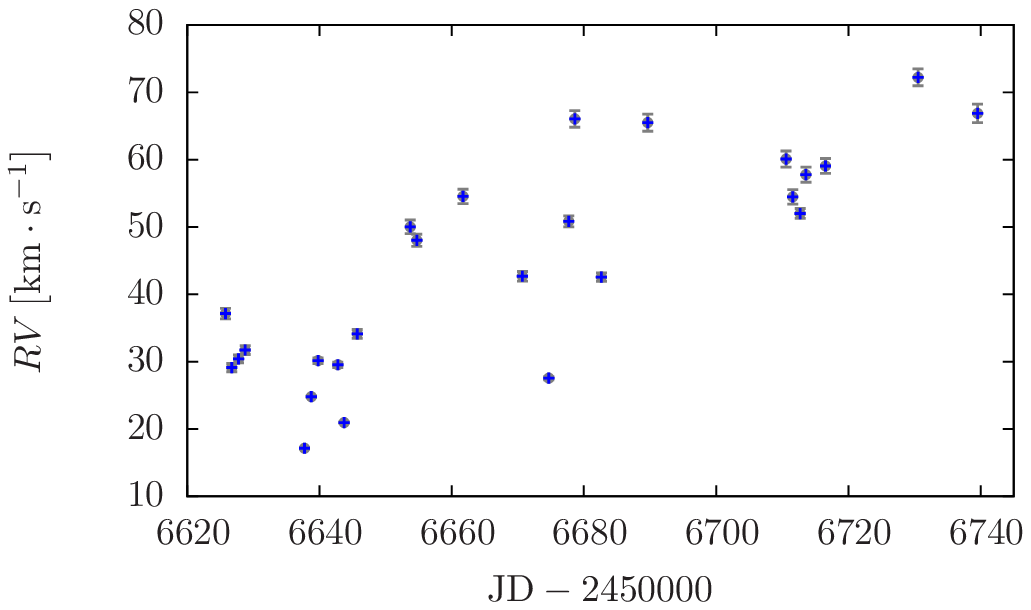}}
  }
  \centerline{
    \resizebox{0.46\textwidth}{!}{\includegraphics{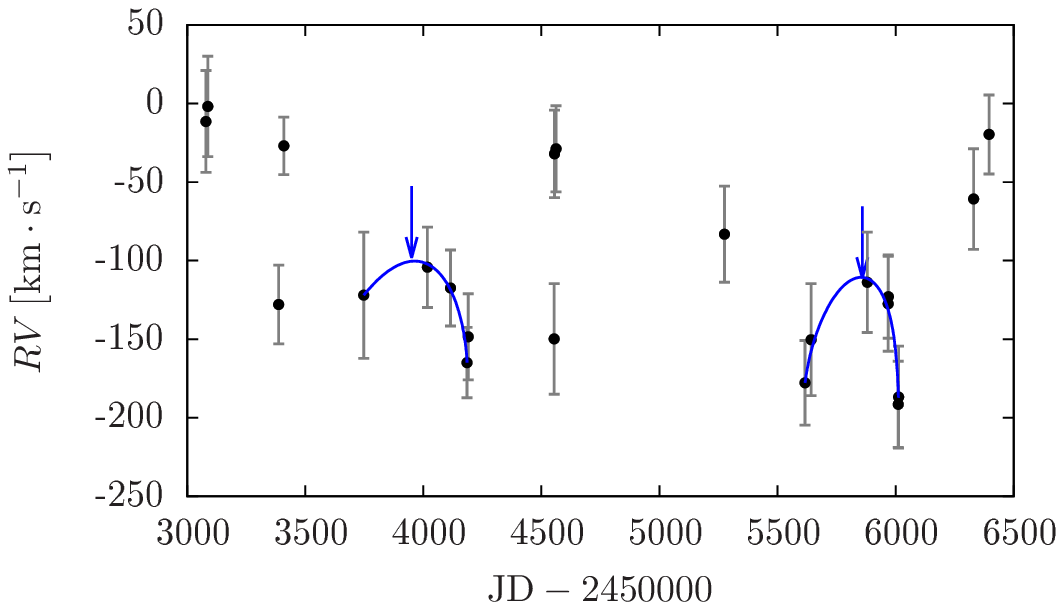}}
    \resizebox{0.46\textwidth}{!}{\includegraphics{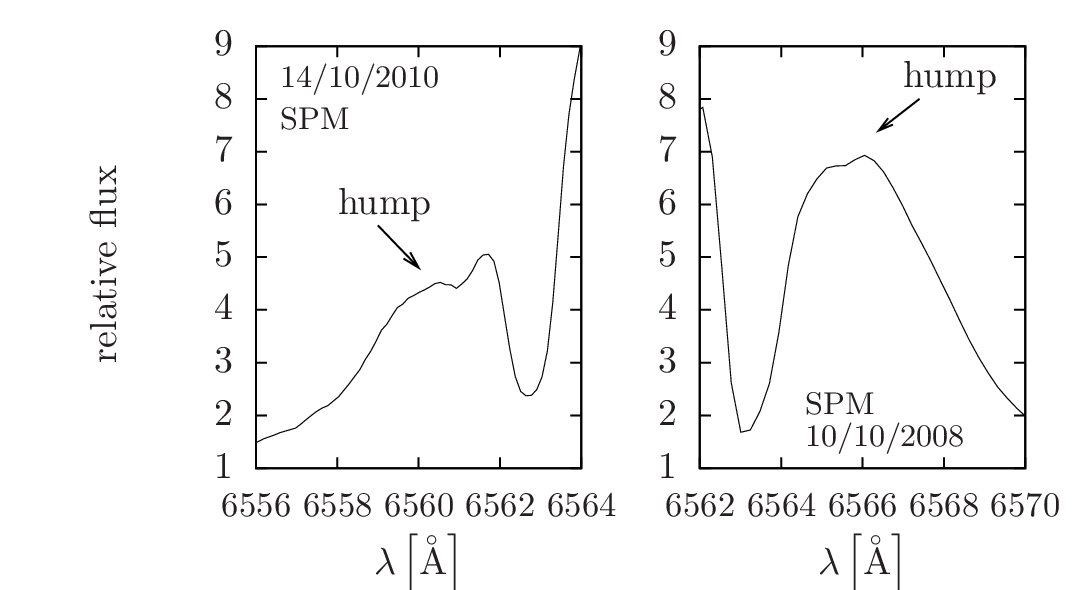}}
  } 
    \caption{$RV$s of the H$\alpha$ features. The right panels show observations obtained at TCO 
        during the last season in detail. $RV$s of the~red peak, the~violet peak, the~central depression, 
        wing, and humps are shown from top to bottom. Parabolic fits to the RV variations of the humps 
        detected during two seasons are shown by the solid lines with arrows. The vertical lines indicate 
        the same epochs as in Fig.~\ref{EWHa}.
    }
    \label{HaRVcomplet}
\end{figure*}

The $RV$ of both peaks (Fig.~\ref{HaRVcomplet}, top two panels) and central depression 
(Fig.~\ref{HaRVcomplet}, central panel) of the~H$\alpha$ line is estimated by the least squares
fitting of the polynomial (Sect.~\ref{analysis},\,\textit{iv}). The hump in the blue peak of 
the H$\alpha$ line was strong enough in some spectra to be measured. The sampling of measurements 
is sufficiently fine to reveal the motion of the humps for two epochs, where the $RV$ of the humps  
(Fig.~\ref{HaRVcomplet}, lowermost panel) can be fit by a~parabolic function (blue lines in this 
figure). This shape indicates the rotation motion of the humps more than the pure expansion, which 
will follow a~linear dependence. The wings of the H$\alpha$ line are measured using the mirror 
method in the relative flux between values 1.05 and 1.4 (Sect.~\ref{analysis},\,\textit{v}).    
The results are plotted in Fig.~\ref{HaRVcomplet} (fourth panel).

The $RV$ of the central depression, peaks, and wings from previous studies are listed in  
Table~\ref{RV_Halpha_tab_st}$^{e)}$. However, the comparison of measurements is not straightforward.
The position of the H$\alpha$ line extremum depends on the resolution because the shapes 
of the peaks and especially the central depression are not symmetric. 

\onltab{
  \begin{table}[!h]
    \caption{$RV$s of the~H$\alpha$}
    \label{RV_Halpha_tab_st}
    \begin{center}
    \begin{tabular}{lllll}
            \hline \hline
                                &      date         & RV                 & Dispersion/   & Ref.\\
                    &                   & (\kms)             & Resolution &  \\
    \hline
    $rv_{\rm{cd}}$   &     1960-Feb-20    & $57 \pm 5 $        & 12.4 \AA/mm & 1 \\
    $rv_{\rm{cd}}$   &     1960-Feb-23    & $52 \pm 5 $        & 12.4 \AA/mm & 1 \\
    $rv_{\rm{cd}}$   &     1967-Jan-27    & $50$               & 9.7  \AA/mm & 2 \\
    $rv_{\rm{cd}}$   &     1987-Feb-09    &  $-2 \pm 2$        & $R=50\,000$       & 3 \\
    $rv_{\rm{cd}}$   &     1987-Mar-22    & $61.4$             & $R=7\,300$        & 4 \\
    $rv_{\rm{cd}}$   &     1995-Jan-11,   & 18                 & $R = 5\,000$      & 5 \\
                    &     1997-Jan-01    &                    &                   & \\
    $rv_{\rm{cd}}$   &     1999-Oct-17    & $60$               & $R = 55\,000$     & 6\\
    $rv_{\rm{cd}}$   &     2007-Oct-04    & $60$               & $R = 55\,000$     & 6\\
   \hline
    $rv_{\rm{v}}$    &     1987-Feb-09    &  $-124 \pm 2$$^{b}$& $R=50\,000$       & 3 \\
                    &                   &  $-50\pm 2$        &   &\\ 
    $rv_{\rm{r}}$    &     1987-Feb-09    &  $76 \pm 2$        & $R=50\,000$       & 3 \\     
   \hline

    \end{tabular}
    \end{center}
    \tablebib{(1)~\citet{Doazan65}; (2)~\citet{Andrillat1972}; 
    (3) \citet{Dachs92}; (4) \citet{Halbedel91}; 
    (5)~\citet{Oudmaijer99}; (6)~\citet{Borges09}.}
    \tablefoot{$rv_{\rm{v}}$, $rv_{\rm{r}}$, $rv_{\rm{cd}}$ $RV$s
        of the~violet, red peak, and central depression;
        \tablefoottext{b}{double peak violet part with almost the~same 
        intensity}
    }
  \end{table}
}
\onltab{
  \begin{table}[!h]
    \caption{Peak separation of the~H$\alpha$ line}
    \label{peak_sep_Halfa_tab_st}
    \begin{center}
    \begin{tabular}{llll}
        \hline \hline
        date                     &  Separation           & Dispersion/       & Ref.\\
                                 &  (\kms)               & Resolution        &  \\
        \hline
        1995-Jan-11, 1997-Jan-01   & 160                   & $R = 5\,000$      & 1 \\
        1999-Oct-17               & $160$                 & $R = 55\,000$     & 2\\
        2007-Oct-04               & $150$                 & $R = 55\,000$     & 2\\
        \hline
    \end{tabular}
    \end{center}
    \tablebib{(1)~\citet{Oudmaijer99};(2)~\citet{Borges09}.
    }
    \end{table}
}

\subsubsection{Correlation between individual quantities of the H$\alpha$ line}
\label{Hacorrsec}

The changes of the observed profile of the H$\alpha$ line are displayed in Figs.~\ref{SHaIIc} -- 
\ref{HaRVcomplet}. However, because of the different temporal steps between individual observations, 
it is possible that some important dependencies between quantities may be suppressed in these diagrams. 
Therefore, we present the most significant values of the Pearson coefficient in Table~\ref{Ha_cor_tab}$^{e)}$.
The most important dependencies are the following:

\renewcommand{\labelenumi}{\roman{enumi})}  
\begin{enumerate} [noitemsep,topsep=0pt,parsep=0pt,partopsep=0pt]
  \item the correlation between fluxes and the absolute value of $EW$; 

  \item	the correlation of peak fluxes themselves, 
        and the flux of the violet peak and the central depression (Fig.~\ref{Cor_FHa}$^{e)}$);       

  \item the strong correlation between $RV$s of peaks and the central depression (Fig.~\ref{Cor_RVHa}$^{e)}$);

  \item the $RV$s of the wings and the violet peak. These have an almost limited value of the Pearson coefficient (0.13), 
        which prevents the use of this quantity for the determination of the role of the wind or binarity;    

  \item the connection between the $EW$ and $RV$s (Fig.~\ref{Cor_RVEWHa}$^{e)}$);

  \item the lack of dependence between the $RV$ of wings and peak $RV$s and fluxes.  

\end{enumerate}

We note that the interpretation of the Pearson coefficient is not always straightforward in this star. 
The situation is complicated here by the episodic gas discharge (Sec.~\ref{expanding_layers}), 
which can distort the observed dependencies during some epochs. This was recognised in several 
correlation diagrams, particularly in dependencies between the $RV$ of the H$\alpha$ red peak and the flux 
of the violet peak, and the $RV$ and flux of the central depression. Based on these values, we were able to 
identify four epochs (Table~\ref{Ha_cor_tab}$^{e)}$) in which the behaviour of the H$\alpha$ line was
substantially different: prior to JD 2452110,  $2452110-2454650$, $2454650-2456600$, and 
subsequent to JD 2456600. Moreover, this classification matches epochs identified in the $RV$
of the central depression and $V/R$ ratio. These four epochs are seen in data from all four observatories 
in the present study.

\onlfig{
  \begin{figure}[!h]
	\resizebox{\hsize}{!}{\includegraphics{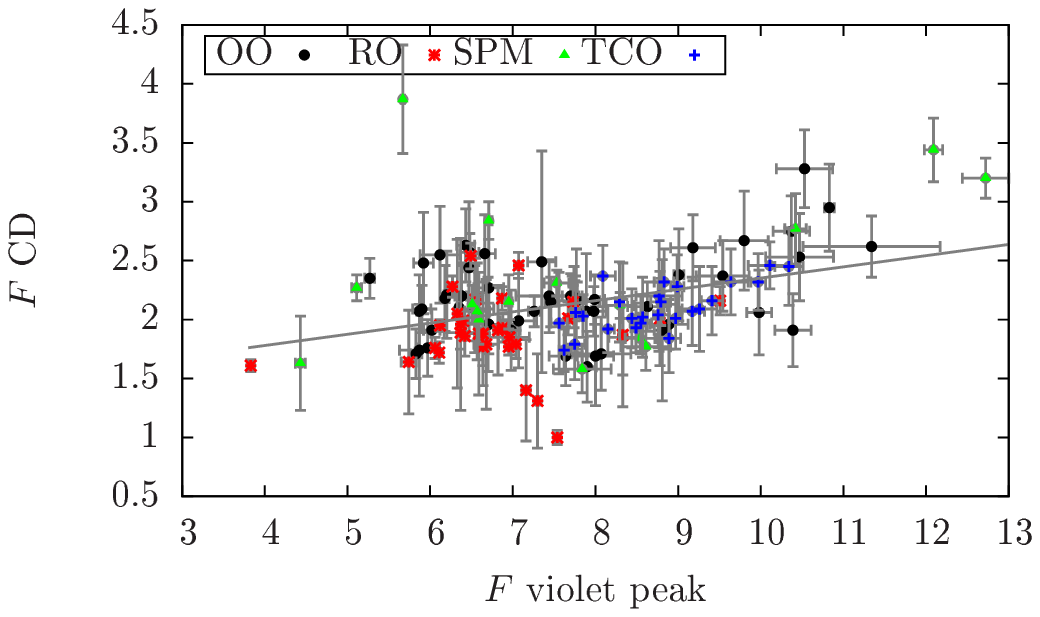}}
	\resizebox{\hsize}{!}{\includegraphics{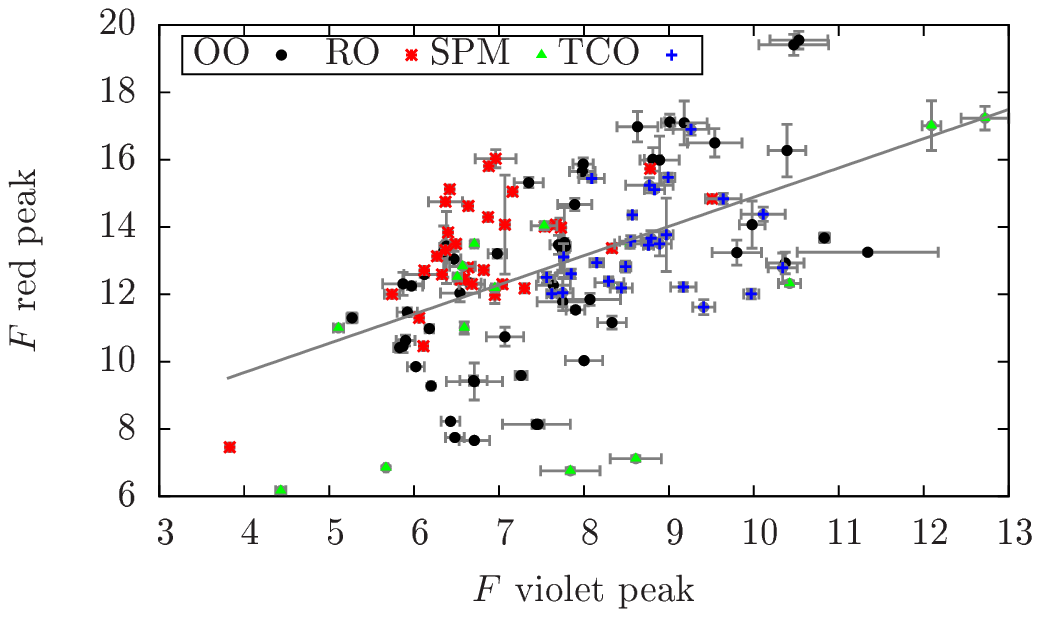}}
	\caption{Correlation between fluxes of H$\alpha$ line.}
	\label{Cor_FHa}
  \end{figure}	
  \FloatBarrier
	}
\onlfig{
  \begin{figure}[!h]
     \resizebox{\hsize}{!}{\includegraphics{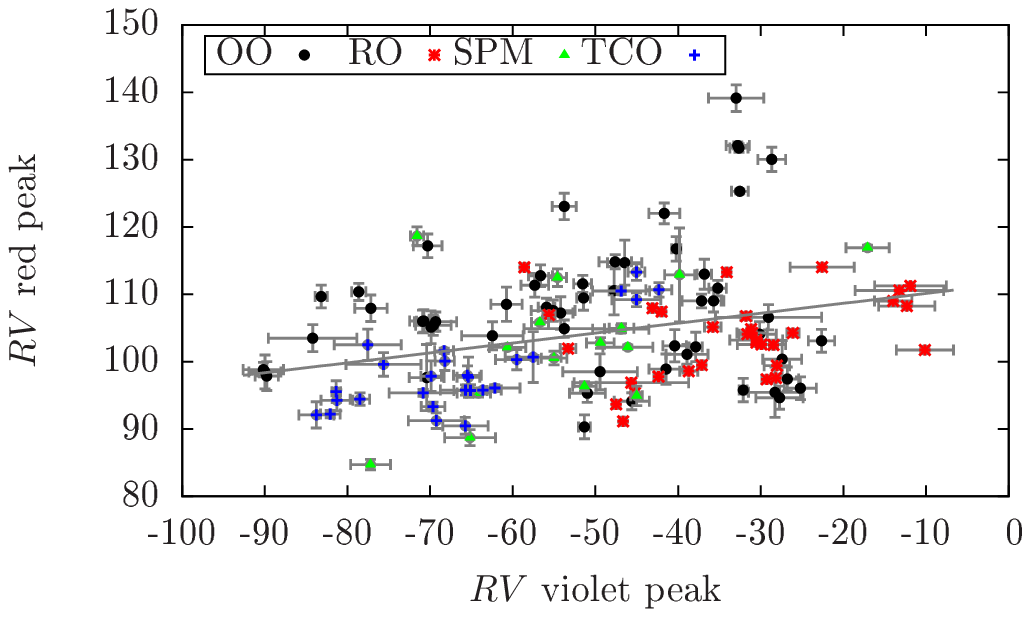}}
     \resizebox{\hsize}{!}{\includegraphics{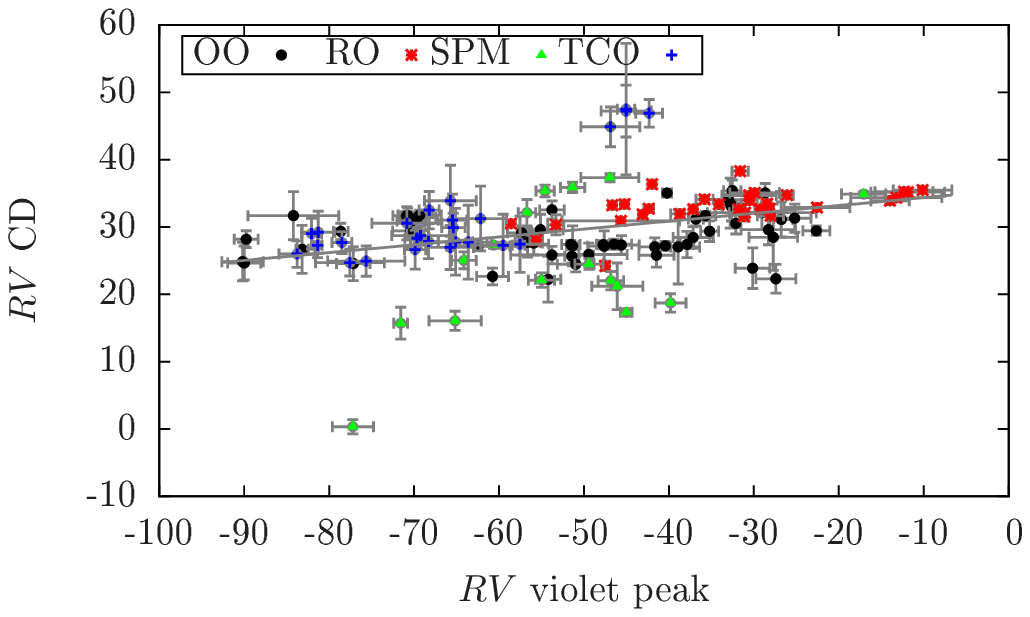}}
     \resizebox{\hsize}{!}{\includegraphics{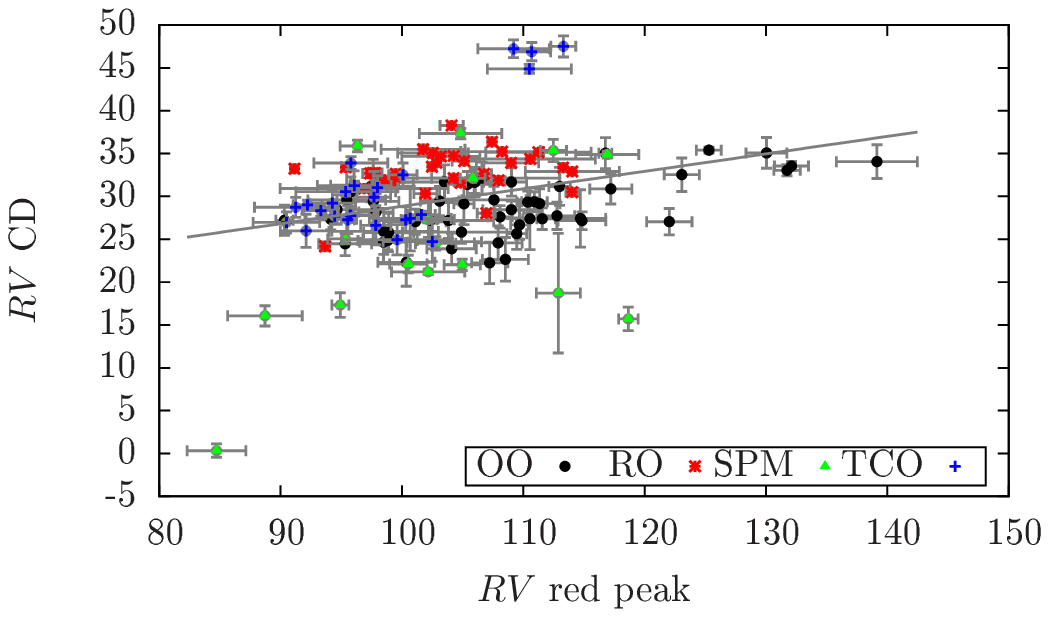}}
     \caption{Correlation between $RV$ of H$\alpha$ line.}
     \label{Cor_RVHa}
  \end{figure}
  \FloatBarrier
	}
\onlfig{
  \begin{figure}[!h]
     \resizebox{\hsize}{!}{\includegraphics{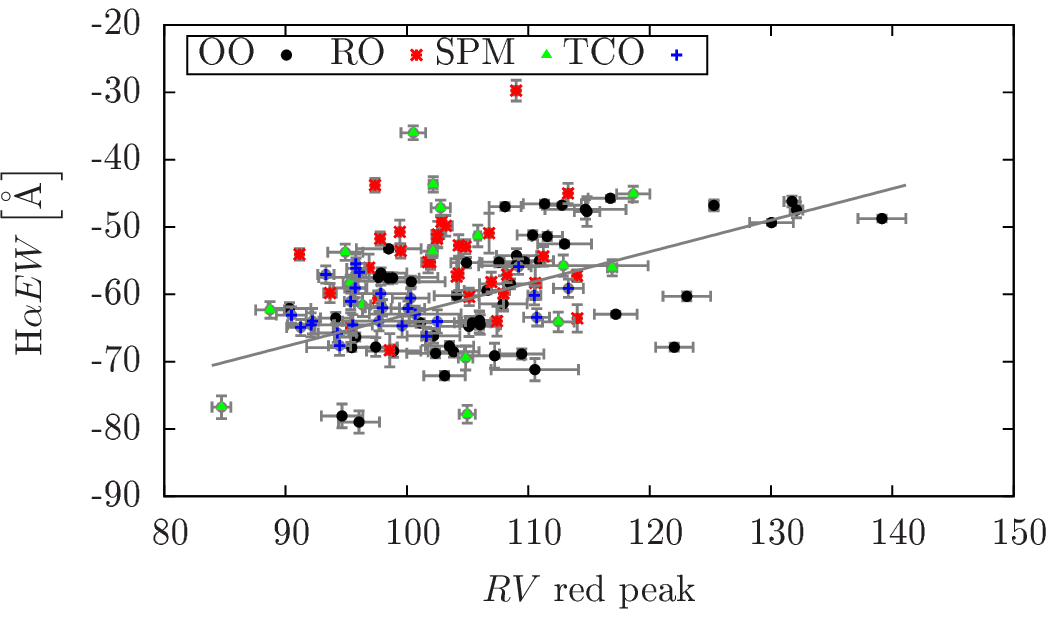}}
     \resizebox{\hsize}{!}{\includegraphics{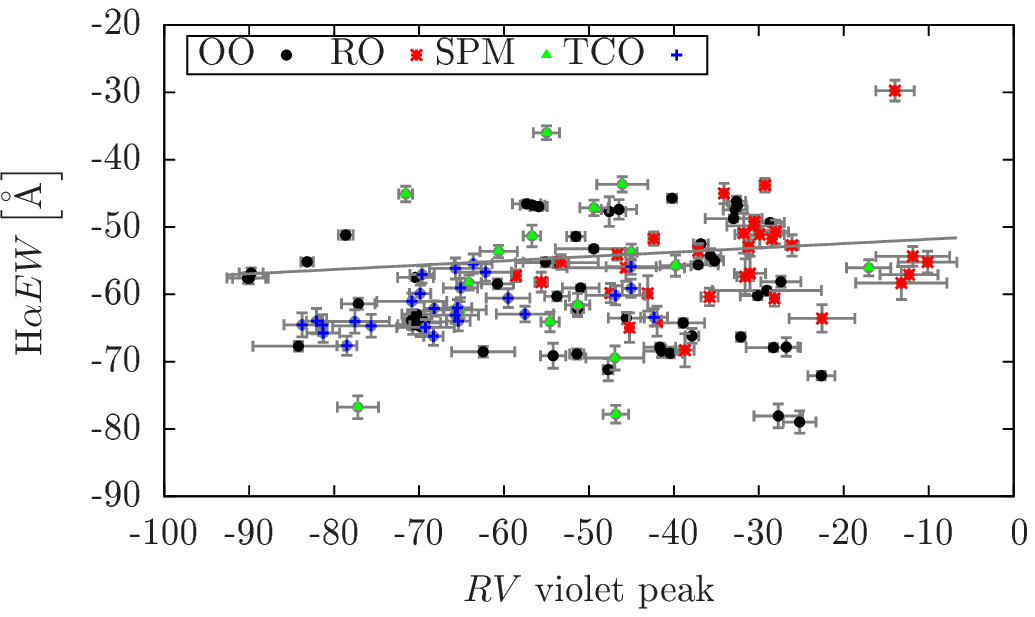}}
     \caption{Correlation between  $EW$ and $RV$ of H$\alpha$ line.}
     \label{Cor_RVEWHa}
  \end{figure}
  \FloatBarrier
     }

\onltab{
   \begin{table*}[!h]
    \caption{
       Pearson correlation coefficients of the individual quantities of the H$\alpha$ line in different epochs.
    }
    \label{Ha_cor_tab}
    \centering
    \begin{tabular}{lrrrr|rr}
    \hline \hline
                JD                                                        &  $2449319-$    & $2452110-$ & $2454650-$ & $2456600-$ & the entire \\ 
                                                                          &  $2452110$     & $2454650$  & $2456600$  & $2456740$  & sample \\ 
    \hfill critical value                                                 & 0.37           & 0.31       & 0.33       & 0.39       & 0.13 \\ 
    \hline
     P($|EW|$, $RV_{\text{\rm{W}}}$)       & 0.09 & -0.22 & -0.07 & 0.69 & -0.04  \\ 
    \hline
     P($|EW|$, $RV_{\rm{VP}}$)            & -0.08 & 0.39 & -0.28 & -0.51 & -0.22  \\ 
    \hline
     P($|EW|$, $F_{\rm{VP}}$)             & -0.26 & 0.84 & 0.12 &  0.34&  0.67\\ 
    \hline
     P($|EW|$, $F_{\rm{CD}}$)             & -0.21 & 0.32 & 0.23 & 0.51 &  0.38 \\ 
    \hline
     P($|EW|$, $F_{RP}$)                 & -0.06 & 0.89 & 0.84 & 0.76 & 0.75  \\ 
    \hline
     P($RV_{\rm{W}}$, $RV_{\rm{VP}}$)     & 0.13 & -0.12 & 0.04 & -0.51  & -0.12  \\ 
    \hline
     P($RV_{\rm{W}}$, $F_{\rm{RP}}$)      & 0.23 & -0.42 & 0.02 & 0.65 & 0.15 \\ 
    \hline
     P($RV_{\rm{VP}}$, $RV_{\rm{CD}}$)    & 0.68 & 0.23 & 0.37 & 0.82 & 0.38 \\ 
    \hline
     P($RV_{\rm{VP}}$, $RV_{\rm{RP}}$)    & 0.32 & -0.08 & 0.46 & 0.78  & 0.31  \\ 
    \hline
     P($RV_{\rm{VP}}$, $F_{\rm{RP}}$)     & -0.19 & 0.38 & -0.42 & -0.66 & -0.07  \\ 
    \hline
     P($F_{\rm{VP}}$, $RV_{\rm{RP}}$)     & -0.18 & 0.02 & -0.18 & 0.55 &  -0.22 \\ 
    \hline
     P($F_{\rm{VP}}$, $F_{\rm{RP}}$)      & 0.62 & 0.62 & 0.07 & 0.22 & 0.54 \\ 
    \hline
     P($RV_{\rm{CD}}$, $F_{\rm{VP}}$)    & 0.50 & -0.14 & 0.40 & 0.28 & -0.07  \\ 
    \hline
     P($RV_{\rm{CD}}$, $RV_{\rm{RP}}$)    & 0.25 & 0.16 & 0.57 & 0.80 & 0.32  \\ 
    \hline
     P($RV_{\rm{RP}}$, $F_{\rm{RP}}$)     & 0.01 & -0.30  & -0.48 & -0.28 & -0.47  \\ 
   \hline
   \end{tabular}
     \tablefoot{
       The dependencies listed here were identified based on the value of Pearson coefficient, and also on the
       correlation diagrams. Notation: $RV$ radial velocity, $EW$ equivalent width, VP violet peak, RP red peak, CD
       central depression, W wings.
     }
   \end{table*}
 }

\subsection{H$\beta$ line}

The H$\beta$ line shows a~more complicated behaviour than the H$\alpha$ line. It had always been observed 
as an absorption line overlapped by two emission peaks. During the ten-year systematic study of 
\cite{Merrill31_ApJ} in 1920s the red peak remained almost constant, while the violet peak showed large 
changes. \cite{Miczaika50} reported observations on 29 January 1949 when the violet part of the line 
almost disappeared. Another three-year systematic study presented by \cite{Doazan65} in the 1960s 
shows a~strong variability of both peaks.

The structure of the H$\beta$ line is often complicated  by the presence of moving humps 
\citep[e.g.][]{Doazan65}. The size of the humps, relative to the line itself, is greater than 
is observed in the H$\alpha$ line. This allows easier detection of the humps and makes the H$\beta$ 
line a~good tracer of the structure of the material close to the central object where the line is forming. 

We present here measurements of the basic parameters of the H$\beta$ line, excluding the $EW$. 
The $EW$ is not measured because the central depression is below the continuum, and therefore the value of 
the $EW$ has no explicit physical meaning.

\subsubsection{Relative fluxes and radial velocities}

\begin{figure*}
  \centerline{ 
    \resizebox{0.5\textwidth}{!}{\includegraphics{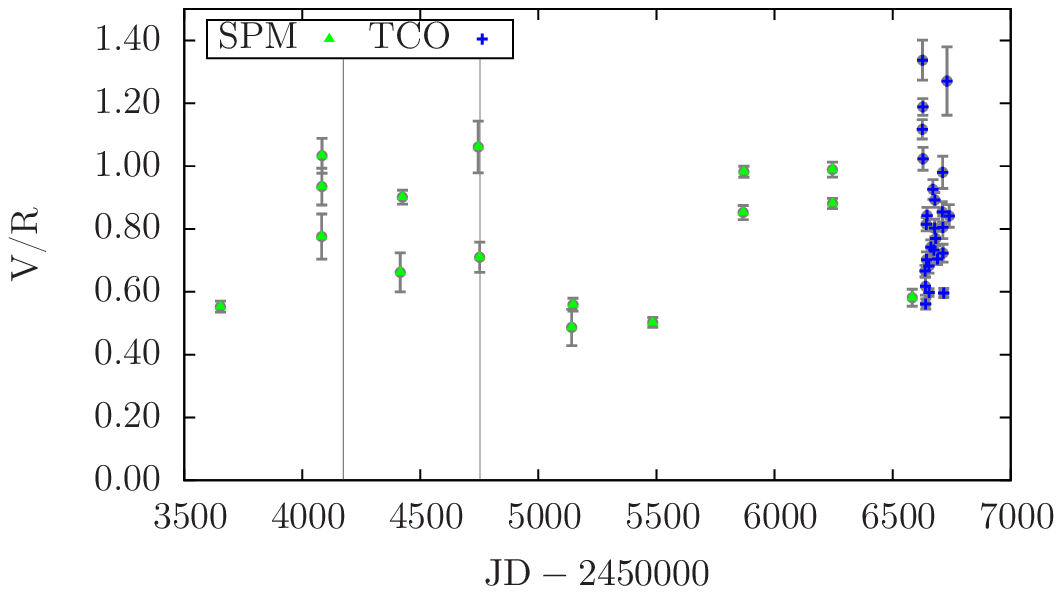}} 
    \resizebox{0.5\textwidth}{!}{\includegraphics{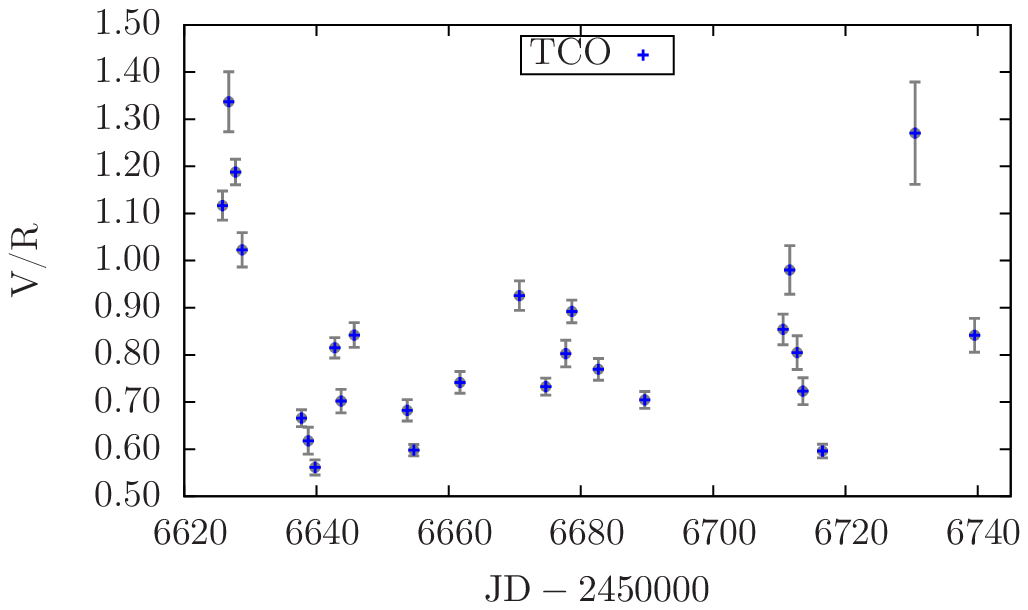}} 
              }
  \centerline{
   \resizebox{0.5\textwidth}{!}{\includegraphics{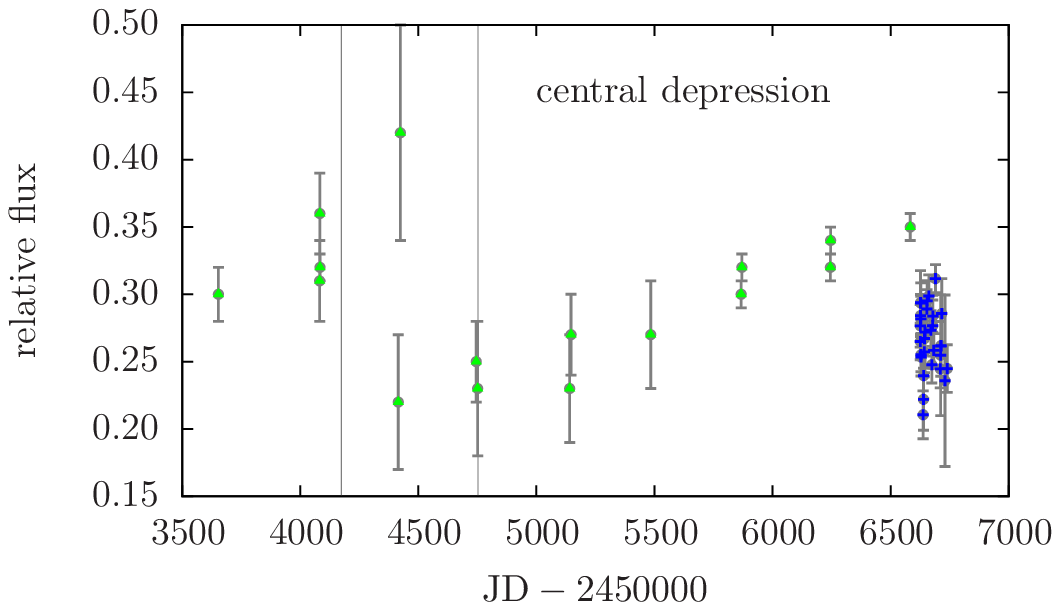}}
   \resizebox{0.5\textwidth}{!}{\includegraphics{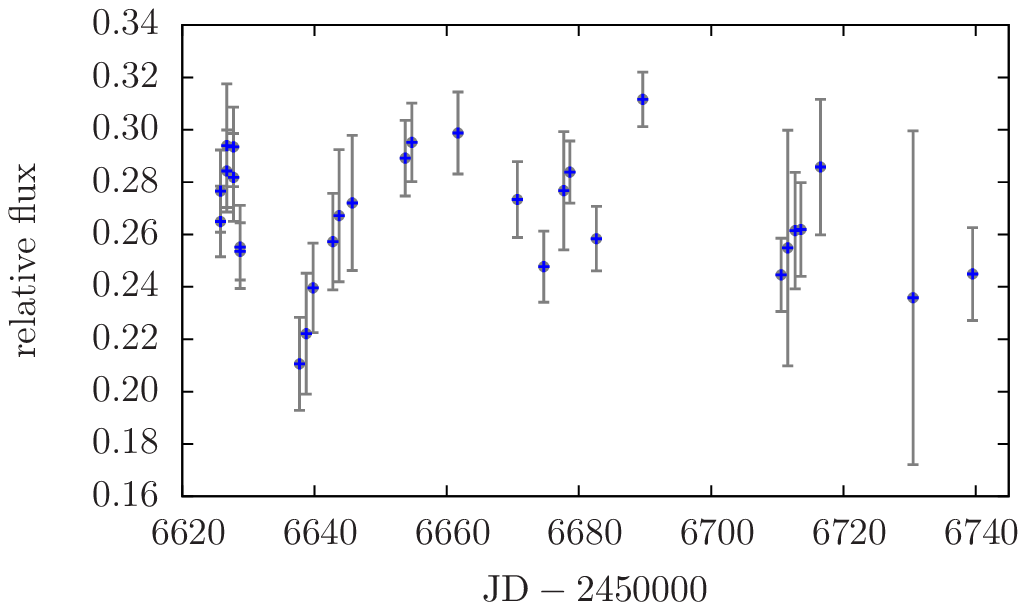}}
             }
   \caption{$V/R$ and relative flux of the central depression of the H$\beta$ line.
           The vertical lines indicate the same epochs as in Fig.~\ref{EWHa}.
           }
   \label{HbVtoR}
\end{figure*}

We measure the relative fluxes and $RV$ of both of the peaks and central depression using a~least squares
polynomial fit (Sect.~\ref{analysis},\,\textit{iv}). The $V/R$ variations and the relative flux 
of the central depression are plotted in Fig.~\ref{HbVtoR}. The relative fluxes of the peaks themselves 
are presented in Fig.~\ref{HbF}$^{e)}$. The results of $RV$ measurements are plotted in Fig.~\ref{HbR}$^{e)}$.

\onlfig{
  \begin{figure*}[!h]
     \centerline{
        \resizebox{0.5\textwidth}{!}{\includegraphics{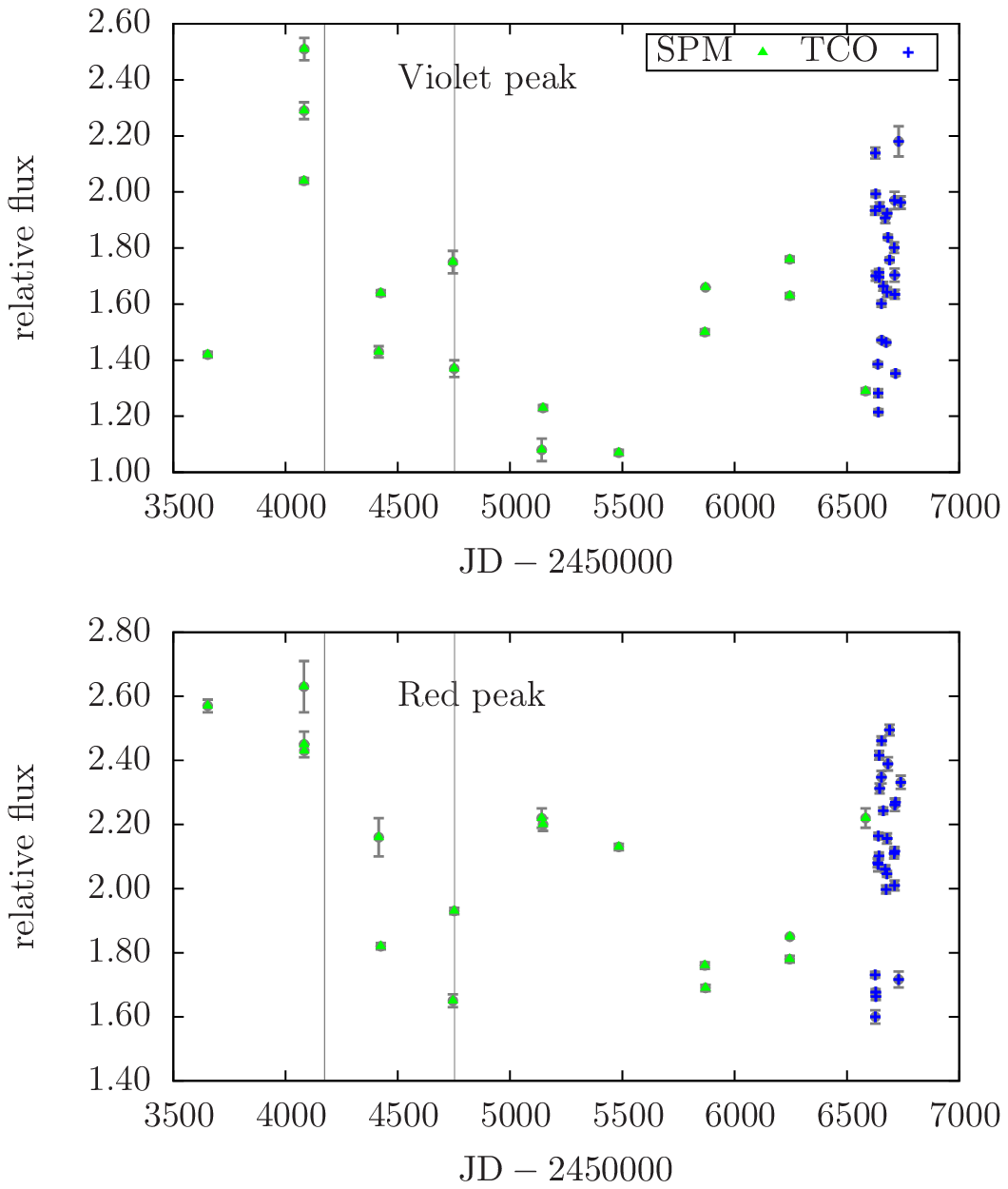}}
        \resizebox{0.5\textwidth}{!}{\includegraphics{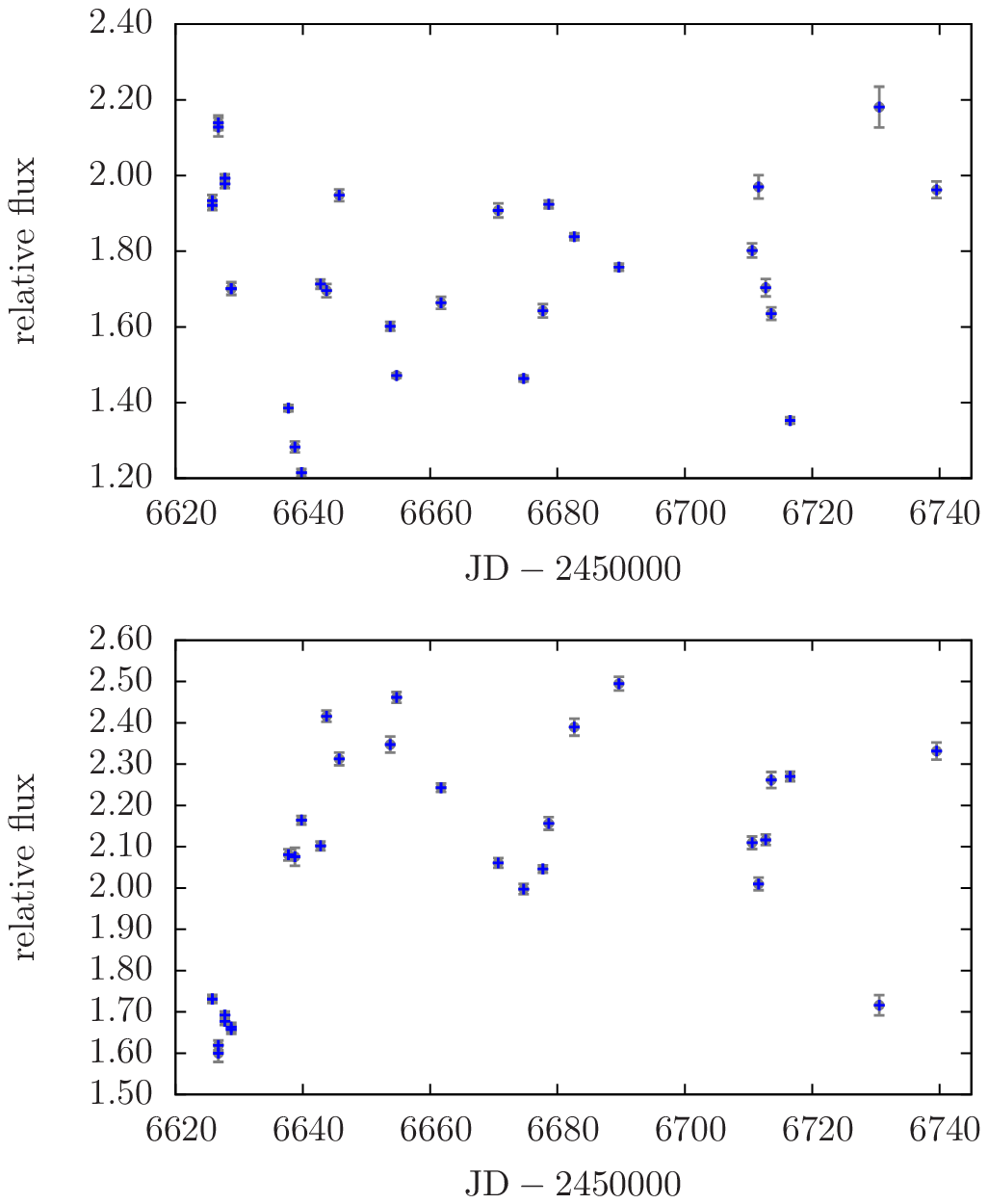}}
      }
      \caption{
       Relative fluxes of the violet and red peaks of the spectral line H$\beta$. Vertical lines show the maximum
       value of $|EW|$ of the H$\alpha$ line and the epoch when $V/R$ of the H$\alpha$ line was greater than one.
        }
    \label{HbF}
    \end{figure*}
  \FloatBarrier
       }

\onlfig{
  \begin{figure*}[!h]
    \centerline{
      \resizebox{0.5\textwidth}{!}{\includegraphics{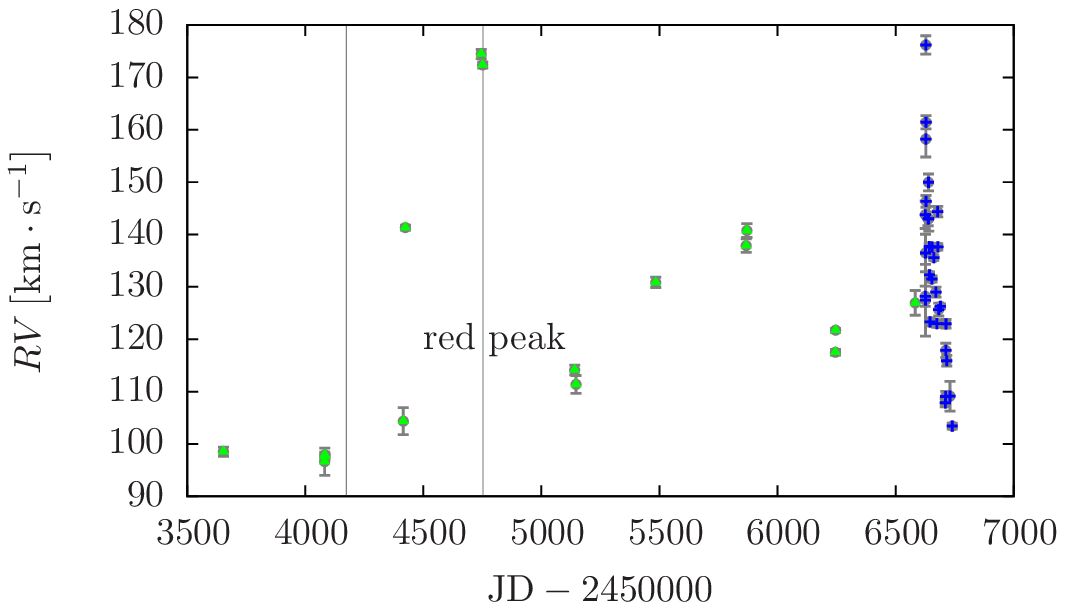}}
      \resizebox{0.5\textwidth}{!}{\includegraphics{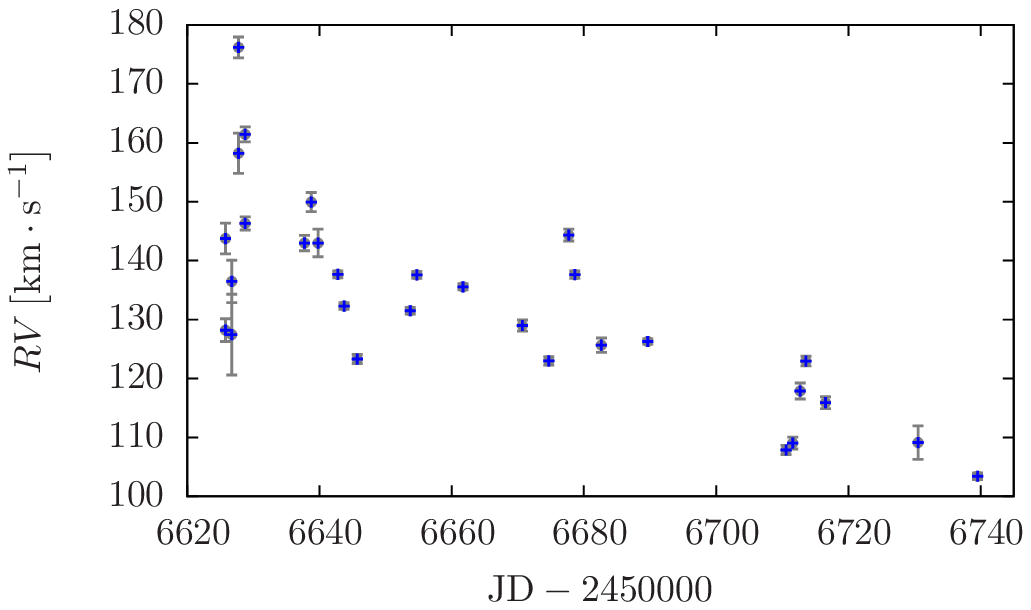}}              
    }
    \centerline{
      \resizebox{0.5\textwidth}{!}{\includegraphics{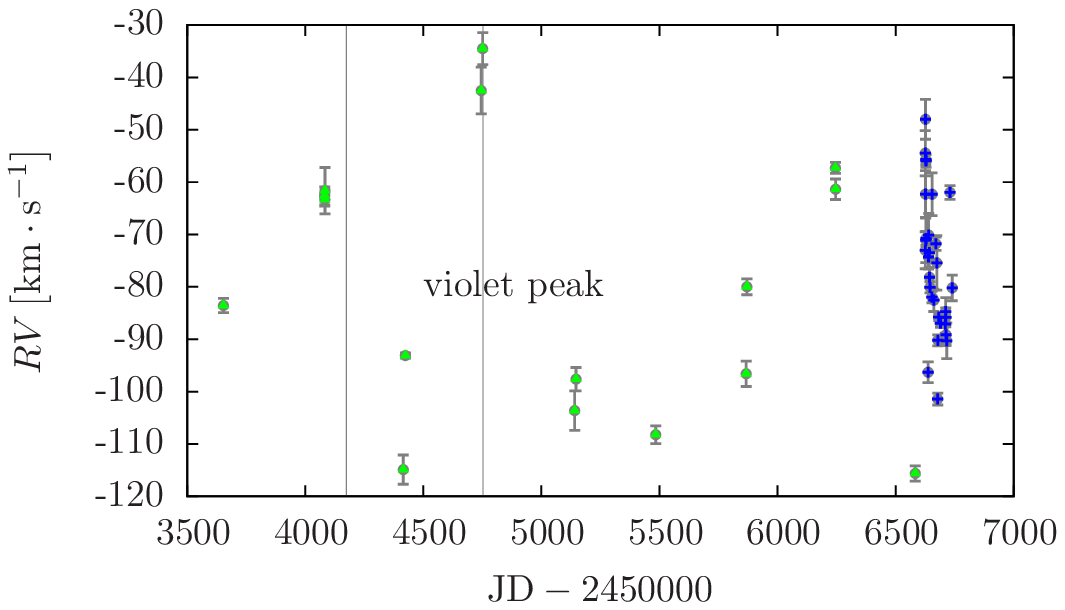}}
      \resizebox{0.5\textwidth}{!}{\includegraphics{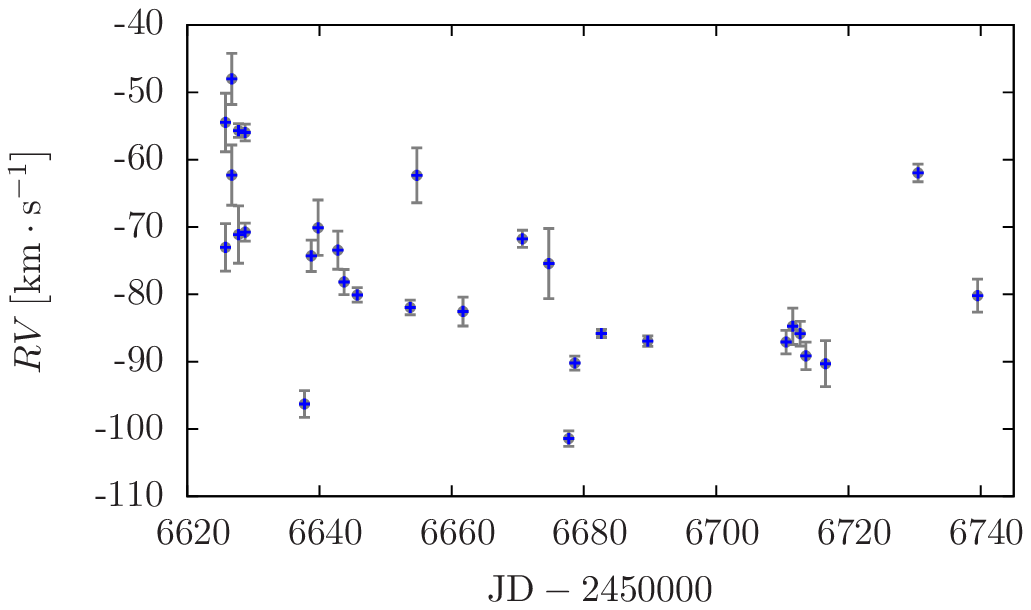}}
    }
    \centerline{
      \resizebox{0.5\textwidth}{!}{\includegraphics{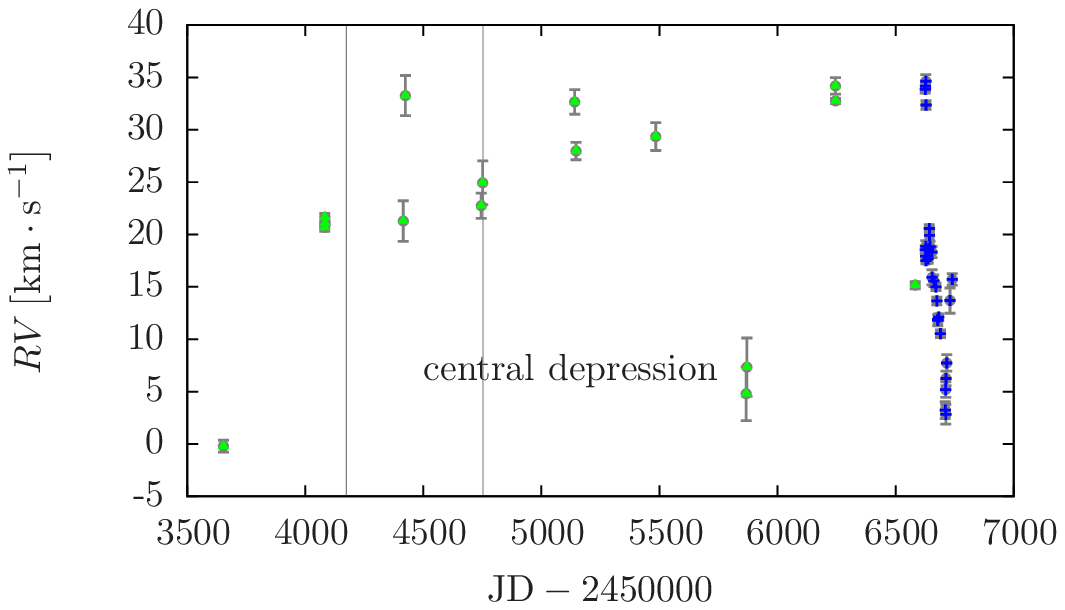}}
      \resizebox{0.5\textwidth}{!}{\includegraphics{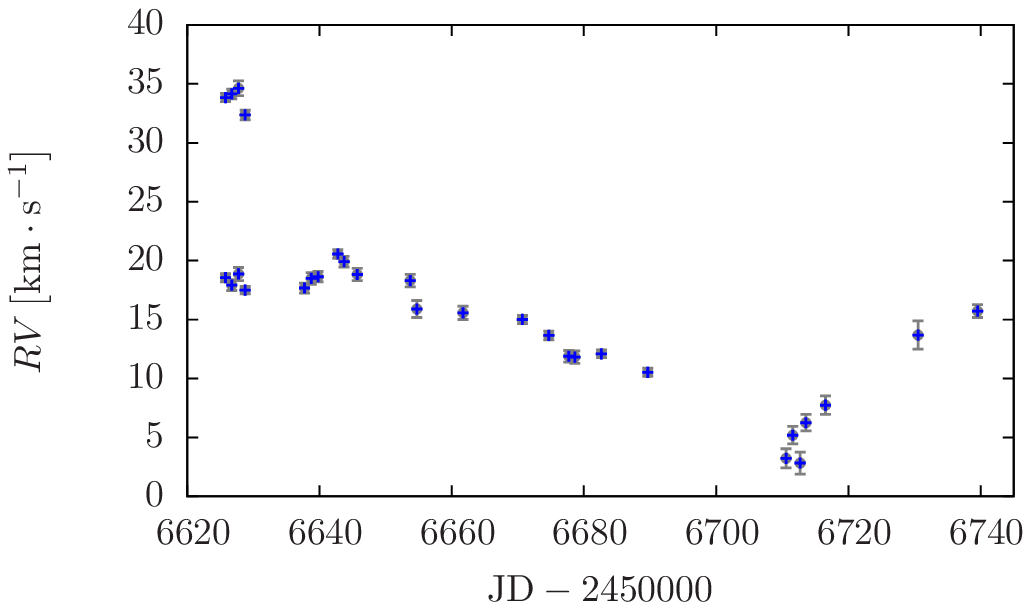}}            
    }
      \caption{$RV$s of the~red peak, the~violet peak, and the~central depression of the H$\beta$ line.}
      \label{HbR}
  \end{figure*}
 \FloatBarrier
      }

\subsubsection{Correlation between H$\beta$ and H$\alpha$ lines}
\label{Hb_Ha_cor_sec}

The correlation between the H$\beta$ and H$\alpha$ lines can be investigated
only on the data from the SPM and TCO observatories. This data set shows 
no strong correlation. However, the maximum of the $RV$ of the central depression occurs between 
these two data samples (see Sect.~\ref{O_Ha_cor}). Therefore, we also analysed the data
from TCO separately. Even if the sample is too small (critical value of $P=0.37$), 
strong dependent relationships between several measured quantities (Fig.~\ref{Hb_Ha_corr}$^{e)}$)
were found:
\renewcommand{\labelenumi}{\roman{enumi})}  
\begin{enumerate} [noitemsep,topsep=0pt,parsep=0pt,partopsep=0pt]
 \item an anti-correlation between the $RV$ of the H$\beta$ red peak and the relative flux of the H$\alpha$ red peak
   ($P_{RV(\rm{H}\beta RP), F(\rm{H}\alpha RP)}$    = $-0.85$);
 \item a correlation between the $RV$ of the H$\beta$ red peak and the $EW$ of the H$\alpha$
      ($P_{RV(\rm{H}\beta RP), EW(\rm{H}\alpha)}$    = $0.63$);
 \item a correlation between the $RV$ of the central depression of the H$\beta$ and the $RV$ of the H$\alpha$ violet peak
   ($P_{RV(\rm{H}\beta CD), RV(\rm{H}\alpha VP)}$    = $0.77$);
 \item a correlation between the relative fluxes of the violet peak of both lines
   ($P_{F(\rm{H}\beta VP), F(\rm{H}\alpha VP)}$    = $0.86$);
 \item suggested dependencies ($P\sim \pm 0.6$), which have to be proved in 
     the future. Because they are important for the study of the system dynamics, we list them:
   ($RV(\rm{H}\beta \, RP), RV(\rm{H}\alpha  \, VP)$); 
   ($RV(\rm{H}\beta \, CD), F(\rm{H}\alpha  \, RP)$); 
   ($RV(\rm{H}\beta \, CD), EW(\rm{H}\alpha)$); 
   ($RV(\rm{H}\beta \, VP), RV(\rm{H}\alpha)  \, RP$). 

\end{enumerate}

\onlfig{
  \begin{figure*}[!h]
    \hspace*{8mm}\scalebox{0.7}{\includegraphics{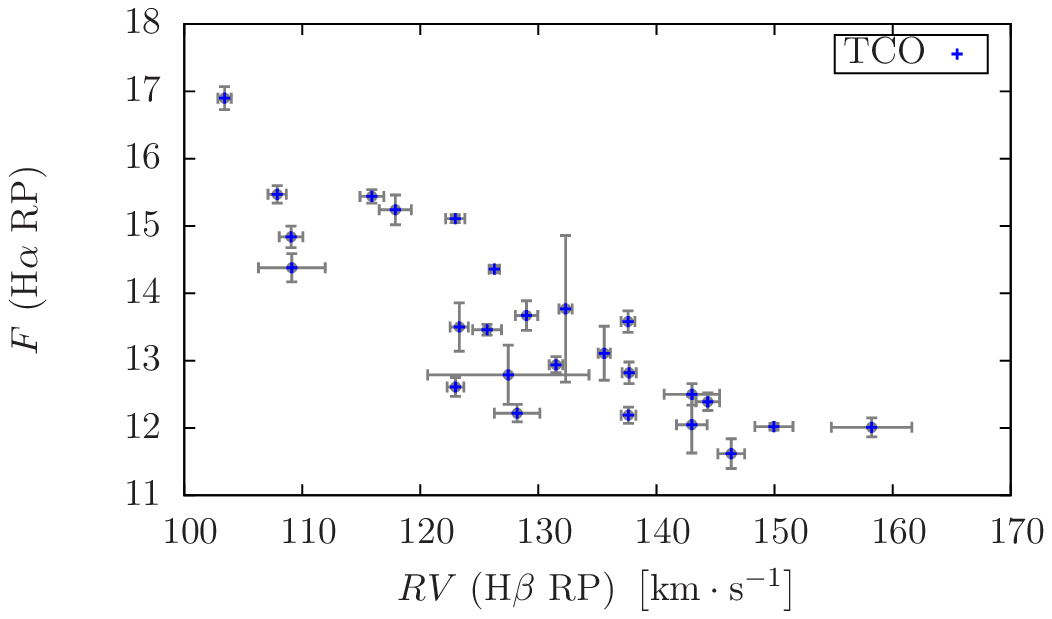}} \hfill
    \scalebox{0.7}{\includegraphics{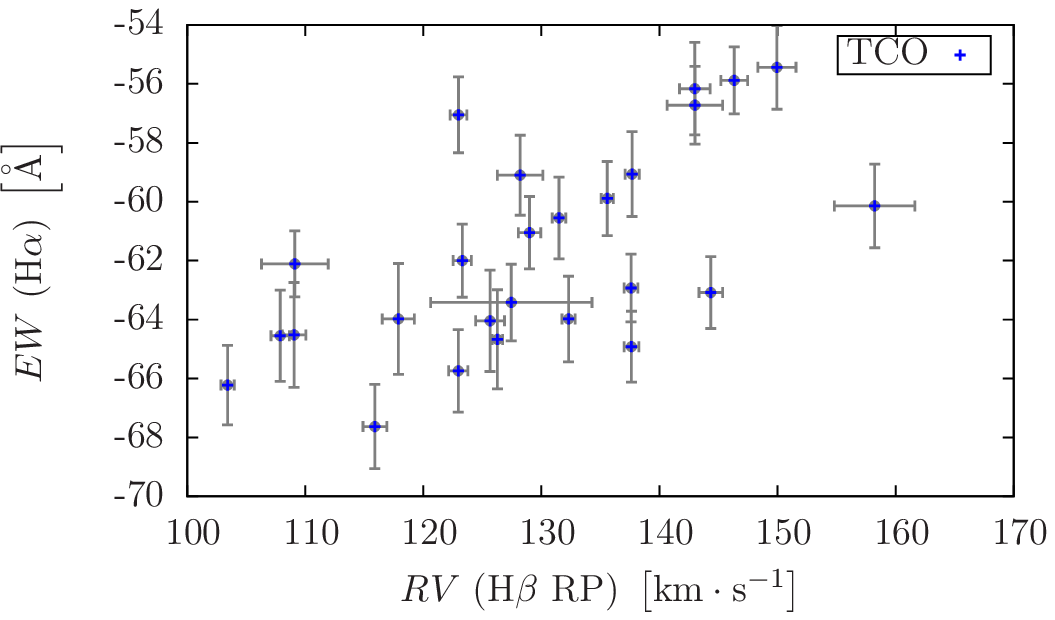}} \\
    \scalebox{0.7}{\includegraphics{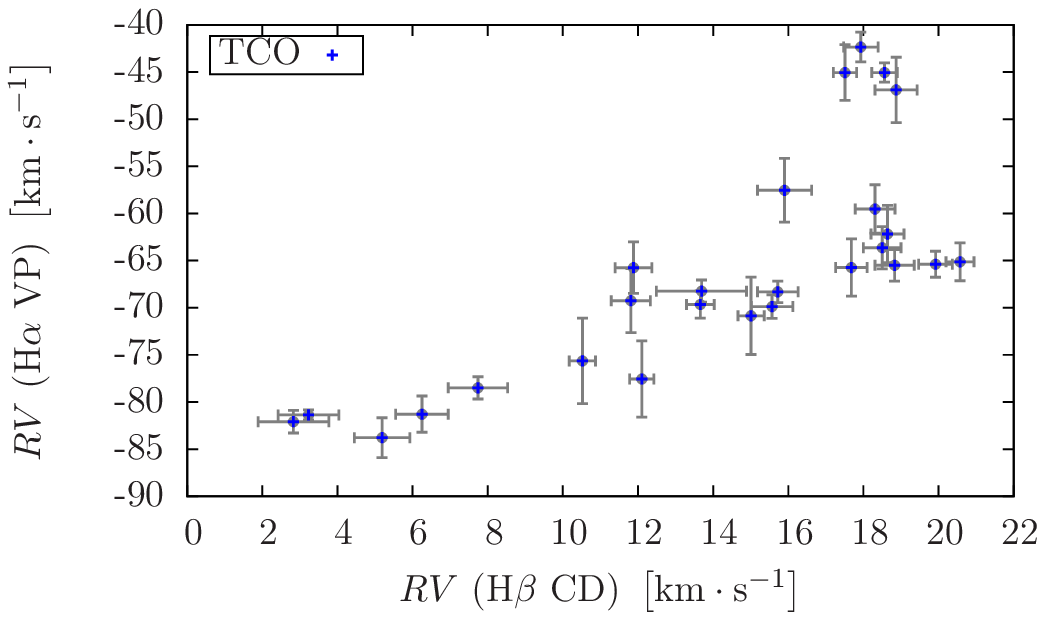}} \hfill
    \scalebox{0.7}{\includegraphics{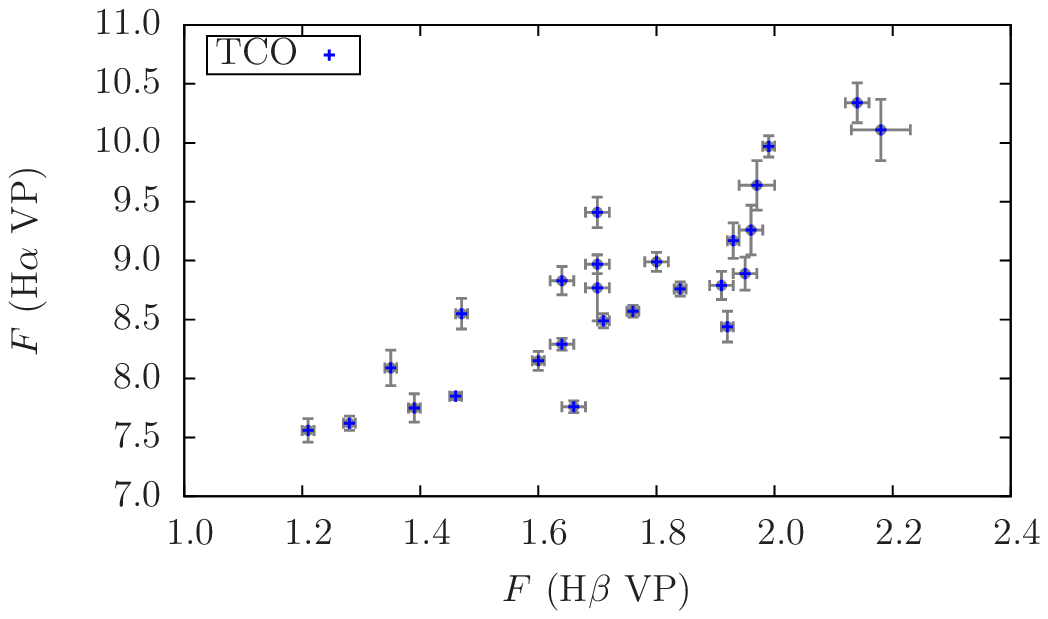}} \\
    \caption{
       Correlation diagrams of H$\alpha$ and H$\beta$ lines. Notation: $RV$ radial velocity, $EW$ equivalent width, 
       VP violet peak, RP red peak, and CD central depression.
        }
    \label{Hb_Ha_corr}
  \end{figure*}
 \FloatBarrier
}

\begin{figure*}
  \centerline{
    \resizebox{0.5\textwidth}{!}{\includegraphics{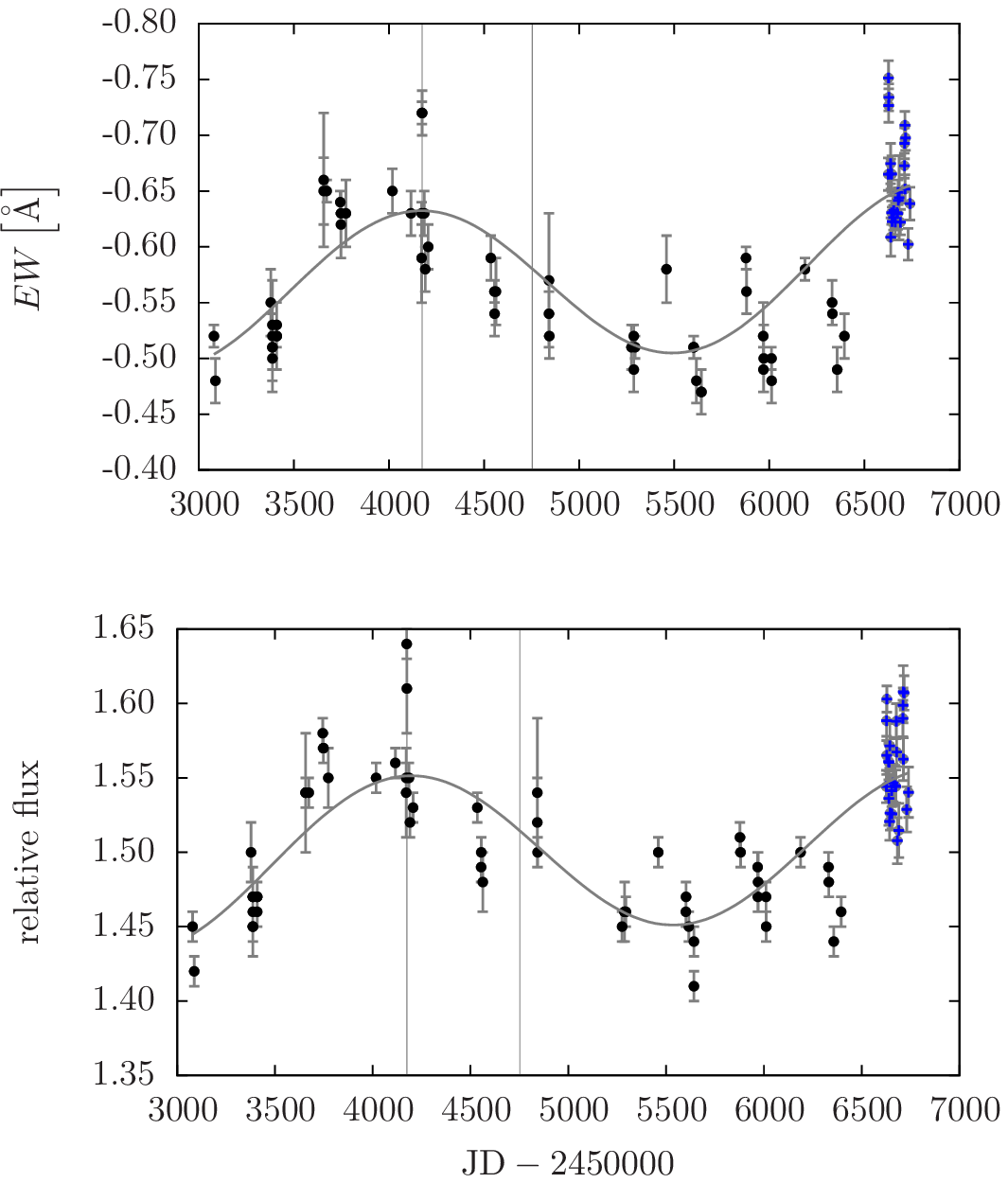}}
    \resizebox{0.5\textwidth}{!}{\includegraphics{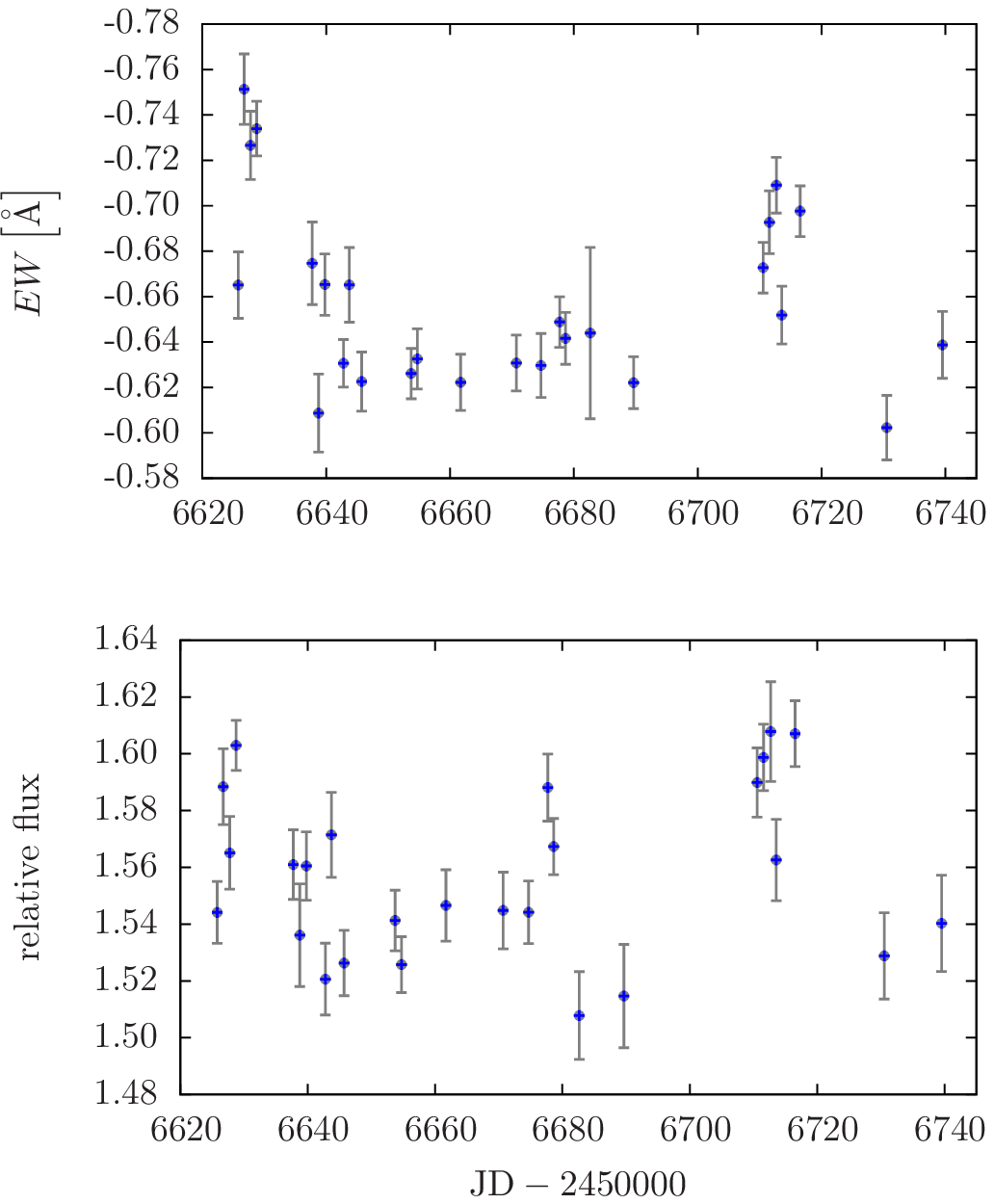}}
  }
  \caption{$EW$ and relative flux of the [\ion{O}{i}] 6300 \AA \, line. 
           The plotted curve is a~fit of two sine functions with fixed periods
           found by the period analysis of the H$\alpha$ $EW$.
    }
  \label{EWOI6300}
\end{figure*}

\begin{figure*}
  \centerline{
    \resizebox{0.5\textwidth}{!}{\includegraphics{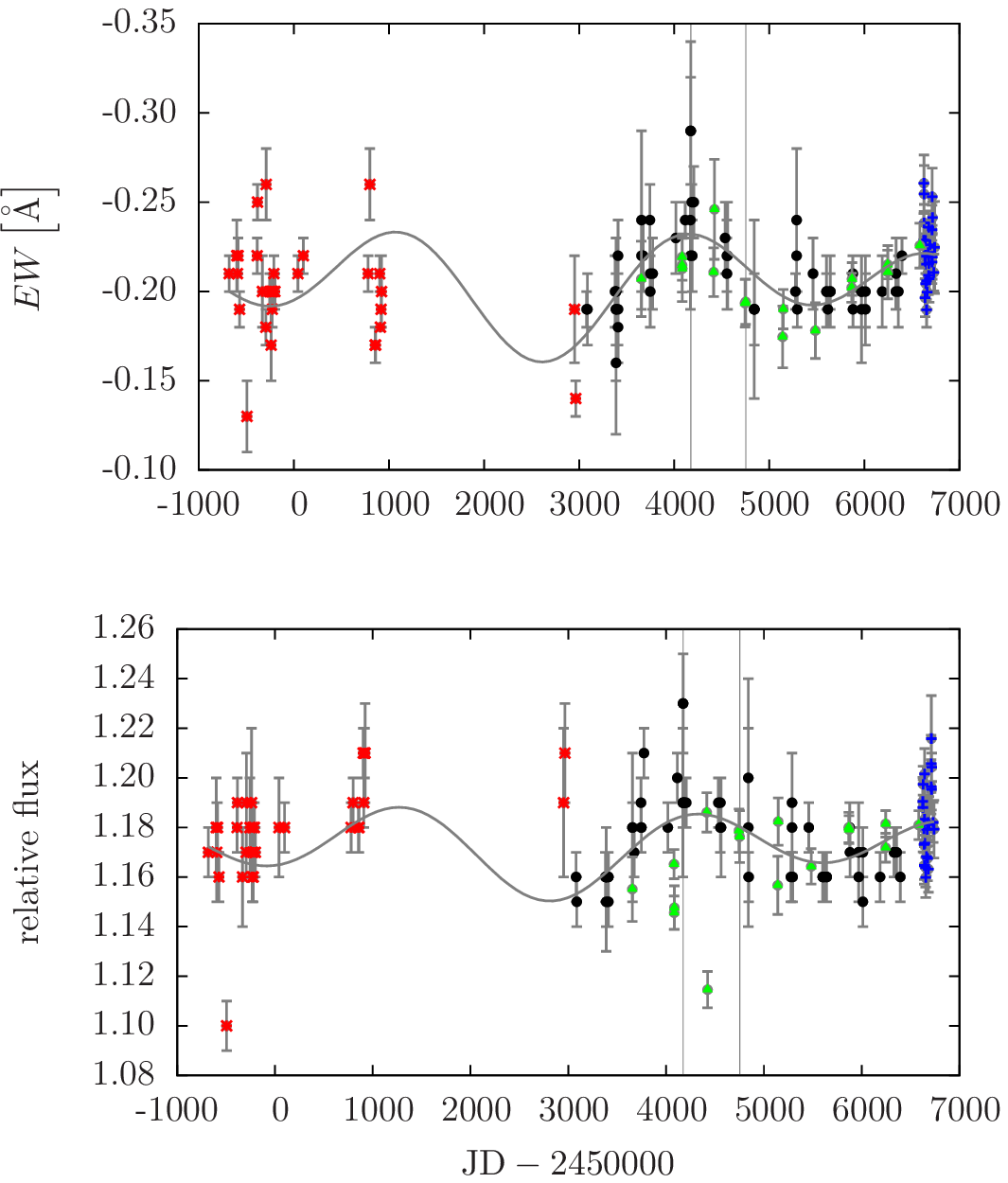}}
    \resizebox{0.5\textwidth}{!}{\includegraphics{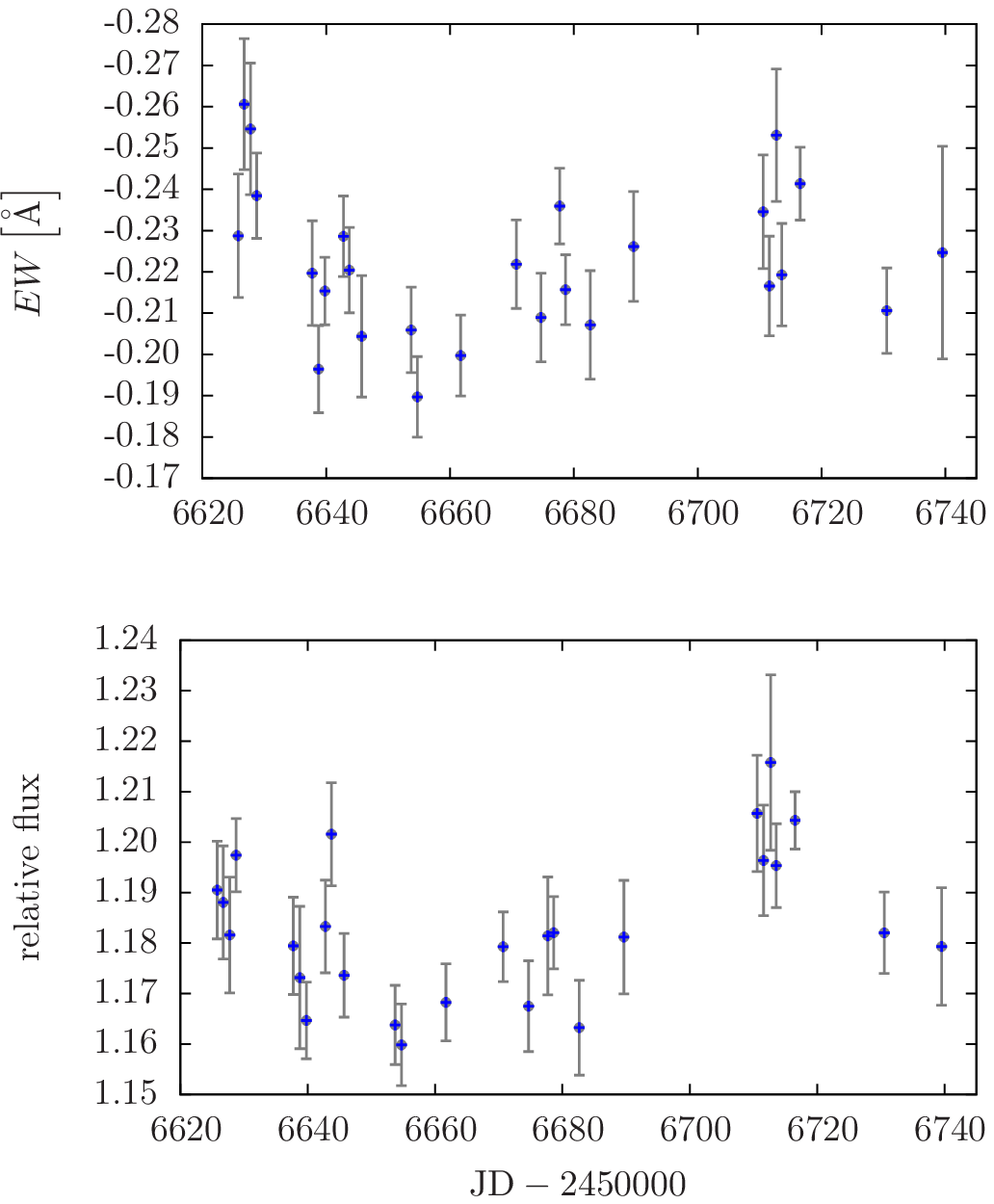}}
  }
  \caption{$EW$ and relative flux of the [\ion{O}{i}] 6364~\AA \, line. The fit of the linear combination 
           of two sine functions is shown. However, the periods were adopted from the analysis of the $EW$
           of the H$\alpha$ line to prove the connection between the line-forming regions. 
    }
    \label{EWOI6364}
\end{figure*}
   
\begin{figure*}
   \centerline{
     \resizebox{0.5\textwidth}{!}{\includegraphics{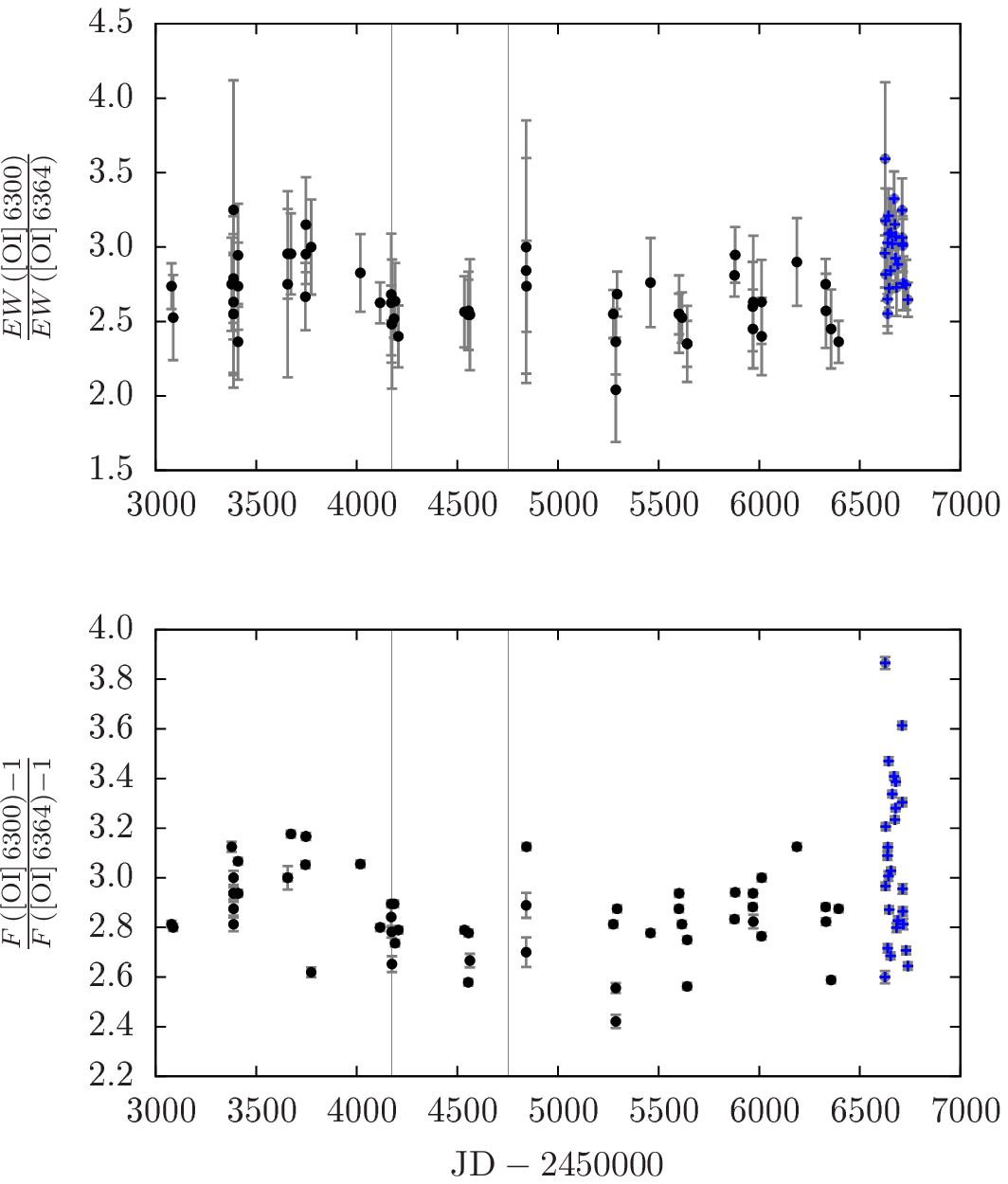}}
     \resizebox{0.5\textwidth}{!}{\includegraphics{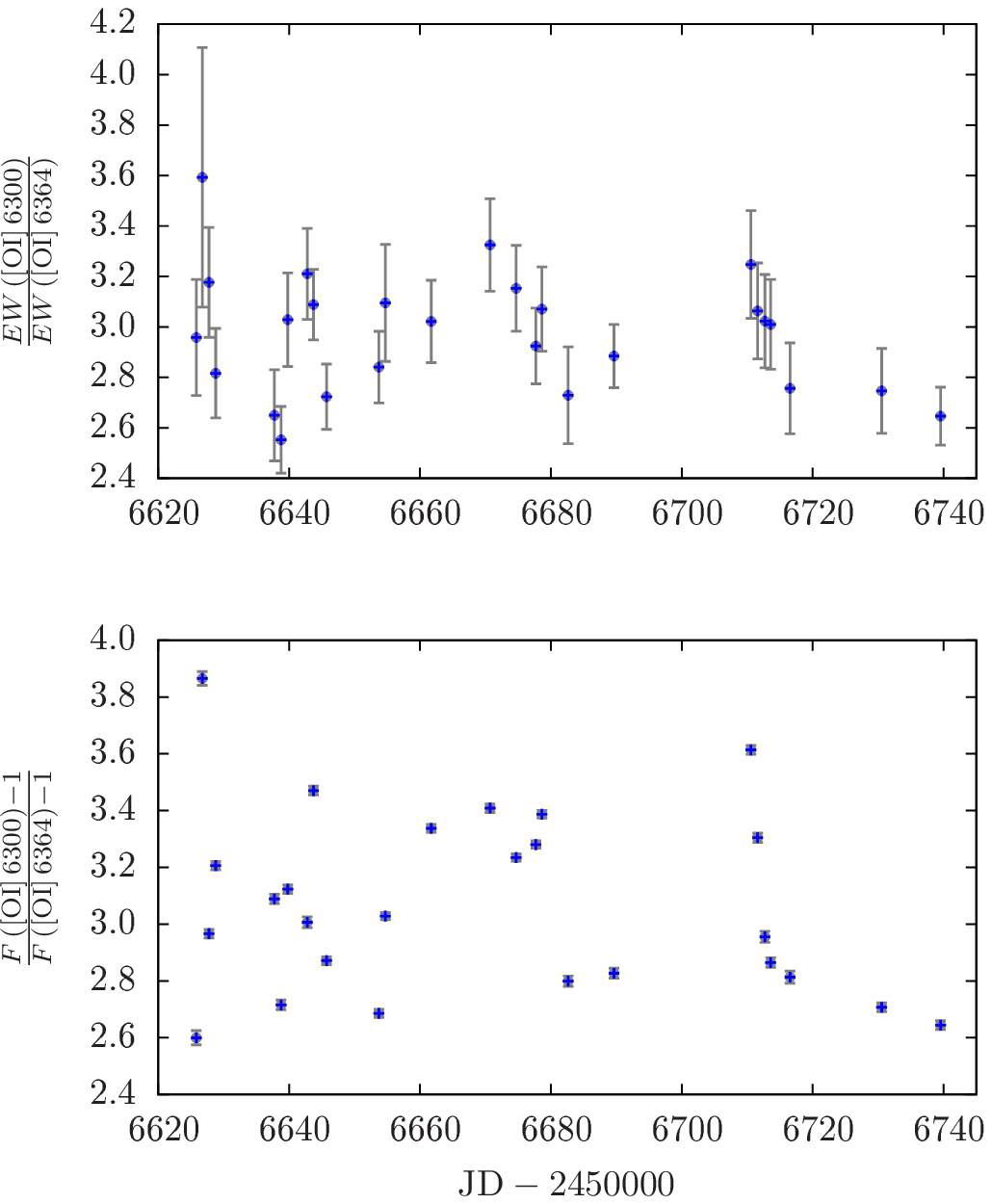}}
   }
   \caption{Ratio of $EW$ (upper panel) and net line fluxes $F$ (lower panel) for the forbidden oxygen lines.}
   \label{OIratio}
\end{figure*}

\begin{figure*}
  \centerline{
    \resizebox{0.5\textwidth}{!}{\includegraphics{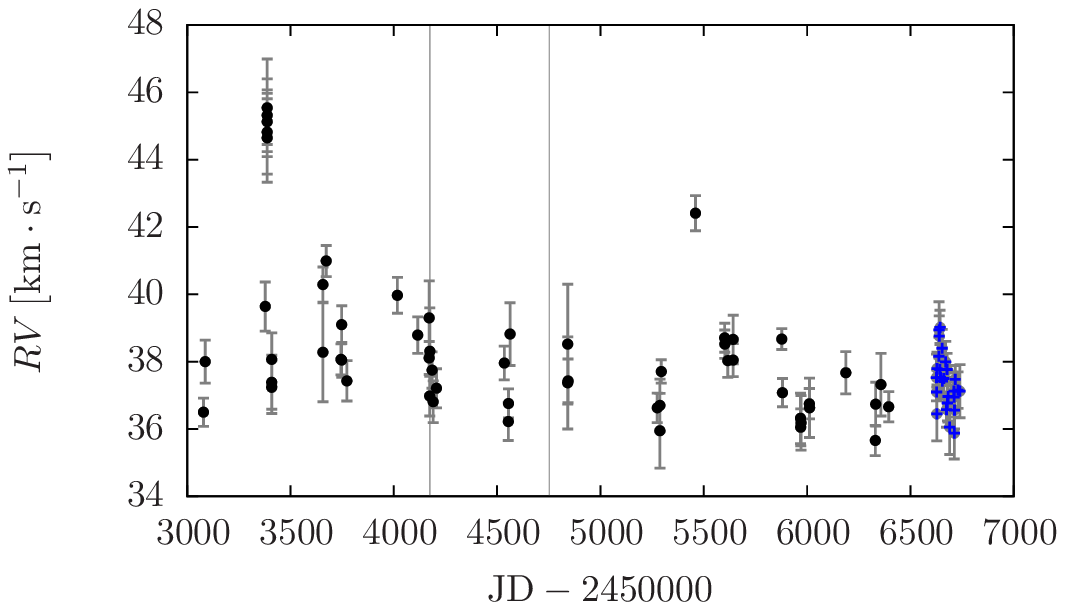}}
    \resizebox{0.5\textwidth}{!}{\includegraphics{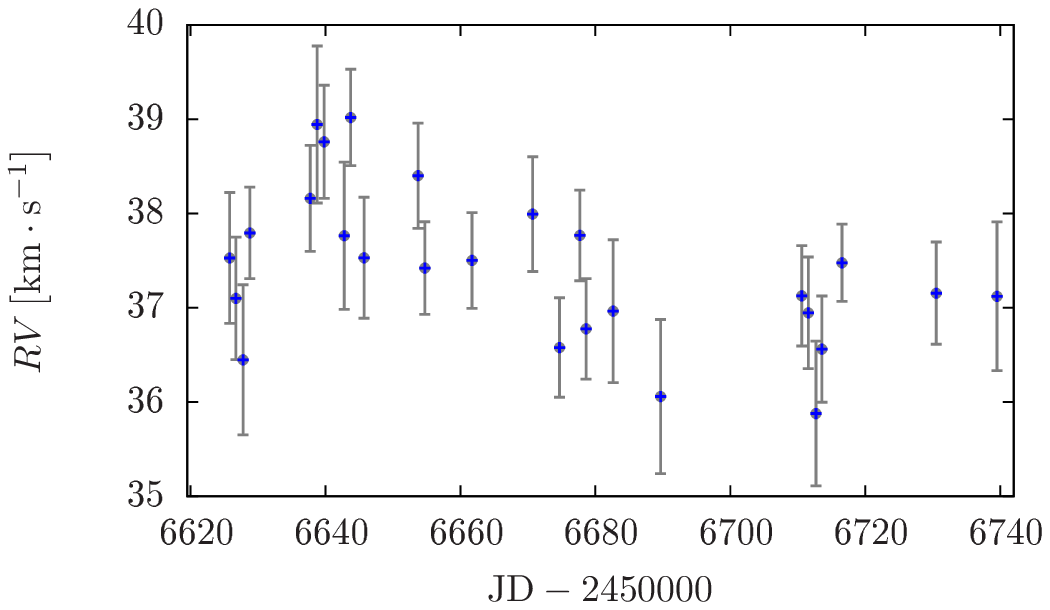}}
  }
    \caption{$RV$ of [\ion{O}{i}] 6300~\AA. The deviated points are the measurement from 
             one night ($JD~2\,453\,387$) with a~pure $S/N$ around sixty. 
    }
    \label{RVOI6300}
\end{figure*}

\onlfig{
  \begin{figure*}[!h]
   \centerline{
      \resizebox{0.5\textwidth}{!}{\includegraphics{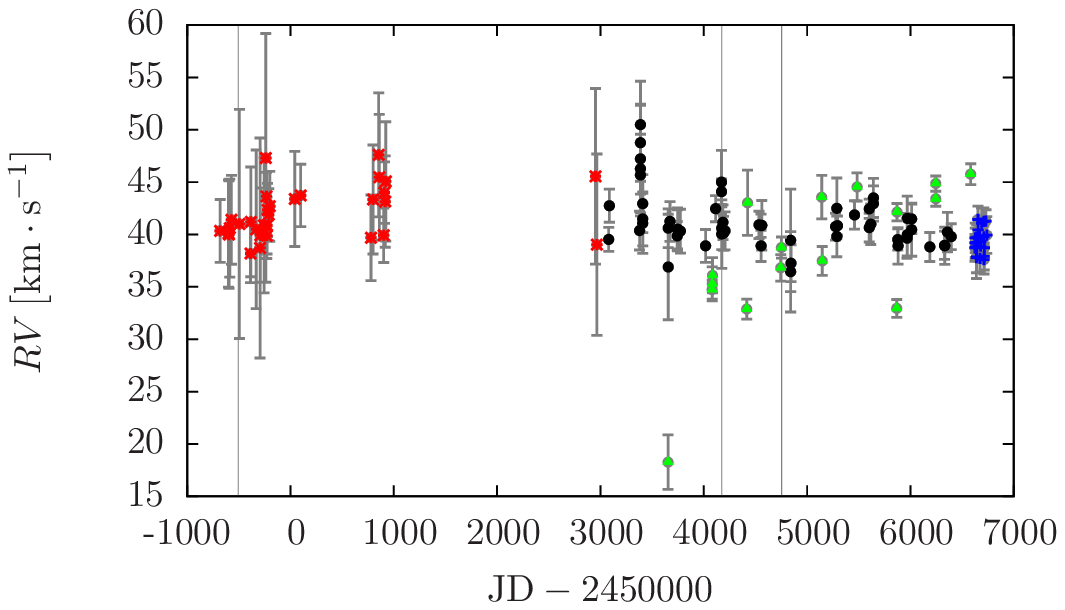}}
      \resizebox{0.5\textwidth}{!}{\includegraphics{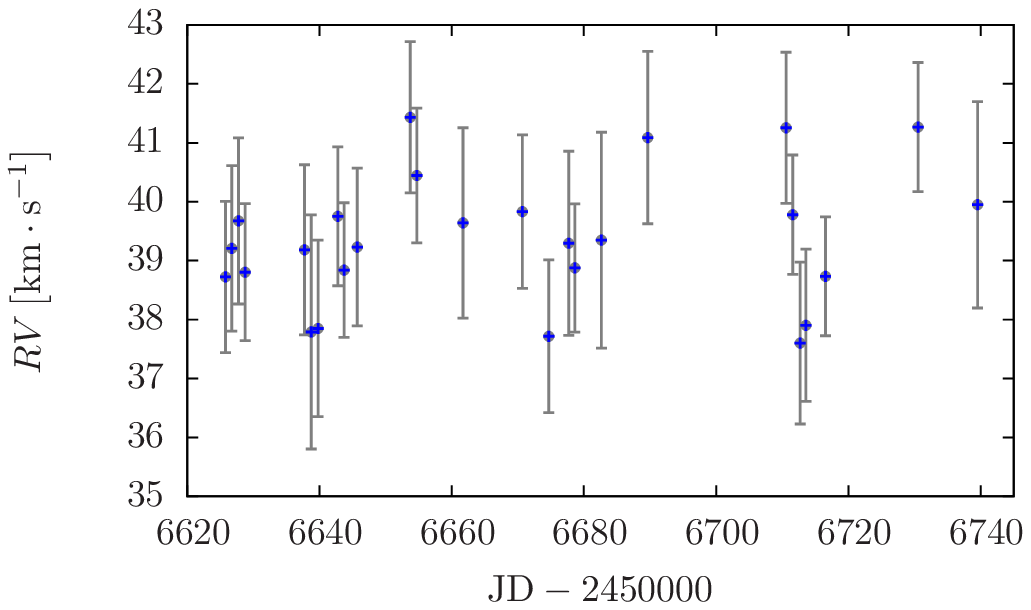}}
      }
   \caption{$RV$ of the  [\ion{O}{i}] 6364~\AA \, line.}
   \label{RVOI6364}
  \end{figure*}
   \FloatBarrier
}

\subsection{[\ion{O}{i}] lines}
\label{OI_section}

The forbidden lines are formed in very low-density media. It is supposed to be in the outer regions, 
which are not significantly affected by the processes that occur close to the star. The forbidden 
lines are good tracers of this matter. Because the [\ion{O}{i}] $\lambda \lambda$ 6300 and 6364~\AA\, 
lines have an identical upper level, its flux ratio in the optically thin static medium is given 
only by the Einstein probability coefficient of the spontaneous emission and frequency of the lines, 
$[F_{\rm{rel}}(\ion{O}{i}(6300)-1]:[F_{\rm{rel}}(\ion{O}{i}(6364)-1] \approx 3:1$. This ratio 
decreases with increasing optical depth up to the limiting value of 1 \citep[e. g.][]{Li92}. 

The variability of the [\ion{O}{i}] $\lambda \lambda$ 6300 and 6364~\AA\, lines 
has not been described well to date. \cite{Merrill31_ApJ} detected only the [\ion{O}{i}] 
6300~\AA\, line in his ten-year study of HD 50138; the [\ion{O}{i}] 6364~\AA\, line was very faint. 
However, any conclusions are somewhat uncertain because photographic plates were not very sensitive 
in the red part of the spectra at that time. Other measurements of these lines are summarised 
in Tables~\ref{RV_O_tab_st}$^{e)}$ and \ref{ew_O_tab_st}$^{e)}$.

We are limited in our study of the long-term behaviour of these lines because the [\ion{O}{i}] 6300~\AA\, 
line is not present in the spectra from RO. To measure the $EW$, relative fluxes, and 
$RVs$ we fit a~Gaussian function (Sect.~\ref{analysis},\,\textit{ii, iii}) to both [\ion{O}{i}] 
$\lambda \lambda$ 6300~\AA\, and 6364 \AA\, lines. The measured values of the $EW$ and relative fluxes are shown in 
Figs.~\ref{EWOI6300} ([\ion{O}{i}] 6300~\AA) and \ref{EWOI6364} ([\ion{O}{i}] 6364~\AA). These quantities 
can serve as tracers of the dynamics of the outer regions. The flux ratio (Fig.~\ref{OIratio}) is close 
to three in our spectra, which allows us to consider these lines as optically thin. Therefore, the change 
of the line shape due to the velocity field will change the relative flux, but the effect on the $EW$ will 
be negligible. Our measurements show that both the $EW$ and relative fluxes follow the same law.

It is not possible to say anything about the behaviour of the $EW$ of these forbidden oxygen lines on 
longer timescales than is present in our data because the published measurements (Table~\ref{ew_O_tab_st}$^{e)}$) 
cover the interval of our observations. 

The $RV$s for [\ion{O}{i}] 6300~\AA\,(Fig.~\ref{RVOI6300}) and for [\ion{O}{i}] 6364~\AA \, 
(Fig.~\ref{RVOI6364}$^{e)}$) are almost constant except for the night of 16 January 2005.
On that night a~series of short exposures (600 and 900~s) was obtained. The $S/N$ 
ratio was low (approximately 60) and rebinning could have played an important role. 

Because the forbidden lines are formed in the outer parts of the circumstellar media, and their 
$RV$s show small variations, it is possible to take the average as a~rough estimate of the $RV$
of the system $rv_{\rm{sys}} = 40 \pm 4$~\kms. This value is obtained from the Ond\v{r}ejov data with 
the exclusion of the night of 16 January 2005. The average of the $RV$s of both [\ion{O}{i}] lines 
is almost identical, $rv_{\text{[\ion{O}{i}] 6300}}=38 \pm 2$~\kms and $rv_{\text{[\ion{O}{i}] 6364}}=41 \pm 3$~\kms. 
To see the behaviour on longer timescales, we summarise published $RV$s of these lines in 
Table~\ref{RV_O_tab_st}$^{e)}$. Unfortunately, only three observations are outside of our time interval. 
One such measurement from outside our sampling interval was from December 1972 \citep{Andrillat1972}.
This low-resolution (40~\AA/mm) spectrum is the only one that shows a~significant difference of 
$RV$s between the two [\ion{O}{i}] lines. All observations show that the $RV$s of the [\ion{O}{i}] lines change only 
slightly around the value, which is stable for almost 70 years. 

To describe the changes of the optical depth of the forbidden oxygen lines, we plot  in 
Fig.~\ref{OIratio} the net flux ratio $(F_{[OI] 6364}~-~1)/(F_{[OI] 6300}-1)$. Since
the flux can be affected by the velocity field we also present the ratio of the $EWs$. The figures 
show that the medium in the forbidden oxygen line forming region has been optically thin for 
more than a~decade.

Even if the optical depth of the [\ion{O}{i}] lines can be considered optically thin
and $RV$s have been constant for a~long time, the lines themselves are not stable (see Figs.~\ref{EWOI6300} 
and \ref{EWOI6364}) as was thought in the past. The detailed connection between the individual lines, 
and hence their line forming regions, is discussed in Sect. \ref{correlation_all}.

\onltab{
  \begin{table}[!h]
    \caption{$RV$s of the~oxygen lines}
        \label{RV_O_tab_st}
        \begin{center}
        \begin{tabular}{lllll}
                \hline \hline
                     line                                       & date         & RV           & Dispersion/       &  Ref.\\
                                                &              & (\kms)       & Resolution                 &  \\
    \hline
    $\rm{[\ion{O}{i}]}$  6300 \AA               & 1945-Oct-24   & 36           &                  & 1 \\
    $\rm{[\ion{O}{i}]}$ 6300 \AA                & 1960-Jan-03   & $34$  & 12 \AA/mm  & 2\\
    $\rm{[\ion{O}{i}]}$ 6364 \AA                & 1960-Jan-03   & $44$  & 12 \AA/mm & 2 \\
    $\rm{[\ion{O}{i}]}$ 6300 \AA                & 1970-Dec      & $22\pm7$     & 40 \AA/mm & 3\\
    $\rm{[\ion{O}{i}]}$ 6364 \AA                & 1970-Dec      & $28 \pm 7$   & 40 \AA/mm &  3 \\
    $\rm{[\ion{O}{i}]}$ 6364 \AA                & 1995-Jan-11,  & $33\pm 5$    & $R = 5\,000$  & 4 \\
                                                & 1997-Jan-01   &              &               & \\
    $\rm{[\ion{O}{i}]}$ 6300 \AA                & 1999-Oct-17   & $38.1$       & $R = 55\,000$ & 5\\
    $\rm{[\ion{O}{i}]}$ 6364 \AA                & 1999-Oct-17   & $37.7$       & $R = 55\,000$ & 5\\
    $\rm{[\ion{O}{i}]}$ 6300 \AA                & 2007-Oct-04   & $38.1$       & $R = 55\,000$ & 5\\
    $\rm{[\ion{O}{i}]}$ 6364 \AA                & 2007-Oct-04   & $37.7$       & $R = 55\,000$ & 5\\
   \hline

        \end{tabular}
        \end{center}
        \tablebib{
         (1)~\citet{Merrill52} (2) \citet{Houziaux60}; 
         (3) \citet{Andrillat1972}; (4)~\citet{Oudmaijer99}; (5)~\citet{Borges09}.
        }  
 \end{table}
}
\onltab{
  \begin{table}[!h]
    \caption{$EW$s of the~oxygen lines}
    \begin{center}
    \begin{tabular}{lllll}
            \hline \hline
              line                                       & date         & EW           & Dispersion/    &  Ref.\\
                                                &              & (\AA)        & Resolution     &  \\
    \hline
    $\rm{[\ion{O}{i}]}$ 6300 \AA                & 1999-Oct-17   & $-0.55$       & $R = 55\,000$ & 1\\
    $\rm{[\ion{O}{i}]}$ 6300 \AA                & 2007-Oct-04   & $-0.58$       & $R = 55\,000$ & 1\\
    \hline
    $\rm{[\ion{O}{i}]}$ 6364 \AA                & 1999-Oct-17   & $-0.20$       & $R = 55\,000$ & 1\\
    $\rm{[\ion{O}{i}]}$ 6364 \AA                & 2007-Oct-04   & $-0.21$       & $R = 55\,000$ & 1\\
      \hline

    \end{tabular}
    \end{center}
    \tablebib{(1)~\citet{Borges09}.
    }
    \label{ew_O_tab_st}
  \end{table}
}

\subsubsection{Correlation between [\ion{O}{i}] and H$\alpha$ lines}
\label{O_Ha_cor}

We find important connections between [\ion{O}{i}] $\lambda$ 6300~\AA \, and H$\alpha$ line
(Fig.~\ref{O_Ha_corr_fig}$^{e)}$ and Table~\ref{O_Ha_corr_tab}$^{e)}$). Moreover, we 
identify three different time epochs determined by the $RV$ of the central depression. 
The same time intervals also show the $V/R$ ratio changes of the H$\alpha$ line. 
Distinct epochs are unambiguously determined when $V/R>1$. Even if the number of observations 
in a~given epoch is small, the size of the changes is significant, indicating 
different behaviour in different epochs similar to LBVs. 

The Pearson correlation coefficients of individual epochs are summarised in Table~\ref{O_Ha_corr_tab}$^{e)}$. 
This table contains only the $EW$ and $RV$ of [\ion{O}{i}]  6300~\AA \, because the results for the relative flux 
are the same thanks to their strong correlation.

Here we present the result for the Ond\v{r}ejov and TCO data: 
\renewcommand{\labelenumi}{\roman{enumi})} 
\begin{enumerate}[noitemsep,topsep=0pt,parsep=0pt,partopsep=0pt]  
 \item a correlation of the $EW$s;
 \item an anti-correlation between the relative flux of the H$\alpha$ red peak and the $EW$/relative flux of [\ion{O}{i}] 6300~\AA; 
 \item a weak correlation between the $RV$ of the H$\alpha$ red peak and the $EW$/relative flux of [\ion{O}{i}] 6300~\AA;
 \item an anti-correlation between the relative flux of the H$\alpha$ violet peak and the $EW$/relative flux of [\ion{O}{i}] 6300~\AA;
 \item no correlation between the $RV$ of the H$\alpha$ violet peak and the $EW$/relative flux of [\ion{O}{i}] 6300~\AA. \\
   We found a~strong dependence in the first epoch determined by $JD=2454650$, when $V/R>1$;
 \item no correlation between the $RV$ of the H$\alpha$ red peak and the $RV$ of [\ion{O}{i}] 6300~\AA; \\
   the data show very different behaviour in different epochs.
\end{enumerate} 

\onltab{
   \begin{table*}[!h]
    \caption{
       Correlation coefficients of the [\ion{O}{i}] 6300~\AA \, and H$\alpha$ lines in individual epochs.
    }
    \label{O_Ha_corr_tab}
    \centering
    \begin{tabular}{lrrr|rr}
    \hline \hline
                                                   JD                  & $2453079-$ & $2454650-$ & $2456600-$ & the entire \\
                                                                       & $2454650$  & $2456600$  & $2456740$  & sample \\
    \hline
     P($EW(\rm{H}\alpha)$, $EW(\text{[\ion{O}{i}]}\, \lambda 6300~\AA)$   &  0.70 & 0.11 & 0.06 & 0.56\\
    \hfill critical value                                                 & 0.35 & 0.38 & 0.37 & 0.22\\
    \hline
     P($F(\rm{H}\alpha RP)$, $EW(\text{[\ion{O}{i}]}\, \lambda 6300~\AA)$ & -0.70  & -0.24 & 0.07 & -0.57 \\
   \hfill  critical value                                                 &  0.35 & 0.38 & 0.37 & 0.22\\
   \hline
    P($RV(\rm{H}\alpha RP)$, $EW(\text{[\ion{O}{i}]}\, \lambda 6300~\AA)$ & 0.08 & 0.39 & 0.40 & 0.49 \\
    \hfill critical value                                                 &  0.35 & 0.38 & 0.37 & 0.22\\
   \hline
   P($F(\rm{H}\alpha VP)$, $EW(\text{[\ion{O}{i}]}\, \lambda 6300~\AA)$   & -0.70 & -0.01 & -0.44 & -0.70 \\
   \hfill  critical value                                                 &  0.35 & 0.38 & 0.37 & 0.22\\
   \hline
   P($RV(\rm{H}\alpha VP)$, $EW(\text{[\ion{O}{i}]}\, \lambda 6300~\AA)$  & -0.73 & 0.10 & -0.34 & 0.00 \\ 
    \hfill  critical value                                                &  0.35 & 0.38 & 0.37 & 0.22\\
   \hline
   P($RV(\rm{H}\alpha RP)$, $RV(\text{[\ion{O}{i}]} \,\lambda 6300~\AA)$  & -0.61  & 0.57 & -0.04 & 0.03 \\
   \hfill critical value                                                  & 0.37  & 0.42  & 0.41 & 0.22\\
   \hline
   \end{tabular}
   \end{table*}
 }

\onlfig{
  \begin{figure*}[!h]
    \includegraphics[width=0.5\textwidth]{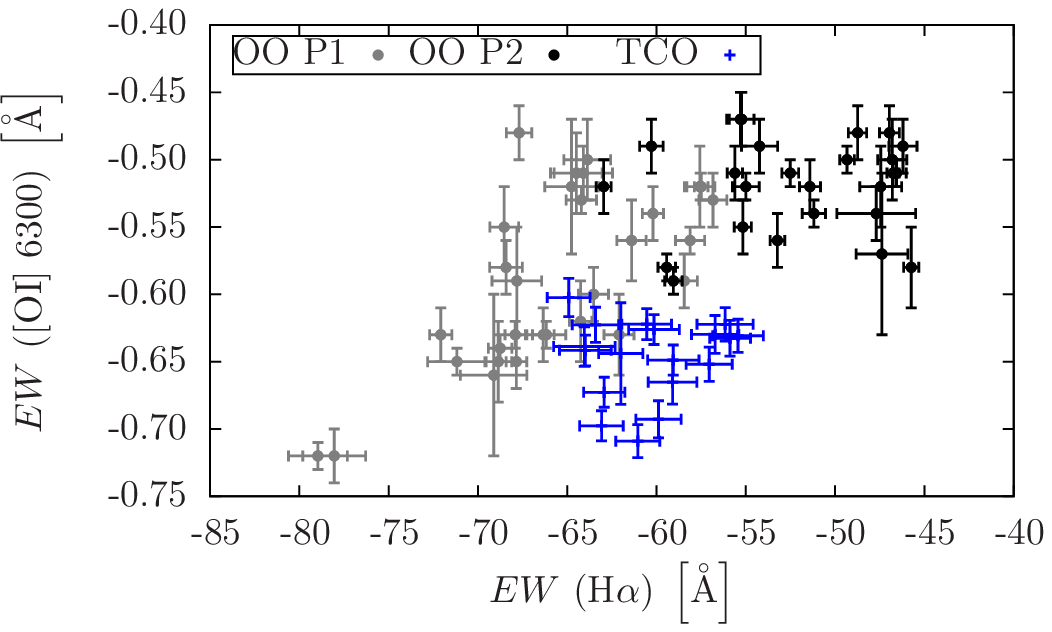}    
    \includegraphics[width=0.5\textwidth]{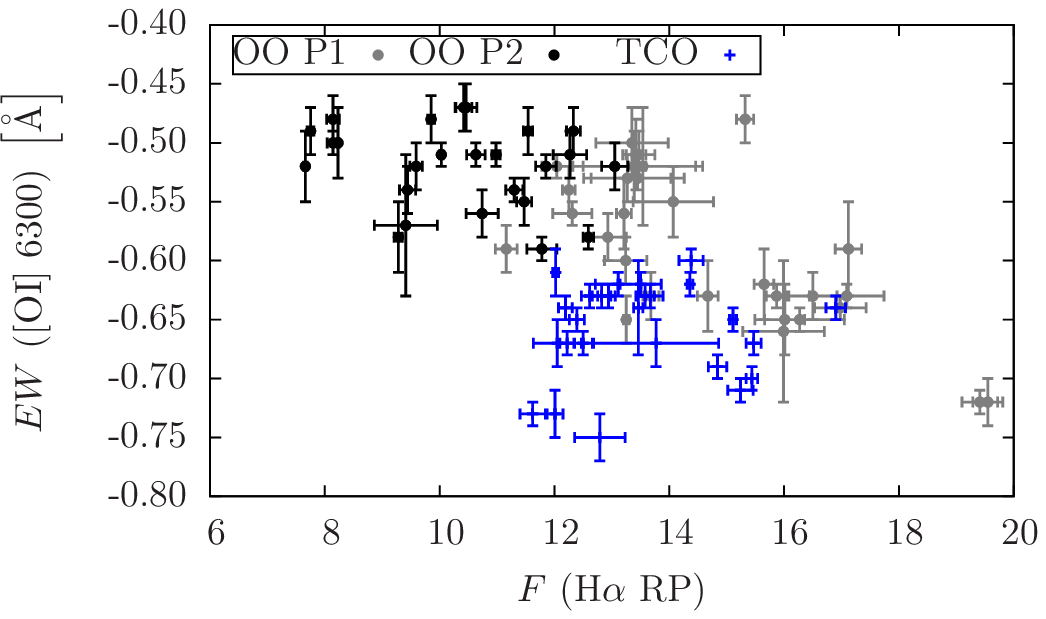}     \\
      \includegraphics[width=0.5\textwidth]{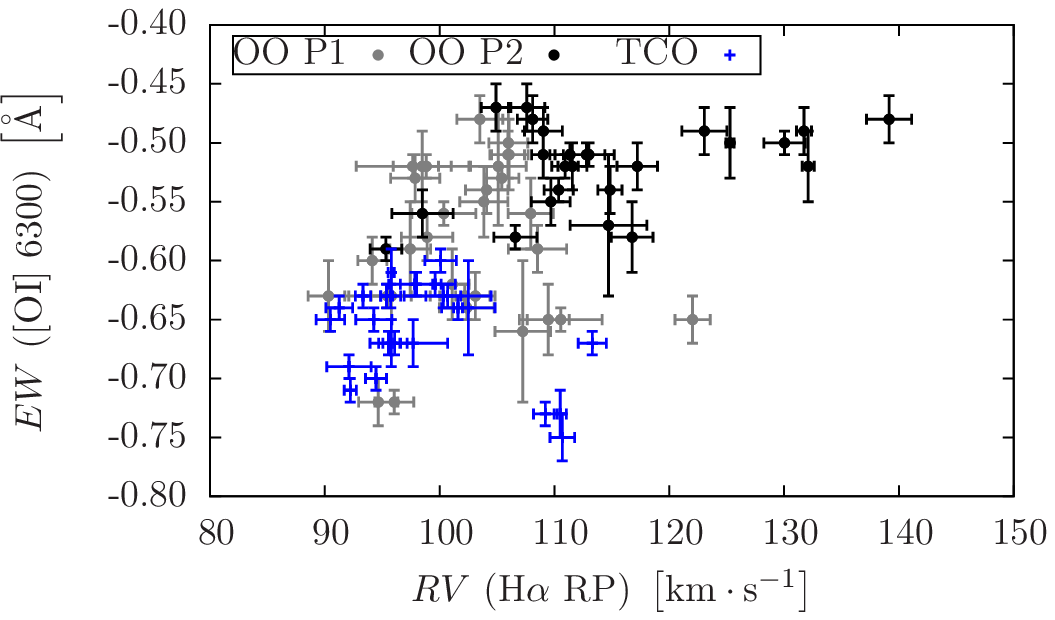} 
    \includegraphics[width=0.5\textwidth]{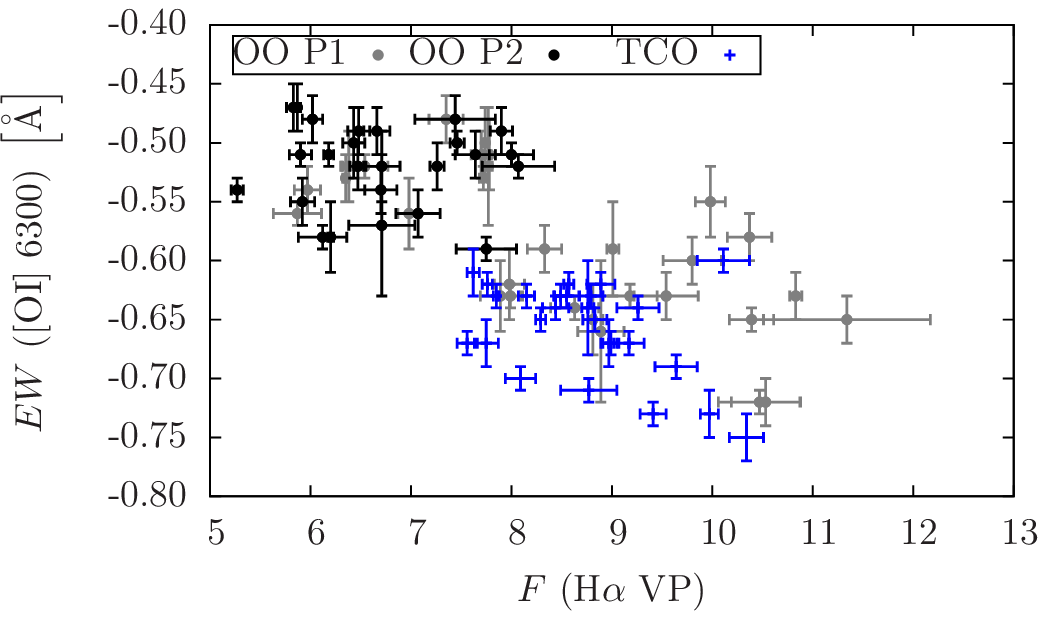} \\
    %
     \includegraphics[width=0.5\textwidth]{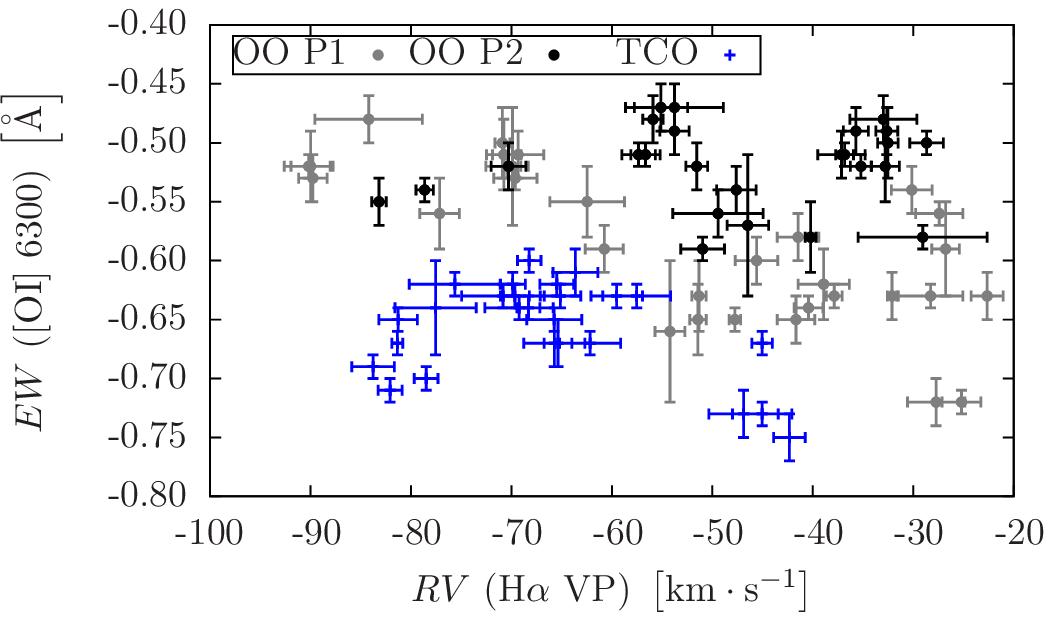} 
    %
     \includegraphics[width=0.5\textwidth]{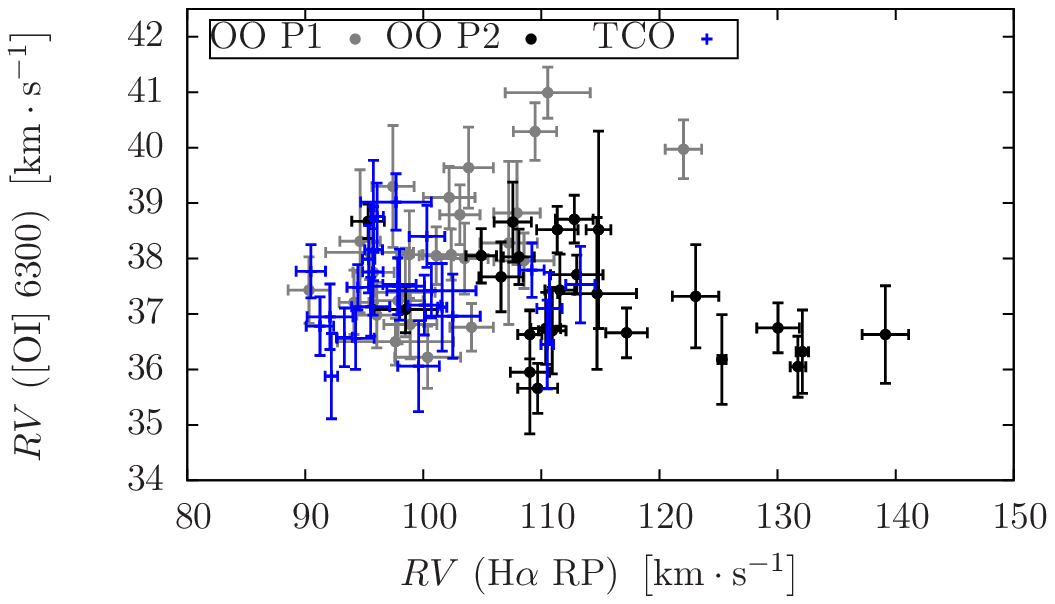} 
     \caption{
        Correlation diagram of H$\alpha$ and [\ion{O}{i}] $\lambda$ 6300~\AA \, lines. 
        Dots consistently denote the measurements of spectra from Ond\v{r}ejov observatory and crosses from TCO observatory
        throughout the paper. The values before the event, when $V/R$ > 1, are in black. 
        Blue indicates the value after another maximum in $V/R$ changes. The epoch between is in grey. 
        To emphasise that the same epochs are also defined by the $RV$s of the central depression, the error 
        bars are coloured according to the $RV$ extrema. The black, grey, and blue are used chronologically. 
        }
      \label{O_Ha_corr_fig}
  \end{figure*}
 \FloatBarrier
}

\subsection{\ion{Si}{ii} $\lambda \lambda$ 6347 and 6371~\AA \, lines}

The \ion{Si}{ii} $\lambda \lambda$ 6347 and 6371~\AA \, lines have always been observed 
in absorption with an occasional appearance of a~red or blue emission component.
The inverse P Cygni profile of these lines was reported by \cite{Borges09} in March 
and October 2007 in the FEROS spectra (R$\sim$55\,000). \cite{Borges12} described 
the rapid night-to-night variability from a~nine-day campaign in February 2011 taken 
with the HERMES spectrograph (R$\sim$ 85\,000).

\begin{figure}
    \resizebox{\hsize}{!}{\includegraphics{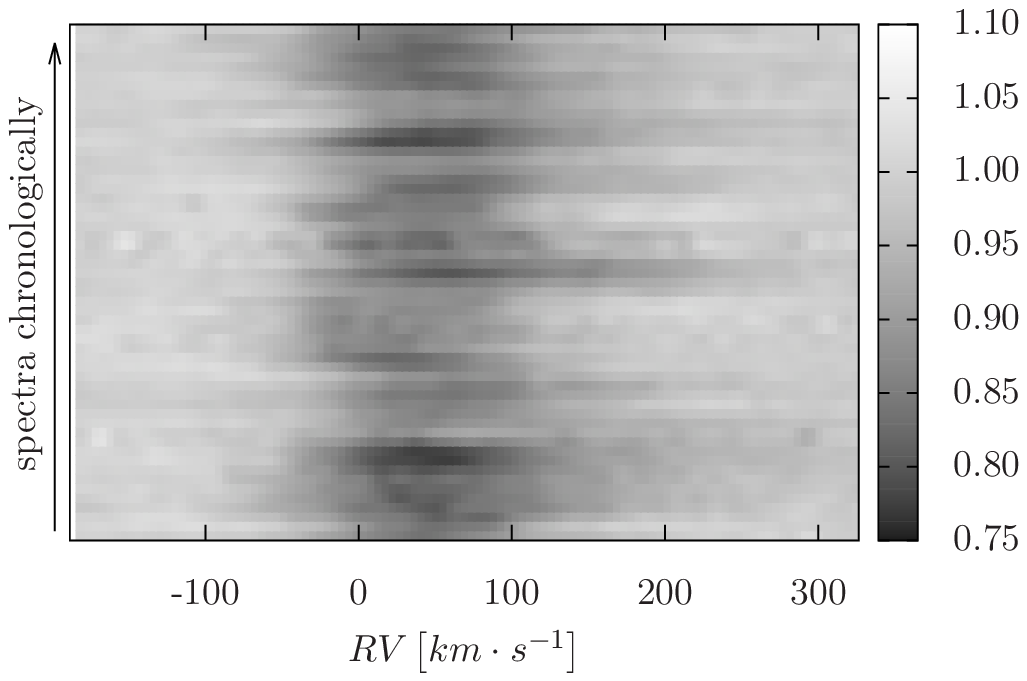}}
    \resizebox{\hsize}{!}{\includegraphics{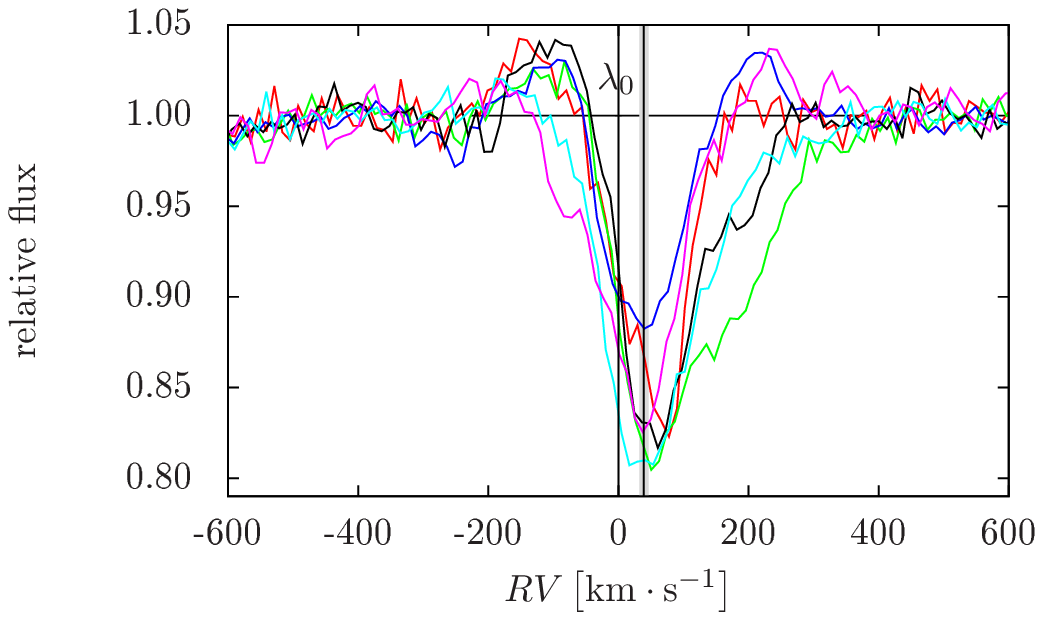}}
    \caption{
      Variations of the \ion{Si}{ii} 6347~\AA\, line. \textit{upper panel:} Grey-scale representation.
      \textit{bottom panel:} Typical profiles (Ond\v{r}ejov data, R$\sim$12500). The sharp 
      vertical line denotes the laboratory wavelength $\lambda_{0}$. The position of $\lambda_{0}$ in the frame 
      connected with the HD~50138 system is also plotted together with its error bar. The line shift is determined by the 
      $RV$ of the system $rv_{\rm{sys}} = 40 \pm 4$~\kms \, based on the $RV$s 
      of the [\ion{O}{i}] $\lambda \lambda$ 6300, 6364~\AA\, lines.
    } 
    \label{Si_prof}
\end{figure}

Most of our spectra showed an absorption line with a blue-shifted emission
wing. After the correction 
for the $RV$ ($rv_{\rm{sys}} = 40 \pm 4$~\kms, Sect.~\ref{OI_section}), we were able to distinguish 
a~real inverse P~Cygni profile, which was observed frequently. With almost the same frequency, a~pure 
absorption line was observed. 
Our present observations have very rarely included either symmetric
emission lines or an absorption line with a red emission wing.
The behaviour of both lines 
is very similar, but the absorption part of the inverse P Cygni profile of the \ion{Si}{ii} 6371~\AA \, 
line is usually shifted further to the red than that of the 6347~\AA\, line. The chosen line profiles 
of the stronger \ion{Si}{ii} 6347 line are plotted in Fig.~\ref{Si_prof} (bottom panel), and its 
variability is shown in its grey-scale representation (Fig.~\ref{Si_prof}, upper panel).

Owing to the complicated structure of these lines, we did not measure the $RV$ and $EW$.
It is important to note, however, that these lines can be crucial for future modelling
because they trace deep layers of the circumstelar media. 
For this reason we provide all of our spectra at the CDS database.

\subsection{\ion{He}{i} $\lambda \lambda$ 5876 and 6678 \AA \, lines}

\begin{figure}
    \resizebox{\hsize}{!}{\includegraphics{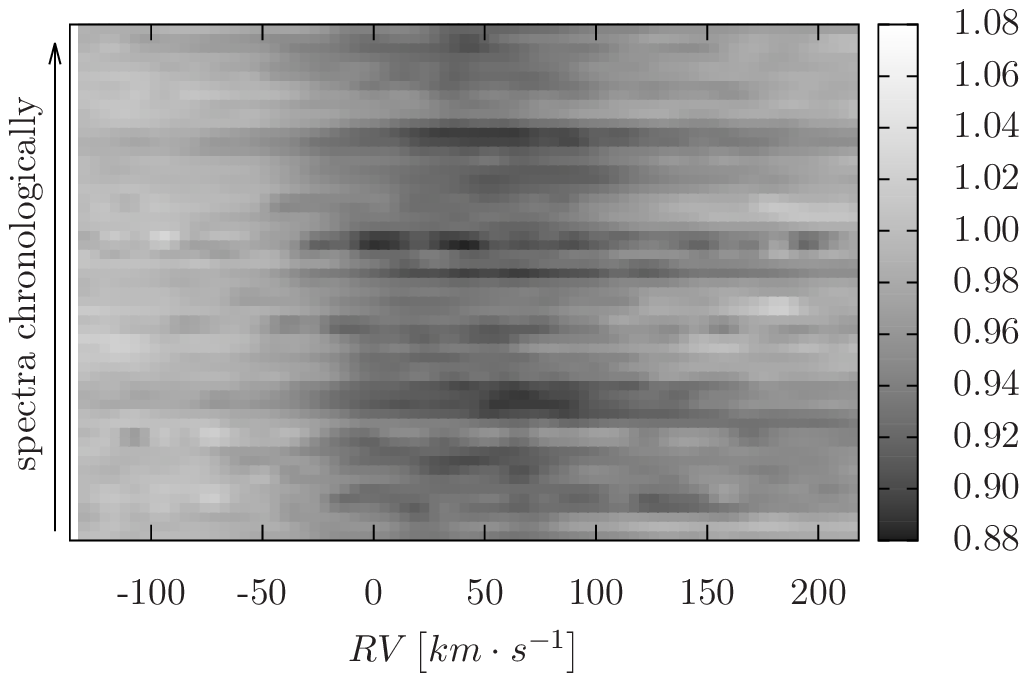}}
    \resizebox{\hsize}{!}{\includegraphics{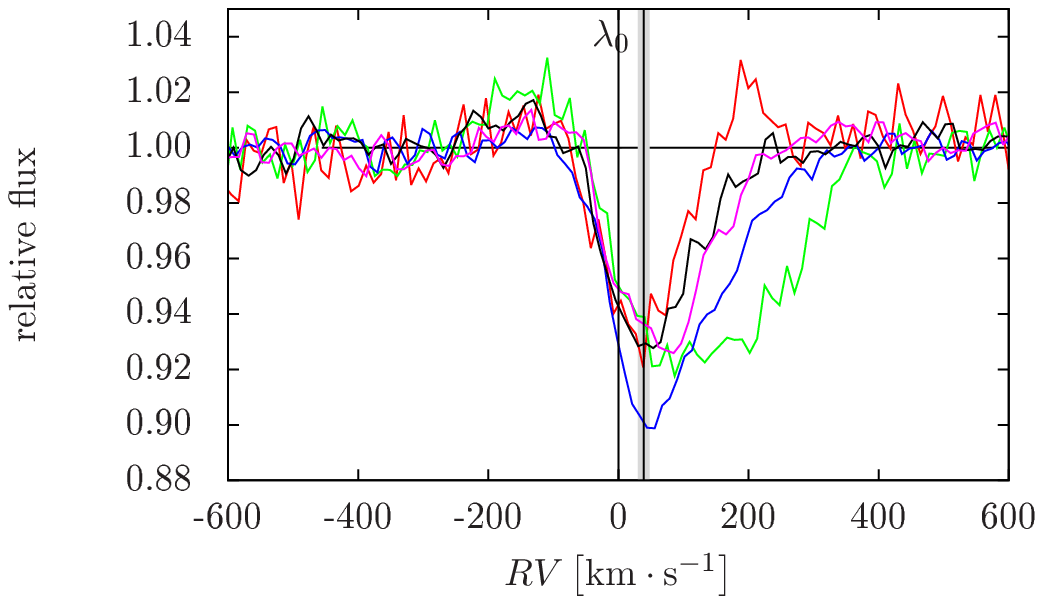}}
    \caption{
      Variations of \ion{He}{i} 6678~\AA \, line. \textit{upper panel:} Grey-scale representation. 
      \textit{bottom panel:} The typical line profiles are plotted from the spectra taken at the
      Ond\v{r}ejov Observatory (R$\sim$ 12\,500). The central wavelength is denoted  
       in the laboratory frame $\lambda_{0}$ and in the frame connected with the object. 
      The shift is determined by the system $RV$ $rv_{\rm{sys}} = 40 \pm 4$~\kms \, obtained from the 
      $RV$s of the [\ion{O}{i}] $\lambda \lambda$ 6300, 6364~\AA\, lines.
    }
    \label{He6678_prof}
\end{figure}

The \ion{He}{i} $\lambda \lambda$ 5876 and 6678~\AA \,lines exhibit behaviour that is similar to
the \ion{Si}{ii} lines. The rapid night-to-night variability is described by \cite{Borges12}. 
During the nine-day campaign of February 2011 with the HERMES spectrograph (R$\sim$~85\,000), 
they observed variations of the absorption profile of the \ion{He}{i} 6678~\AA \, line.
Our data (Fig.~\ref{He6678_prof}) show a~pure absorption profile, usually red-shifted, 
but more frequently an inverse P~Cygni profile (relative to the system radial velocity $rv_{\rm{sys}}$, 
Sect.~\ref{OI_section}). The absorption line showed an symmetric emission wing only once.

\begin{figure}
    \resizebox{\hsize}{!}{\includegraphics{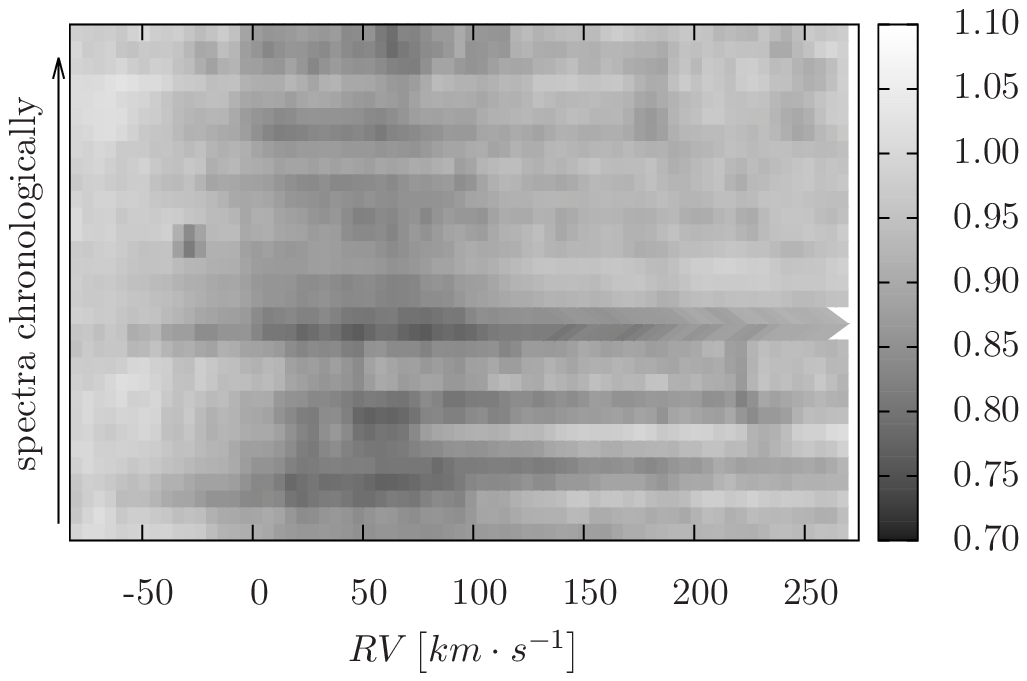}}
    \resizebox{\hsize}{!}{\includegraphics{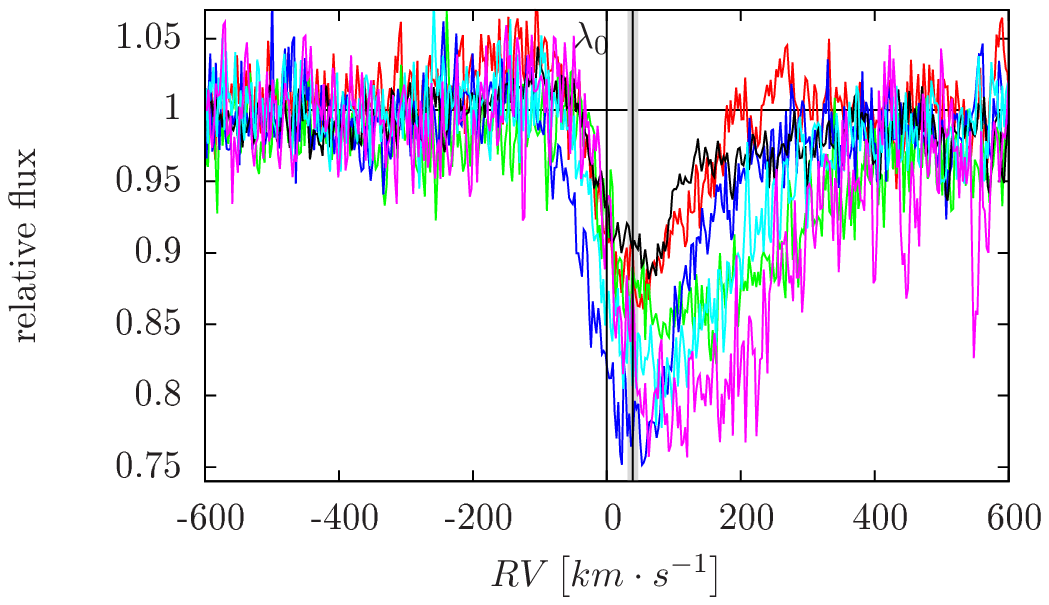}}
    \caption{Variations of the \ion{He}{i} 5876~\AA \, line. 
      \textit{upper panel:} Grey-scale representation. 
      \textit{bottom panel:} The typical line profiles were selected from the spectra taken at SPM 
      (R$\sim$18\,000). The vertical lines denote the central wavelength at 
      laboratory frame ($\lambda_{0}$, sharp line) and at the frame connected with HD~50138. 
      The line shift and its error bar are determined by the $RV$
      of the system $rv_{\rm{sys}} = 40 \pm 4$~\kms \, based on the $RV$s
      of the [\ion{O}{i}] $\lambda \lambda$ 6300, 6364~\AA\, lines.
    }
    \label{He5876_prof}
\end{figure}

The second line, \ion{He}{i} 5876~\AA, has been studied more frequently because of its position 
close to the NaD doublet. The line also shows a~strong variability. An inverse P~Cygni profile 
was reported by \cite{Bopp93} in spectra taken in 1992 and by \cite{Borges09} in the FEROS spectra 
from 27 October 1999. The absorption with a~weak violet wing was described by
\cite{Jaschek98} in spectra taken between 1989 and 1996 at Haute Provence observatory. Asymmetric absorption 
was noticed by \cite{Grady96} in January 1995. A~detailed study of the variability of the \ion{He}{i} 5876~\AA \, line
was done by \cite{Pogodin97} on the data from  15 - 18 March 1994 taken at the CAT telescope using 
the CES spectrograph. During these four nights the line showed a~pure absorption profile,
double-peaked emission, and an inverse P Cygni profile. We were able to study this line in the spectra from 
RO. Mostly we detected an inverse P Cygni profile (relative to the system radial velocity $rv_{\rm{sys}}$, 
Sect.~\ref{OI_section}), and we detected a~pure absorption a~few times.
The changes of the line-profile shape are shown 
by a~grey-scale representation in Fig.~\ref{He5876_prof}.

\subsection{\ion{Fe}{ii} $\lambda \lambda$ 6381 and 6384~\AA \, lines }

These two lines are too faint to be measured accurately in our spectra, but a~few notes on the
long-term behaviour of these lines can be useful for the future study of HD~50138. We always observe
both lines in emission, with a~slight variability in intensity and position. The  variability 
of the stronger \ion{Fe}{ii} 6384~\AA \, line is shown in the grey-scale representation in 
Fig.~\ref{Fe_col}$^{e)}$. When the line reached its minimum in intensity, the minimum intensity 
was also reached in the \ion{Si}{ii} $\lambda \lambda$ 6347, 6371~\AA, \ion{He}{i}  6678~\AA \, 
lines, and the red peak of the H$\alpha$ line.

\onlfig{
   \begin{figure}[!h]
        \resizebox{\hsize}{!}{\includegraphics{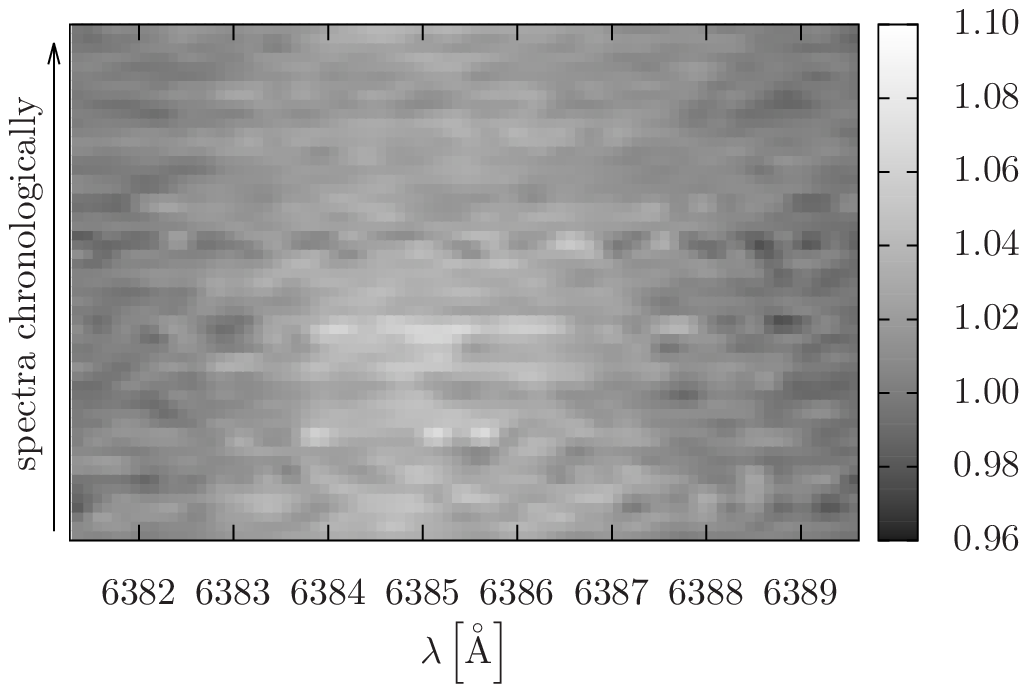}}
        \resizebox{\hsize}{!}{\includegraphics{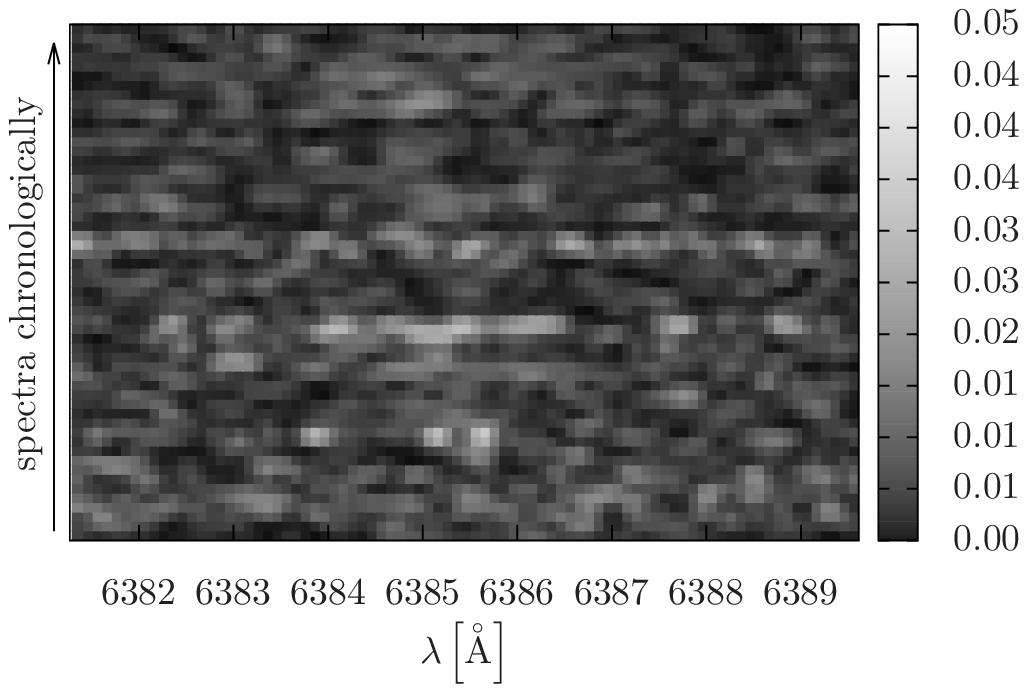}}
        \caption{Variability of \ion{Fe}{ii}  6384~\AA \, line.
          \textit{upper panel:}  Grey-scale representation. 
	  \textit{bottom panel:} Absolute variance from average.}
        \label{Fe_col}
    \end{figure}
 \FloatBarrier
	}

\subsection{Correlation between individual lines and events}
\label{correlation_all}

\begin{figure}
  \resizebox{\hsize}{!}{\includegraphics{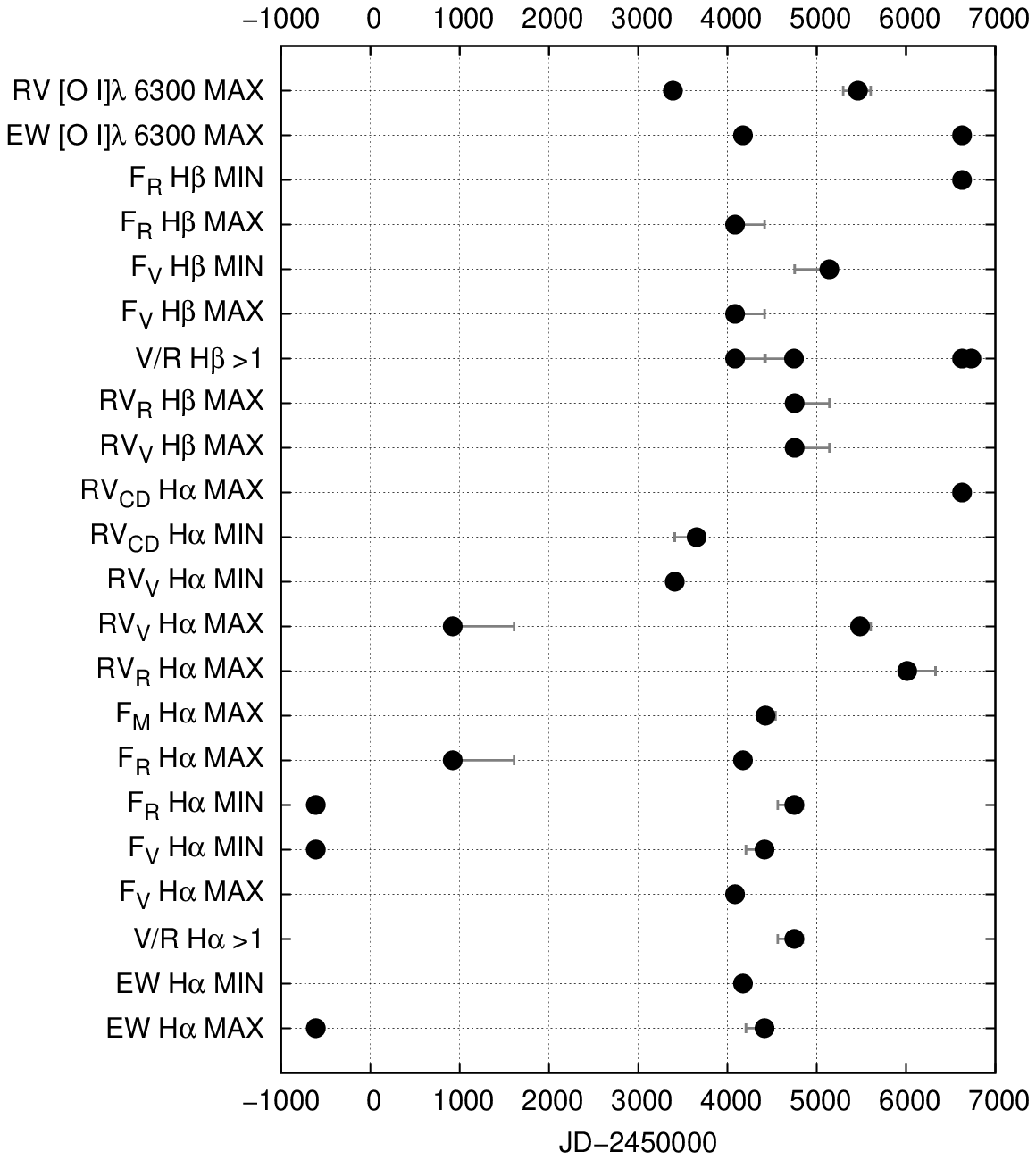}}
    \caption{Timeline of characteristic events in the spectra. $RV$, $EW$, and $F$ denote radial velocity,
     $EW$, and flux, respectively. Subscripts $R$, $V$, and $CD$ refer to the red peak,
     violet peak and central depression of the line profile. 
    }
    \label{events}
\end{figure}

To trace the important events through our observations, we plot the maxima
and minima of the individual measured quantities in the time diagram (Fig.~\ref{events}).  
Moreover, the maximum and minimum of the H$\alpha \, EW$ and maximum of the $V/R$ ratio
are indicated in all the relevant figures (vertical grey lines).
%

In Sects.~\ref{Hacorrsec}, \ref{Hb_Ha_cor_sec}, and \ref{O_Ha_cor} we discussed the correlations
between individual quantities in detail. Our analysis indicates that the behaviour of the system
was different in different epochs. This kind of variability is seen in other spectral studies 
such as studies of LBVs.

The number of measurements in the individual epochs is not sufficient (from 25 to 35) 
for satisfactory statistics. The measured data were separated by the $RV$ of the central depression,
which is connected with the speed of the wind. The difference between individual epochs is too high
to be simply statistical error (Tables~\ref{Ha_cor_tab}$^{e)}$ and \ref{O_Ha_corr_tab}$^{e)}$). 
The largest deviation was detected in the period from $JD$ 2\,453\,079 to 2\,454\,650, which is 
the first half of the uniform Ond\v{r}ejov data set. Moreover, the end of this period is determined 
by the maximum of the $V/R$ changes in the H$\alpha$ line ($V/R>1$). Figure~\ref{events} shows that 
H$\alpha$ $V/R>1$ occurred at the same time as the minimum of the H$\alpha$ red peak, the maximum 
of the H$\beta$ peak's $RV$s, and the maximum of the $V/R$ of the H$\beta$. Therefore, 
the different behaviour of the object in different epochs should be taken as one of the 
important criteria in the modelling of HD~50138 system.

It was impossible to describe the behaviour of the \ion{He}{i} 6678~\AA \, and \ion{Si}{ii} 
$\lambda \lambda$ 6347, 6371~\AA \, lines in a~simple way because of their complicated 
line profiles. However, these lines show a~similar behaviour. The most important event 
occurred on $JD\, 2\,454\,554.50$ when these three lines showed an extremely wide and deep red wing. 
At the same time, the $RV$ of the central depression of the H$\alpha$ line reached its most negative value.

\subsection{Long-term and short-term periodicity}

The period analysis was done using the \texttt{scargle} and \texttt{HEC27} codes 
(Sect.~\ref{analysis},\,\textit{vi}). We searched for a~long-term periodicity 
in a~time interval of a~thousand days and for a~short-term periodicity after data detrending 
(by dynamical averaging or long-term period subtraction). 

The most important periods are revealed in the $EW$ of the H$\alpha$ line.
The combination of the two most significant peaks of the power spectrum (Fig.~\ref{EW_Ha_periods}$^{e)}$, top panel)
fit the measured values well (Fig.~\ref{EWHa}; upper panel) and give the period estimates of
$3\,200 \pm 500$ and $5\,000 \pm 500$~days. The same periods are found 
 in the $RV$ of the violet peak. The violet peak data also show an additional 
period of $1000\pm600$~days. The red peak and central depression variability 
supports this result ($P=3\,000 - 4\,000$~days). It is possible to identify a~period 
of $3\,200$~days in the $EW$s of [\ion{O}{i}]~6300~\AA. Because of the reduced data set for 
the [\ion{O}{i}]~6300~\AA \, line, we use the exact values of periods found
in the H$\alpha$ $EW$ dependence to fit the [\ion{O}{i}]~6300~\AA \, values. 
The resulting fit is shown in Fig.~\ref{EWOI6300} for $EW$s and for relative fluxes. 

After subtraction of the global trend (Fig.~\ref{EW_Ha_periods}, bottom panel),
a~period of $300 \pm 50$~days was found in the $EW$ of the H$\alpha$ line.
Changes in the $V/R$ of H$\alpha$ show a~period of $600\pm 50$~days; 
a~less significant period of approximately $50\pm10$~days was found only by \texttt{HEC27}. 
This value is similar to the value of $60\pm10$~days, which is shown by the
$RVs$ of the H$\alpha$ red peak and central depression. \texttt{HEC27} indicates the shortest period 
of $20\pm5$~days in the $RV$s of the H$\alpha$ peaks and central depression.

To search for short-term periods, we use the TCO data set because its time coverage is better.
This allows us to detect a~periodicity of the order of days to tens of days. We found no
distinct period. The $RV$ of the H$\alpha$ red peak suggests 20 and 35 days
periods and the central depression of the H$\beta$ line 9 and 33 days, 
but they are not able to fit the data well.

\onlfig{
 \begin{figure}[!h]
    \resizebox{0.5\textwidth}{!}{\includegraphics{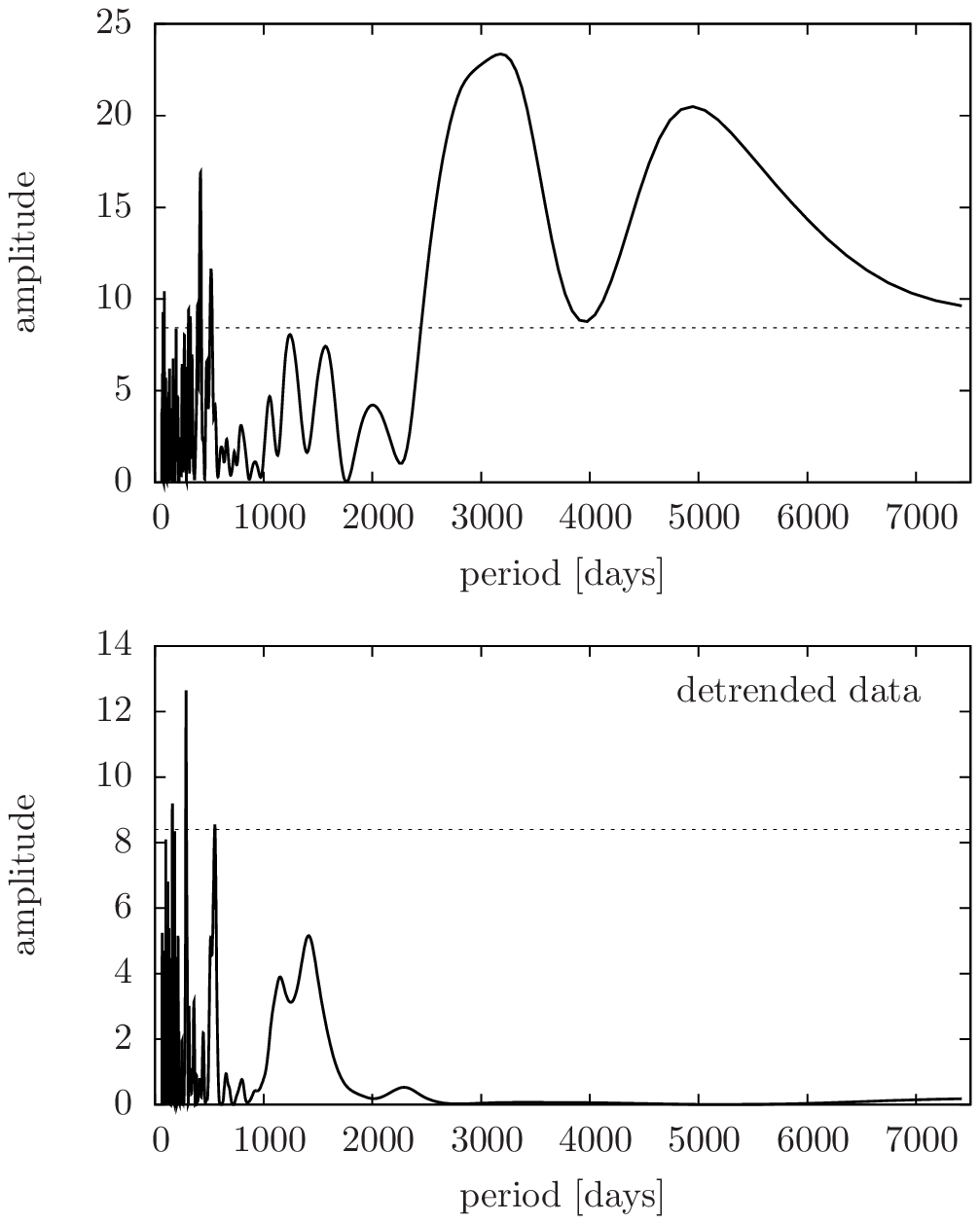}}
   %
   \caption{
     Power spectra of the H$\alpha$ $EW$s. \textit{upper panel:} Periodogram of the measured values of $EW$s.
     \textit{bottom panel:} Periodogrogram after the 
     removal
     of the long-term trends by the dynamical averaging.
   }
  \label{EW_Ha_periods}
\end{figure}
 \FloatBarrier
 }


\section{Discussion} 
\label{diskuze}

Our spectra of HD~50138 showed a~strong variability of this object. A~detailed discussion of the influence 
of the individual physical phenomena on the spectral line profiles is presented in this section, 
but first it is necessary to briefly describe the $EW$ variability in the context of the photometry.

\subsection{EW variations vs. photometry}

Changes in the line $EW$s result not only from changes in the lines themselves, but also 
from the variations of the continuum radiation. This effect must be taken into account
in the interpretation of the $EW$ variability. 

In this regard it is essential to consider the very important work of \cite{Halbedel91}. 
She photometrically monitored HD~50138 for five years, from 1985 to 1991. During the 
observing campaign, the brightness of the star remained almost constant. The maximum
variation in the $V$--band was $0.11$\,mag. The observations were also carried out eleven days 
before the discovery of a~new shell phase by \citet{Andrillat91} and forty days after that. No 
significant photometric changes were found. The photometric variability from another
observational campaign (Sect.~\ref{uvod}) was also found only on small scales. This indicates
that the observed line variability is connected mainly with the line emissivity and opacity 
of given transitions. 

\subsection{System dynamics vs. correlation of measured quantities}

The relationships between individual measured line parameters
contain the information about the dynamics of the circumstellar matter,
which makes them very important for the construction of the model of HD~50138.

Most of the strong correlations discussed here can be qualitatively explained by 
changing the speed of the expansion velocity -- the correlation between the $RV$s of the
H$\alpha$ peaks and the central depression (Sect.~\ref{Hacorrsec} \textit{iii}, Fig.~\ref{Cor_RVHa}$^{e)}$),
the $RV$ of the H$\beta$ red peak and the flux of the H$\alpha$ red peak (Sect.~\ref{Hb_Ha_cor_sec} \textit{i}, 
Fig.~\ref{Hb_Ha_corr}$^{e)}$, upper left panel), the $RV$s of the H$\beta$ central depression and the H$\alpha$ 
violet peak (Sect.~\ref{Hb_Ha_cor_sec} \textit{iii}, Fig.~\ref{Hb_Ha_corr}$^{e)}$, bottom left panel),
and the flux of the H$\alpha$ red peak and the $EW$ and relative flux of the [\ion{O}{i}] $\lambda$ 6300~\AA \, 
line (Sect.~\ref{O_Ha_cor} \textit{ii}, Fig.~\ref{O_Ha_corr_fig}$^{e)}$, right panel, upper graph).

The effect of changing line opacities and emissivities is visible on the relative fluxes 
of both peaks and central depression  of the H$\alpha$ line (Sect.~\ref{Hacorrsec} \textit{ii}, 
Fig.~\ref{Cor_FHa}$^{e)}$), and the $EW$s of the H$\alpha$ and [\ion{O}{i}] $\lambda$ 6300~\AA \, line
(Sect.~\ref{O_Ha_cor} \textit{i},  Fig.~\ref{O_Ha_corr_fig}$^{e)}$, left panel, upper graph).

Important for the determination of the region of the forbidden line formation can be the dependence between 
the $RV$s of the H$\alpha$ red peak and [\ion{O}{i}] (Sect.~\ref{O_Ha_cor} \textit{vi},  
Fig.~\ref{O_Ha_corr_fig}$^{e)}$, right panel, bottom graph). The fact that they are not correlated
could be explained by changing the speed of the outflow, but with the assumption 
that the forming region of the forbidden lines is far from the central object, 
whose projection is almost negligible in this region. 

Other correlations found in the present study have no straightforward interpretation
since they are affected by several processes. Nevertheless, the well-established correlations 
confirm that the material outflow is present in the system, but it is unclear 
whether they are caused by a~variable wind or by an expansion of the disc.

A~rough guess of the system geometry is shown in Fig.~\ref{events} summarising the timeline
of the maxima and minima of individual quantities. Particulary, we point out that the maxima
and minima correspond to the $V/R$ of the H$\beta$ line greater than one.
This behaviour could not be explained by a~simple binary model with a~symmetric disc. 
A~highly distorted disc with a~spiral arm fits the situation better (Fig.~\ref{model_disc}$^{e)}$). 
The line-profile variations are affected by orbital and precession period \citep{Foulkes04,Regaly11}. 
Similar geometry show some short periodic cataclysmic variables, SU UMa stars, or some of the transient low-mass X-ray
binaries.
\onlfig{
  \begin{figure}[!h]
    \begin{center}
      \includegraphics[width=0.3\textwidth]{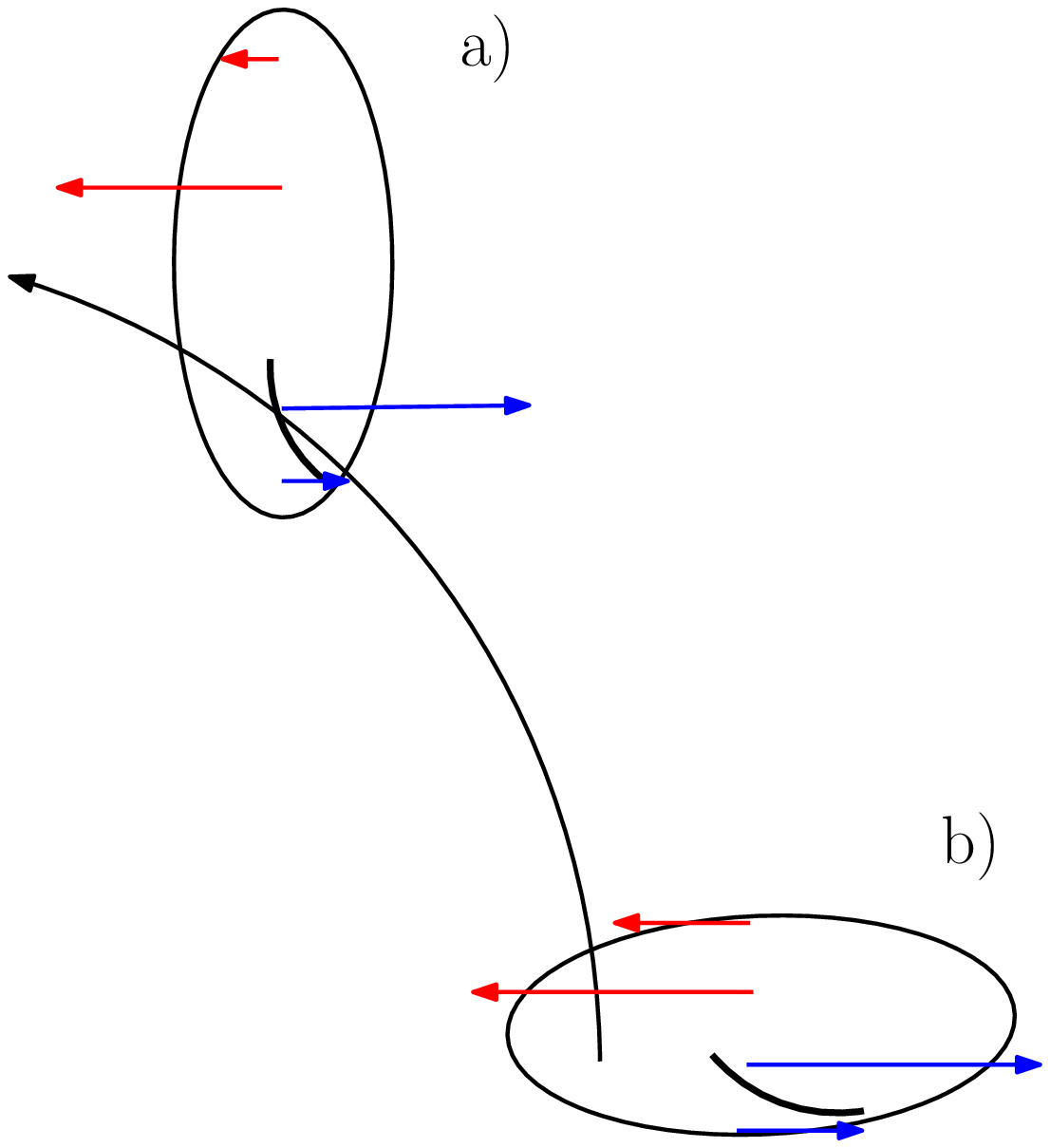}
    \end{center} 
    \caption{
     Possible system geometry corresponding to H$\beta$ $V/R>1$. \textit{a)} JD $\sim$ 2454000: $RV(\rm{H}\alpha(\rm{RP}))$ shows
     the local minimum, $F(\rm{H}\alpha (\rm{RP}))$ maximum, and $RV(\rm{H}\alpha(\rm{VP}))$ minimum; \textit{b)} JD $\sim$ 2454800:
     $RV(\rm{H}\alpha(\rm{RP}))$ reaches the local maximum, and  $F(\rm{H}\alpha (\rm{RP}))$  minimum. 
     The straightforward inclusion of other measured quantities is not possible. To determine them and prove the model,
     the radiative transfer based on hydrodynamic simulations is needed. 
                       }
           \label{model_disc}
  \end{figure}
 \FloatBarrier
   }

\subsection{Wind}

Clear evidence of outflowing material is the detection of a~P~Cygni profile. This line-profile shape was 
observed, for example in the \ion{Mg}{ii} resonance doublet in IUE spectra \cite[$1978-1982$;][]{Hutsemekers85}. We did not detect 
a~P~Cygni profile in any of the lines in our optical spectra. On the other hand, there is indirect 
evidence supporting material outflow. The central depression of the H$\alpha$ line is always blue-shifted 
with respect to the system velocity defined by the forbidden oxygen lines. Signatures of the wind can also be seen
in Fig.~\ref{HaRVcomplet}. On the right-hand side of the figure there is a~clear correlation of the 
$RV$s of both peaks and the central depression (Table~\ref{Ha_cor_tab}$^{e)}$). A~possible wind effect can be seen 
in the $RV$ variations of the violet peak, which are twice as large as those of the red peak and the central depression.

However, the situation in HD~50138 is more complicated. Previous studies, 
as well as our observations, show signatures of material inflow and expanding layers.

\subsection{Expanding layers}
\label{expanding_layers}

Expanding layers with periodic changes of the expansion velocity were found by \cite{Doazan65}.
Almost two decades later \cite{Hutsemekers85} detected a~slowly expanding decelerating layer 
in IUE spectra ($1978-1982$).

Our long-term observations show that the model of the system has to be even more complex. 
The $RV$s of the central depression and red peak of the H$\alpha$ line (Fig.~\ref{HaRVcomplet}) show, 
after the global trend subtraction, changes with a~period of about 60 days. This is probably the same 
phenomenon described by \cite{Doazan65} in the analysis of spectra (75 \AA/mm, and 100 \AA/mm) taken in 
$1960-1963$ at Observatoire de Haute-Provence. She found a~period of 50 days, but her observing
campaign was too short to describe long-term behaviour. Our data not only reveal the presence of 
long-term changes, but also indicate that their origin is different. A~few correlation
diagrams (mostly the $RV$ of the red peak vs. flux of the violet peak of the H$\alpha$ line,
and flux and $RV$ of the central depression of the H$\alpha$ line) show inhomogeneous structure. 
It is possible to break down these diagrams based on the individual epochs in the diagram
of the $RV$ of the central depression (Fig.~\ref{HaRVcomplet}). In these epochs,  
the correlations and anti-correlations are different. It is noteworthy that splitting 
the observations into four sub-intervals significantly 
lowers the number of measurements for the statistics in individual sub-intervals. To demonstrate the size
of this effect, we present the correlation coefficients for the dependence of flux of the violet peak 
and $EW$ of the H$\alpha$ line (Table~\ref{Ha_cor_tab}$^{e)}$).

\subsection{Material infall}

Material infall has also been observed in HD~50138. The most distinctive signature of this
event is the detection of inverse P~Cygni profiles. We found them in the \ion{Si}{ii} 
$\lambda \lambda$ 6347, 6371~\AA, and \ion{He}{i} $\lambda \lambda$  5876, 6678~\AA \, lines
in several spectra. To determine if the absorption is red- or blue-shifted with respect to the 
HD~50138 system, we used the $RV$s of the [\ion{O}{i}] $\lambda \lambda$~6300, 6364~\AA \, lines.

The inverse P~Cygni profile in various lines was detected several times. \cite{Andrillat91} reported
this line-profile shape at \ion{O}{i} 7773~\AA \, on  29 December 1990. A~few months 
later, on 10 April 1991  \cite{Bopp93} detected the inverse P~Cygni profile in \ion{He}{i} 5876~\AA. 
This profile of this line was also observed by \cite{Pogodin97} in March 1994. 
\cite{Grady96} found an inverse P~Cygni profile in the \ion{C}{iv} 
resonance doublet $\lambda \lambda$~1548, 1551~\AA \, in the ultraviolet.

\subsection{Rotating disc} 

Our data show the presence of a~rotating disc. We detected moving humps in both the red and violet peaks.
Detection of a~hump in the violet peak is more frequent, which is probably only due to its
shape. The red peak is steeper, therefore small deformations are more difficult to detect. The 
$RV$s of the well-defined humps in the violet peak are plotted in Fig.~\ref{HaRVcomplet}. We were able 
to follow some temporal dependencies indicated by the blue lines in Fig.~\ref{HaRVcomplet}. 
These polynomial fits show that the inhomogenities revolve around the central object. 

Evidence in support of a~rotating disc were found by comparing peak separations of the Balmer lines.
Even if the line-profile shape is significantly distorted by material outflow we argue that H$\alpha$ 
and H$\beta$ lines in several spectra imply a~disc-like structure. The peak separation is larger for
the H$\beta$ line. This proves that the velocity is higher for the region closer to the star as in a~rotating disc.

\subsection{Binary hypothesis}

The binary origin of HD~50138 is the most favourable scenario not only for this object, 
but also for other FS~CMa stars. It naturally explains a~large amount of the circumstellar matter
that can be created due to mass-transfer between the two stars. On the other hand, direct conclusive 
and incontrovertible evidence of the presence of the secondary component is still missing.
\renewcommand{\labelenumi}{\roman{enumi})}  
\begin{enumerate}[noitemsep,topsep=0pt,parsep=0pt,partopsep=0pt] 
 \item 
   Analysis of the \textit{radial velocities} is not straightforward in this case
   because the line-profile shape is changed by the circumstellar matter suppressing 
   the effect of the binary motion. We present here the most extensive set of measurements of 
   $RV$s with the best temporal coverage. However, the periodogramme analysis does not show  
   a~distinctive period. Observations obtained between 1920 and 1930 also did not show any simple 
   periodicity \citep{Merrill31_ApJ}. The variations of $RV$s measured between 1960 and 1963 
   were interpreted by \cite{Doazan65} as changes in the expansion velocity. Unfortunately, the other 
   studies are based on only a~small amount of data and are not appropriate for the period analysis. 

 \item 
   Evidence of the binary nature of HD~50138 was found by
   \cite{Baines06} in the \textit{spectroastrometric} observations.

 \item The \textit{polarimetric measurements} of \cite{Bjorkman94} show signatures of 
   binarity. On the other hand, these observations can also be explained by the presence 
   of orbiting dust. 

 \item The detection of the \textit{central quasi-emission peak} (Fig.~\ref{cqepHa}$^{e)}$) 
   can support the binary hypothesis. This weak emission appears almost at the systemic
   velocity, which indicates that the emission can originate near the Lagrangian point L1. However, 
   this explanation is not unique. \cite{Hanuschik95} showed that this emission is the
   radiative transfer effect through the circumstellar disc, which is optically thin
   in the continuum and thick in the line center. A~detailed discussion is presented in
   \cite{Rivinius_hab}.

 \item A~possible explanation of why the secondary companion has not been
    detected is that it is much fainter than the visible B-type companion.
    Comparison with models of non-conservative evolution of binary systems  \citep{Rensbergen08} 
    shows that the initially more massive companion may
    lose a~significant fraction of its mass and become much fainter than the
    initially less massive companion while still filling its Roche lobe. Since
    at this time the mass reversal has occurred, the hot companion has been spun
    up and become more massive. Therefore, it may not be easy to detect its orbital motion. 

\end{enumerate}

\subsection{Excitation waves}
\label{excitation_waves}

Our observations show that the $RV$s of the [\ion{O}{i}] lines are constant but their $EWs$ and 
fluxes change. This phenomenon can be explained by the presence of excitation waves in the media. 
This is inconsistent with photometric observations.

\subsection{Transit-time damping}

An alternative explanation of the observed phenomena described above (Sect.~\ref{excitation_waves}) 
is offered by a~magnetic field, particularly the transit-time damping effect \citep{Suzuki06}. They show that 
propagating fast waves through the rotating stellar wind that includes a~magnetic
field leads to the transfer of the magnetic energy into the plasma. 

The conditions necessary for this effect are found in B[e] stars, namely \textit{i)} rotation 
of the circumstellar matter, which leads to the creation of spiral arms; \textit{ii)}
low density, to achieve the condition of a~collisionless plasma; and \textit{iii)}
a~very extended volume that generates enough radiation so that it can be detected.

As the angle between the wave propagation vector and magnetic field changes due to the 
rotation of the circumstellar matter, the size of this effect grows and shrinks.
Differing amounts of energy are transformed from the magnetic field into the plasma. Especially
important are electrons, which travel quickly through the media. The different group
velocity changes the electron density in different areas. The different energy and density 
of electrons change the level population in the different areas of the media and leads 
to the changing of the line intensity. A~distant observer detects changes of the line intensity, 
but with no changes in the $RV$ and the brightness; this is exactly what is observed in the [\ion{O}{i}] lines.

However, our observations themselves do not prove that transit-time damping plays a~significant role 
in the circumstellar media of the B[e] stars. To show this, finer time coverage and 
more precise photometry are needed, as well as simultaneous observations in UV and 
visible because of the L$\beta$ pumping of the \ion{O}{i} 3d $^{3}D^{0}$ energy level. 
Decay from this level also affects [\ion{O}{i}] $\lambda \lambda$ 6300, 6364~\AA.

\subsection{Post-merger systems} 

\cite{Fuente15} found two FS~CMa stars in the open clusters Mercer 20 and 70. This 
allows them to restrict the age of these stars to the interval from 3.5 to 6.5 Myr.
This excludes pre- as well as post-main sequence scenarios. They were not able to detect
any signature of the secondary component in the spectra, which indicates that if 
the stars are binaries, the luminosity ratio must be high. Alternatively, the stars could 
be post-merger stellar objects. The current or former binary nature of FS~CMa type stars
is supported by the fact that both stars were detected in the crowded regions of the clusters.

\subsection{HD~50138 among others FS~CMa stars}

The variability of FS~CMa  stars has not been well established, but it is possible 
to make some statements. HD~50138 is noteworthy among FS~CMa stars because of the very
small variability of its brightness. On the other hand, the spectral variability is large, 
as it is in MWC~342 \citep{Kucerova13}. They detected periodicity of $RV$ of the central depression 
(4.3 years), $EW$ (769 days and 2.1 years), and $V/R$ changes (4 years) which are shorter but 
better defined than in HD~50138.


\section{Conclusions} 
\label{conclusion}

We obtained and analysed the spectra of the B[e] star of the FS~CMa type HD~50138
over the past twenty years. Because of the huge amount of circumstellar matter,
we have no direct information about the central object. Therefore a~description of the variability 
is crucial in the study of such objects.

The present measurements show a~previously unexpected behaviour. 
The forbidden oxygen lines are strongly correlated with the H$\alpha$ line. 
Therefore, their region of formation may be located closer to each other than previously thought. 
The double-peaked Balmer line profiles and especially the moving humps detected in both the H$\alpha$
and H$\beta$ lines proved that there is a~disc-like structure of the circumstellar gas around the star. 

We used different methods to search for a~periodicity of the measured quantities, but
we detected no explicit period. We only found modest signatures of a~regular periodicity 
of about 34 days only during one season from TCO (from 29 November 2013 to 23 March 2014). 
Even if the corresponding peak in the power spectra of the $RV$s of the H$\alpha$ red peak
and H$\beta$ central depresion is not very significant and the fit of measured values shows 
large deviations, the variations on this timescale are probably real. A~period of about 30 
days can also be detected in the spectra obtained from December 1920 to February 1930 by \cite{Merrill31_ApJ}.
Such a~timescale is compatible with a~possible orbital period of a~binary system with 
current or previous mass transfer between the components \cite[e.g.][]{MWC728}.

Long-term variations can be described by a~combination of two periods of 8.8 and 13.7 years.
Considering the results of previous studies of long-term variability of other FS~CMa stars \citep{Polster12,Kucerova13}, 
one can conclude that quasi-periodicity on timescales of several years seems to be a~typical signature of 
these objects.

The detrending of the H$\alpha$ $V/R$ changes allows a~variability to be detected of the order of 50 days, 
which is consistent with the results of \citet{Doazan65}. The presence of the period on this timescale 
is confirmed by the $RV$s of the red peak and central depression of the H$\alpha$ line,
which exhibit a~period of 60~days after the removal of a~global trend.

Our long-term study shows that the object passed through several episodes of very different spectral 
patterns. Confirmation of the irregular behaviour of the object has important consequences. It 
suppresses signatures of the periodic spectral changes caused by orbital motion and complicates 
the confirmation (or rejection) of the binary nature of FS~CMa stars. The coincidence of the extrema of 
measured quantities points to an asymmetric disc around the HD~50138 system. The variations of 
line profiles that arise from an eccentric disc are affected by orbital motion of the gas, possible 
precession of the disc, as well as the changes of its structure. With regard to the real data distribution, 
the determination of the orbital and precession periods is very complicated. Moreover, the found periods 
differ for prograde and retrograde disc precession. These are the reasons why the binary nature of HD~50138 
could not be excluded based on our data. On the other hand, the irregular behaviour of HD~50138 can also 
be explained by a~new hypothesis about the nature of FS~CMa stars as post-merger systems \citep{Fuente15}.
Other ideas about the nature of FS~CMa stars do not seem to be able to explain all the observed properties 
by including only one or two phenomena and lead to complicated scenarios. 

Based on our observations and previously published results, we summarise the features that
have to be described by a~physically consistent model:
\renewcommand{\labelenumi}{\roman{enumi})}  
\begin{enumerate}[noitemsep,topsep=0pt,parsep=0pt,partopsep=0pt]
 \item rotating regions; 
 \item material infall;
 \item material outflow;
 \item episodic mass ejection;
 \item different behaviour in different epochs;
 \item constant brightness vs. strong line-profile variability;
 \item no regular periodicity;
 \item quasi-periodic behaviour.
\end{enumerate}\vspace{4pt}

HD~50138 is a~perfect laboratory for the study of stellar/binary evolution, and of various 
other physical phenomena such as excitation waves, (magneto-)hydrodynamical waves, and complex 
non-LTE effects. This star offers the unique opportunity to support the study of such 
complicated objects as $\eta$ Car because some of the observed properties are similar 
to the LBVs. Moreover, it is bright, and close enough to obtain good interferometry 
and spectra with high temporal and spectral resolution.

\begin{acknowledgements}
We would like to thank the referee for his comments, which helped to improve the paper.
We thank Steven N. Shore for his valuable and inspiring remarks and discussions. 
We also would like to thank Petr Harmanec for his advice
and Karen Bjorkman for providing her code Scargle.
We thank the Ritter observing team for taking, reducing, and sharing their data, and
Pavel Mayer, Jana Alexandra Nemravov\'a, Andrea Budovi\v{c}ov\'{a}, and Martin Netolick\'{y} for taking some of the 
Ond\v{r}ejov spectra. \\

This research was financially supported by the 
Czech-Mexican project
CONACYT/14/001 of the Academy of Sciences of the Czech Republic, UNCE 204020
and P209/10/0715 of Czech Science Foundation, and PAPIIT grant IN-100614. 
The Astronomical Institute of the Czech Academy of
Science is supported by the project RVO 67985815.
We acknowledge the use of the cluster {\sc Sunquake} maintained by
the Solar Department of Astronomical Institute of the Czech Academy of
Science in Ond\v{r}ejov. 
\end{acknowledgements}


\bibliographystyle{aa}   
\bibliography{26290}  

\begin{thebibliography}{69}
\expandafter\ifx\csname natexlab\endcsname\relax\def\natexlab#1{#1}\fi

\bibitem[{{Allen}(1973)}]{Allen73}
{Allen}, D.~A. 1973, \mnras, 161, 145

\bibitem[{{Alvarez} \& {Schuster}(1981)}]{Alvarez81}
{Alvarez}, M. \& {Schuster}, W.~J. 1981, \rmxaa, 6, 163

\bibitem[{{Andrillat} \& {Fehrenbach}(1982)}]{Andrillat82}
{Andrillat}, Y. \& {Fehrenbach}, C. 1982, \aaps, 48, 93

\bibitem[{{Andrillat} \& {Houziaux}(1972)}]{Andrillat1972}
{Andrillat}, Y. \& {Houziaux}, L. 1972, \apss, 18, 324

\bibitem[{{Andrillat} \& {Houziaux}(1991)}]{Andrillat91}
{Andrillat}, Y. \& {Houziaux}, L. 1991, \iaucirc, 5164, 3

\bibitem[{{Baines} {et~al.}(2006){Baines}, {Oudmaijer}, {Porter}, \&
  {Pozzo}}]{Baines06}
{Baines}, D., {Oudmaijer}, R.~D., {Porter}, J.~M., \& {Pozzo}, M. 2006, \mnras,
  367, 737

\bibitem[{{Bjorkman} {et~al.}(1998){Bjorkman}, {Miroshnichenko}, {Bjorkman},
  {Meade}, {Babler}, {Code}, {Anderson}, {Fox}, {Johnson}, {Weitenbeck},
  {Zellner}, \& {Lupie}}]{Bjorkman98}
{Bjorkman}, K.~S., {Miroshnichenko}, A.~S., {Bjorkman}, J.~E., {et~al.} 1998,
  \apj, 509, 904

\bibitem[{{Bjorkman} \& {Schulte-Ladbeck}(1994)}]{Bjorkman94}
{Bjorkman}, K.~S. \& {Schulte-Ladbeck}, R.~F. 1994, in Astronomical Society of
  the Pacific Conference Series, Vol.~62, The Nature and Evolutionary Status of
  Herbig Ae/Be Stars, ed. P.~S. {The}, M.~R. {Perez}, \& E.~P.~J. {van den
  Heuvel}, 74

\bibitem[{{Bopp}(1993)}]{Bopp93}
{Bopp}, B.~W. 1993, Information Bulletin on Variable Stars, 3834, 1

\bibitem[{{Borges Fernandes}(2010)}]{Borges10}
{Borges Fernandes}, M. 2010, in Revista Mexicana de Astronomia y Astrofisica,
  vol. 27, Vol.~38, Revista Mexicana de Astronomia y Astrofisica Conference
  Series, 98--99

\bibitem[{{Borges Fernandes} {et~al.}(2009){Borges Fernandes}, {Kraus},
  {Chesneau}, {Domiciano de Souza}, {de Ara{\'u}jo}, {Stee}, \&
  {Meilland}}]{Borges09}
{Borges Fernandes}, M., {Kraus}, M., {Chesneau}, O., {et~al.} 2009, \aap, 508,
  309

\bibitem[{{Borges Fernandes} {et~al.}(2012){Borges Fernandes}, {Kraus},
  {Nickeler}, {De Cat}, {Lampens}, {Pereira}, \& {Oksala}}]{Borges12}
{Borges Fernandes}, M., {Kraus}, M., {Nickeler}, D.~H., {et~al.} 2012, \aap,
  548, A13

\bibitem[{{Borges Fernandes} {et~al.}(2011){Borges Fernandes}, {Meilland},
  {Bendjoya}, {Domiciano de Souza}, {Niccolini}, {Chesneau}, {Millour},
  {Spang}, {Stee}, \& {Kraus}}]{Borges11}
{Borges Fernandes}, M., {Meilland}, A., {Bendjoya}, P., {et~al.} 2011, \aap,
  528, A20

\bibitem[{{Briot}(1981)}]{Briot81}
{Briot}, D. 1981, \aap, 103, 5

\bibitem[{{Carciofi} \& {Bjorkman}(2006)}]{Carciofi06}
{Carciofi}, A.~C. \& {Bjorkman}, J.~E. 2006, \apj, 639, 1081

\bibitem[{{Carciofi} \& {Bjorkman}(2008)}]{Carciofi08}
{Carciofi}, A.~C. \& {Bjorkman}, J.~E. 2008, \apj, 684, 1374

\bibitem[{{Corporon} \& {Lagrange}(1999)}]{Corporon99}
{Corporon}, P. \& {Lagrange}, A.-M. 1999, \aaps, 136, 429

\bibitem[{{Cur{\'e}}(2004)}]{Cure04}
{Cur{\'e}}, M. 2004, \apj, 614, 929

\bibitem[{{Dachs} {et~al.}(1992){Dachs}, {Hummel}, \& {Hanuschik}}]{Dachs92}
{Dachs}, J., {Hummel}, W., \& {Hanuschik}, R.~W. 1992, \aaps, 95, 437

\bibitem[{{de la Fuente} {et~al.}(2015){de la Fuente}, {Najarro}, {Trombley},
  {Davies}, \& {Figer}}]{Fuente15}
{de la Fuente}, D., {Najarro}, F., {Trombley}, C., {Davies}, B., \& {Figer},
  D.~F. 2015, \aap, 575, A10

\bibitem[{{de Winter} {et~al.}(2001){de Winter}, {van den Ancker}, {Maira},
  {Th{\'e}}, {Djie}, {Redondo}, {Eiroa}, \& {Molster}}]{Winter01}
{de Winter}, D., {van den Ancker}, M.~E., {Maira}, A., {et~al.} 2001, \aap,
  380, 609

\bibitem[{{Doazan}(1965)}]{Doazan65}
{Doazan}, V. 1965, Annales d'Astrophysique, 28, 1

\bibitem[{{Ellerbroek} {et~al.}(2015){Ellerbroek}, {Benisty}, {Kraus},
  {Perraut}, {Kluska}, {le Bouquin}, {Borges Fernandes}, {Domiciano de Souza},
  {Maaskant}, {Kaper}, {Tramper}, {Mourard}, {Tallon-Bosc}, {ten Brummelaar},
  {Sitko}, {Lynch}, \& {Russell}}]{Ellerbroek15}
{Ellerbroek}, L.~E., {Benisty}, M., {Kraus}, S., {et~al.} 2015, \aap, 573, A77

\bibitem[{{Foulkes} {et~al.}(2004){Foulkes}, {Haswell}, {Murray}, \&
  {Rolfe}}]{Foulkes04}
{Foulkes}, S.~B., {Haswell}, C.~A., {Murray}, J.~R., \& {Rolfe}, D.~J. 2004,
  \mnras, 349, 1179

\bibitem[{{Grady} {et~al.}(1996){Grady}, {Perez}, {Talavera}, {Bjorkman}, {de
  Winter}, {The}, {Molster}, {van den Ancker}, {Sitko}, {Morrison}, {Beaver},
  {McCollum}, \& {Castelaz}}]{Grady96}
{Grady}, C.~A., {Perez}, M.~R., {Talavera}, A., {et~al.} 1996, \aaps, 120, 157

\bibitem[{{Halbedel}(1991)}]{Halbedel91}
{Halbedel}, E.~M. 1991, Information Bulletin on Variable Stars, 3585, 1

\bibitem[{{Hanuschik}(1995)}]{Hanuschik95}
{Hanuschik}, R.~W. 1995, \aap, 295, 423

\bibitem[{{Harrington} \& {Kuhn}(2007)}]{Harrington07}
{Harrington}, D.~M. \& {Kuhn}, J.~R. 2007, \apjl, 667, L89

\bibitem[{{Houziaux}(1960)}]{Houziaux60}
{Houziaux}, L. 1960, Journal des Observateurs, 43, 217

\bibitem[{{Houziaux} \& {Andrillat}(1976)}]{Houziaux76}
{Houziaux}, L. \& {Andrillat}, Y. 1976, in IAU Symposium, Vol.~70, Be and Shell
  Stars, ed. A.~{Slettebak}, 87

\bibitem[{{Humason} \& {Merrill}(1921)}]{Humanson21}
{Humason}, M.~L. \& {Merrill}, P.~W. 1921, \pasp, 33, 112

\bibitem[{{Hutsemekers}(1985)}]{Hutsemekers85}
{Hutsemekers}, D. 1985, \aaps, 60, 373

\bibitem[{IAUS~70{, Conti}(1976)}]{zavorky_Conti}
IAUS~70{, Conti}, P.~S. 1976, in IAU Symposium, Vol.~70, Be and Shell Stars,
  ed. A.~{Slettebak}, 447

\bibitem[{{Jaschek} \& {Andrillat}(1998)}]{Jaschek98}
{Jaschek}, C. \& {Andrillat}, Y. 1998, \aaps, 128, 475

\bibitem[{{Kilkenny} {et~al.}(1985){Kilkenny}, {Whittet}, {Davies}, {Evans},
  {Bode}, {Robson}, \& {Banfield}}]{Kilkenny85}
{Kilkenny}, D., {Whittet}, D.~C.~B., {Davies}, J.~K., {et~al.} 1985, South
  African Astronomical Observatory Circular, 9, 55

\bibitem[{{Kor{\v c}{\'a}kov{\'a}} \& {Kub{\'a}t}(2005)}]{Korcakova05}
{Kor{\v c}{\'a}kov{\'a}}, D. \& {Kub{\'a}t}, J. 2005, \aap, 440, 715

\bibitem[{{Kuan} \& {Kuhi}(1975)}]{Kuan75}
{Kuan}, P. \& {Kuhi}, L.~V. 1975, \apj, 199, 148

\bibitem[{{Ku{\v c}erov{\'a}} {et~al.}(2013){Ku{\v c}erov{\'a}}, {Kor{\v
  c}{\'a}kov{\'a}}, {Polster}, {Wolf}, {Votruba}, {Kub{\'a}t}, {{\v S}koda},
  {{\v S}lechta}, \& {K{\v r}{\'{\i}}{\v z}ek}}]{Kucerova13}
{Ku{\v c}erov{\'a}}, B., {Kor{\v c}{\'a}kov{\'a}}, D., {Polster}, J., {et~al.}
  2013, \aap, 554, A143

\bibitem[{{Lamers} {et~al.}(1998){Lamers}, {Zickgraf}, {de Winter}, {Houziaux},
  \& {Zorec}}]{Lamers98}
{Lamers}, H.~J.~G.~L.~M., {Zickgraf}, F.-J., {de Winter}, D., {Houziaux}, L.,
  \& {Zorec}, J. 1998, \aap, 340, 117

\bibitem[{{Li} \& {McCray}(1992)}]{Li92}
{Li}, H. \& {McCray}, R. 1992, \apj, 387, 309

\bibitem[{{Marston} \& {McCollum}(2008)}]{Marston08}
{Marston}, A.~P. \& {McCollum}, B. 2008, \aap, 477, 193

\bibitem[{{Mel'nikov}(1997)}]{Melnikov97}
{Mel'nikov}, S.~Y. 1997, Astronomy Letters, 23, 799

\bibitem[{{Merrill}(1931)}]{Merrill31_ApJ}
{Merrill}, P.~W. 1931, \apj, 73, 348

\bibitem[{{Merrill}(1952)}]{Merrill52}
{Merrill}, P.~W. 1952, \apj, 116, 501

\bibitem[{{Merrill} {et~al.}(1925){Merrill}, {Humason}, \&
  {Burwell}}]{Merrill25}
{Merrill}, P.~W., {Humason}, M.~L., \& {Burwell}, C.~G. 1925, \apj, 61, 389

\bibitem[{{Miczaika}(1950)}]{Miczaika50}
{Miczaika}, G.~R. 1950, Astronomische Nachrichten, 279, 19

\bibitem[{{Miroshnichenko} \& {Zharikov}(2015)}]{Miroshnichenko_prehled15}
{Miroshnichenko}, A. \& {Zharikov}, S. 2015, http://arxiv.org/abs/1508.06298;
  submitted in EAS Publication Series; Physics Of Evolved Star, June 8-12 2015,
  Nice, France [\eprint[arXiv]{1508.06298}]

\bibitem[{{Miroshnichenko}(2007)}]{Miroshnichenko07-FS_CMa}
{Miroshnichenko}, A.~S. 2007, \apj, 667, 497

\bibitem[{{Miroshnichenko} {et~al.}(2015){Miroshnichenko}, {Zharikov},
  {Danford}, {Manset}, {Kor{\v c}{\'a}kov{\'a}}, {K{\v r}{\'{\i}}{\v c}ek},
  {{\v S}lechta}, {Omarov}, {Kusakin}, {Kuratov}, \& {Grankin}}]{MWC728}
{Miroshnichenko}, A.~S., {Zharikov}, S.~V., {Danford}, S., {et~al.} 2015, \apj,
  809, 129

\bibitem[{{Miroshnichenko} {et~al.}(2013){Miroshnichenko}, {Zharikov},
  {Manset}, {Rossi}, \& {Polcaro}}]{Miroshnichenko13}
{Miroshnichenko}, A.~S., {Zharikov}, S.~V., {Manset}, N., {Rossi}, C., \&
  {Polcaro}, V.~F. 2013, Central European Astrophysical Bulletin, 37, 57

\bibitem[{{Monnier} {et~al.}(2009){Monnier}, {Tuthill}, {Ireland}, {Cohen},
  {Tannirkulam}, \& {Perrin}}]{Monnier09}
{Monnier}, J.~D., {Tuthill}, P.~G., {Ireland}, M., {et~al.} 2009, \apj, 700,
  491

\bibitem[{{Oudmaijer} \& {Drew}(1999)}]{Oudmaijer99}
{Oudmaijer}, R.~D. \& {Drew}, J.~E. 1999, \mnras, 305, 166

\bibitem[{{Pogodin}(1997)}]{Pogodin97}
{Pogodin}, M.~A. 1997, \aap, 317, 185

\bibitem[{{Polster} {et~al.}(2012){Polster}, {Kor{\v c}{\'a}kov{\'a}},
  {Votruba}, {{\v S}koda}, {{\v S}lechta}, {Ku{\v c}erov{\'a}}, \&
  {Kub{\'a}t}}]{Polster12}
{Polster}, J., {Kor{\v c}{\'a}kov{\'a}}, D., {Votruba}, V., {et~al.} 2012,
  \aap, 542, A57

\bibitem[{{Press} \& {Teukolsky}(1988)}]{Press}
{Press}, W.~H. \& {Teukolsky}, S.~A. 1988, Computers in Physics, 2, 77

\bibitem[{{Pych}(2004)}]{Pych}
{Pych}, W. 2004, \pasp, 116, 148

\bibitem[{{Reg{\'a}ly} {et~al.}(2011){Reg{\'a}ly}, {S{\'a}ndor}, {Dullemond},
  \& {Kiss}}]{Regaly11}
{Reg{\'a}ly}, Z., {S{\'a}ndor}, Z., {Dullemond}, C.~P., \& {Kiss}, L.~L. 2011,
  \aap, 528, A93

\bibitem[{{Rivinius}(2005)}]{Rivinius_hab}
{Rivinius}, T. 2005, Habilitationsschrift, Ruprecht-Karls-Universi\"{a}t,
  Heidelberg

\bibitem[{{Savage} {et~al.}(1978){Savage}, {Wesselius}, {Swings}, \&
  {The}}]{Savage78}
{Savage}, B.~D., {Wesselius}, P.~R., {Swings}, J.~P., \& {The}, P.~S. 1978,
  \apj, 224, 149

\bibitem[{{Steffen}(1990)}]{Steffen}
{Steffen}, M. 1990, \aap, 239, 443

\bibitem[{{Stellingwerf}(1978)}]{Stellingwerf78}
{Stellingwerf}, R.~F. 1978, \apj, 224, 953

\bibitem[{{Suzuki} {et~al.}(2006){Suzuki}, {Yan}, {Lazarian}, \&
  {Cassinelli}}]{Suzuki06}
{Suzuki}, T.~K., {Yan}, H., {Lazarian}, A., \& {Cassinelli}, J.~P. 2006, \apj,
  640, 1005

\bibitem[{{Swings} \& {Allen}(1971)}]{Swings71}
{Swings}, J.~P. \& {Allen}, D.~A. 1971, \apjl, 167, L41

\bibitem[{{van Rensbergen} {et~al.}(2008){van Rensbergen}, {De Greve}, {De
  Loore}, \& {Mennekens}}]{Rensbergen08}
{van Rensbergen}, W., {De Greve}, J.~P., {De Loore}, C., \& {Mennekens}, N.
  2008, \aap, 487, 1129

\bibitem[{{Vink} {et~al.}(2002){Vink}, {Drew}, {Harries}, \&
  {Oudmaijer}}]{Vink02}
{Vink}, J.~S., {Drew}, J.~E., {Harries}, T.~J., \& {Oudmaijer}, R.~D. 2002,
  \mnras, 337, 356

\bibitem[{{Vollmann} \& {Eversberg}(2006)}]{Vollmann}
{Vollmann}, K. \& {Eversberg}, T. 2006, AN, 327, 862

\bibitem[{{Votruba} {et~al.}(2007){Votruba}, {Feldmeier}, {Kub{\'a}t}, \&
  {R{\"a}tzel}}]{Votruba07}
{Votruba}, V., {Feldmeier}, A., {Kub{\'a}t}, J., \& {R{\"a}tzel}, D. 2007,
  \aap, 474, 549

\bibitem[{{Yudin} \& {Evans}(1998)}]{Yudin98}
{Yudin}, R.~V. \& {Evans}, A. 1998, \aaps, 131, 401

\bibitem[{{Zsarg{\'o}} {et~al.}(2008){Zsarg{\'o}}, {Hillier}, \&
  {Georgiev}}]{Zsargo08}
{Zsarg{\'o}}, J., {Hillier}, D.~J., \& {Georgiev}, L.~N. 2008, \aap, 478, 543

\end{thebibliography}

\clearpage

\end{document}